\documentclass[prd,%
preprint,%
tightenlines,superscriptaddress,showpacs,%
nofootinbib,eqsecnum,amsfonts,amsmath]{revtex4}
\usepackage{bm}
\usepackage{graphicx}

\begin{document}

\title{Dimensional regularization of the third post-Newtonian
dynamics\\
of point particles in harmonic coordinates}

\date{November 17, 2003}

\author{Luc Blanchet} \email{blanchet@iap.fr}
\affiliation{${\mathcal{G}}{\mathbb{R}}
\varepsilon{\mathbb{C}}{\mathcal{O}}$, FRE 2435-CNRS, Institut
d'Astrophysique de Paris, 98$^{\text{bis}}$ boulevard Arago, F-75014
Paris, France}

\author{Thibault Damour}
\email{damour@ihes.fr}
\affiliation{Institut des Hautes \'Etudes Scientifiques,
35 route de Chartres, F-91440 Bures-sur-Yvette, France}

\author{Gilles \surname{Esposito-Far\`ese}} \email{gef@iap.fr}
\affiliation{${\mathcal{G}}{\mathbb{R}}
\varepsilon{\mathbb{C}}{\mathcal{O}}$, FRE 2435-CNRS, Institut
d'Astrophysique de Paris, 98$^{\text{bis}}$ boulevard Arago, F-75014
Paris, France}

\begin{abstract}
Dimensional regularization is used to derive the equations of motion
of two point masses in harmonic coordinates. At the third
post-Newtonian (3PN) approximation, it is found that the dimensionally
regularized equations of motion contain a pole part [proportional to
$(d-3)^{-1}$] which diverges as the space dimension $d$ tends to
$3$. It is proven that the pole part can be renormalized away by
introducing suitable shifts of the two world-lines representing the
point masses, and that the same shifts renormalize away the pole part
of the ``bulk'' metric tensor $g_{\mu\nu} (x^{\lambda})$. The ensuing,
finite renormalized equations of motion are then found to belong to
the general parametric equations of motion derived by an extended
Hadamard regularization method, and to uniquely determine the
3PN ambiguity parameter $\lambda$ to be: $\lambda = - 1987/3080$. This
value is fully consistent with the recent determination of the
equivalent 3PN ``static ambiguity'' parameter, $\omega_s = 0$, by a
dimensional-regularization derivation of the Hamiltonian in
Arnowitt-Deser-Misner coordinates. Our work provides a new, powerful
check of the consistency of the dimensional regularization method
within the context of the classical gravitational interaction of point
particles.
\end{abstract}


\preprint{gr-qc/0311052}

\maketitle

\section{Introduction}
\label{Intro}
\subsection{Relation to previous works}
\label{Relation}

The problem of motion, one of the cardinal problems of Einstein's
gravitation theory, has received continuous attention over the years.
The early, classic works of Lorentz-Droste, Eddington-Clark,
Einstein-Infeld-Hoffmann, Fock, Papapetrou and others led to a good
understanding of the equations of motion of $N$ bodies at the first
post-Newtonian (1PN) approximation\,\footnote{As usual the $n$PN order
refers to the terms of order $1/c^{2n}$ in the equations of motion.}
(see, \textit{e.g.}, \cite{Damour:1986ny} for a general review of the
problem of motion). In the 1970's, an important series of works by a
Japanese group \cite{Ohta:1973,Ohta:1974a,Ohta:1974b} led to a nearly
complete control of the problem of motion at the second post-Newtonian
(2PN) approximation. Then, in the early 80's, motivated by the
observation of secular orbital effects in the Hulse-Taylor binary
pulsar PSR1913+16, several groups solved the two-body problem at the
2.5PN level (while completing on the way the derivation of the 2PN
dynamics) \cite{Bel:1981be,Damour:1981bh,Damour:1982,Damour:1982wm,%
Schafer:1985rd,Schafer:1986,Damour:1985,Kopejkin:1985} (for more recent
work on the 2.5PN dynamics see
\cite{Blanchet:1998vx,Itoh:2001np,Pati:2002ux}).

In the late 90's, motivated by the aim of deriving high-accuracy
templates for the data analysis of the upcoming international network
of interferometric gravitational-wave detectors, two groups embarked on
the derivation of the equations of motion at the third post-Newtonian
(3PN) level. One group used the Arnowitt-Deser-Misner (ADM) Hamiltonian
approach
\cite{Jaranowski:1998ky,Jaranowski:1999ye,Damour:1999cr,Damour:2000kk,%
Damour:2000ni} and worked in a corresponding ADM-type coordinate
system. Another group used a direct post-Newtonian iteration of the
equations of motion in harmonic coordinates
\cite{Blanchet:2000nv,Blanchet:2000ub,Blanchet:2000nu,Blanchet:2000cw,%
deAndrade:2000gf,Blanchet:2002mb}. The end results of these two
approaches have been proved to be physically equivalent
\cite{Damour:2000ni,deAndrade:2000gf}. However, both approaches, even
after exploiting all symmetries and pushing their
Hadamard-regularization-based methods to the maximum of their
possibilities, left undetermined \textit{one and only one}
dimensionless parameter: $\omega_s$ in the ADM approach and $\lambda$
in the harmonic-coordinates one. The unknown parameters in both
approaches are related by
\begin{equation}
\lambda=-\frac{3}{11}\omega_s-\frac{1987}{3080}\,,
\label{lambdaomegas}
\end{equation}
as was deduced from the comparison between the invariant energy
functions for circular orbits in the two approaches
\cite{Blanchet:2000nv}, and from two independent proofs of the
equivalence between the two formalisms for general orbits
\cite{Damour:2000ni,deAndrade:2000gf}. The appearance of one (and only
one) unknown parameter in the equations of motion is quite striking;
it is related with the choice of the regularization method used to
cure the self-field divergencies of point particles. Both lines of
works
\cite{Jaranowski:1998ky,Jaranowski:1999ye,Damour:1999cr,Damour:2000kk,%
Damour:2000ni} and
\cite{Blanchet:2000nv,Blanchet:2000ub,Blanchet:2000nu,Blanchet:2000cw,
deAndrade:2000gf,Blanchet:2002mb} regularized the self-field
divergencies by some version of the Hadamard regularization
method. The second line of work defined an extended version of the
Hadamard regularization \cite{Blanchet:2000nu,Blanchet:2000cw}, which
permitted a self-consistent derivation of the 3PN equations of motion,
but its use still allowed for the presence of arbitrary parameters in
the final equations. On the other hand, the Hadamard regularization
also yielded some arbitrary parameters in the gravitational radiation
field of point-particle binaries at the 3PN order, the most important
of which being the parameter $\theta$ entering the binary's energy
flux \cite{Blanchet:2001aw,Blanchet:2001ax}.

Let us notice that the regularization (when dealing with point
particles) and the renormalization (needed when dealing either with
point particles or with extended bodies\,\footnote{In the case of
extended compact bodies, the gravitational self-energy (divergent when
the radii of the bodies formally tend to zero) must be renormalized
into the definition of the mass.}) of self-field effects has
recurrently plagued the general relativistic problem of motion. Even at
the 1PN level, early works often contained incorrect treatments of
self-field effects (see, \textit{e.g.}, Section 6.14 of
\cite{Damour:1986ny} for a review). At the 2PN level, the self-field
divergencies are more severe than at the 1PN level. For instance, they
caused Ref. \cite{Ohta:1974a} to incorrectly evaluate the ``static''
(\textit{i.e.}, velocity-independent) part of the 2PN two-body
Hamiltonian. The first correct and complete evaluation of the 2PN
dynamics has been obtained by using the Riesz analytical continuation
method \cite{Riesz}. (See \cite{Damour:1982wm,Damour:1985} for a
detailed discussion of the evaluation of the static 2PN two-body
Hamiltonian.) In brief, the Riesz analytical continuation method
consists of replacing the problematic delta-function stress-energy
tensor of a set of point particles $y_a^{\mu} (s_a)$,
\begin{equation}
T^{\mu\nu} (x) = \sum_a m_a \, c^2 \int ds_a \,
\frac{dy_a^{\mu}}{ds_a} \, \frac{dy_a^{\nu}}{ds_a} \,
[-g(y_a)]^{-1/2} \,
\delta^{(4)} (x^{\lambda} - y_a^{\lambda} (s_a))\,,
\label{Tdelta4}
\end{equation}
[where $ds_a^2 = -g_{\mu\nu} (y_a^{\lambda}) \, dy_a^{\mu} \,
dy_a^{\nu}$, $g \equiv \det g_{\mu\nu}$] by an auxiliary, smoother
source
\begin{equation}
T_A^{\mu\nu} (x) = \sum_a m_a \, c^2 \int ds_a \,
\frac{dy_a^{\mu}}{ds_a} \, \frac{dy_a^{\nu}}{ds_a} \, [-g(y_a)]^{-1/2}
\, Z_A^{(4)} (x^{\lambda} - y_a^{\lambda} (s_a))\,.
\label{TRiesz}
\end{equation}
[Actually, in the implementation of \cite{Damour:1982wm}, one works
with ${\cal T}^{\mu\nu} (x) \equiv \vert g(x)\vert\, T^{\mu\nu} (x)$.]
In Eq.~(\ref{TRiesz}) the four-dimensional delta function entering
Eq.~(\ref{Tdelta4}) has been replaced by the
Lorentzian\,\footnote{$Z_A^{(d+1)}$ is the Lorentzian version of the
Euclidean kernel $\delta_{\alpha}^{(d)}$ discussed in Appendix
\ref{Formulae}.} four-dimensional Riesz kernel $Z_A^{(4)} (x-y)$, which
depends on the complex number $A$. When the real part of $A$ is large
enough the source $T_A^{\mu\nu} (x)$ is an ordinary function of
$x^{\mu}$, which is smooth enough to lead to a well-defined iteration
of the harmonically relaxed Einstein field equations, involving no
divergent integrals linked to the behavior of the integrands when
$x^{\mu} \rightarrow y_a^{\mu}$. One then analytically continues $A$
down to $0$, where the kernel $Z_A^{(4)} (x-y)$ tends to $\delta^{(4)}
(x-y)$. The important point is that it has been shown
\cite{Damour:1982wm} that all the integrals appearing in the 2PN
equations of motion are meromorphic functions of $A$ which admit a
smooth continuation at $A = 0$ (without poles). It was also shown there
that the formal construction based on (\ref{TRiesz}) does generate, at
the 2.5PN level, the metric and equations of motion of $N$ ``compact''
bodies (\textit{i.e.}, bodies with radii comparable to their
Schwarzschild radii).

The Riesz analytic continuation method just sketched works within a
normal 4-dimensional space-time (as recalled by the superscript $(4)$
in (\ref{TRiesz})). However, it was mentioned in \cite{Damour:1980}
that the same final result (at the 2.5PN level) is obtained by
replacing $Z_A^{(4)} (x-y)$ by $Z_0^{(4-A)} (x-y) \equiv
\delta^{(4-A)} (x-y)$, \textit{i.e.}, by formally considering
delta-function sources in a space-time of complex dimension $4-A$. In
other words, at the 2.5PN level, the Riesz analytic continuation
method is equivalent to the \textit{dimensional regularization}
method.\footnote{Dimensional regularization was invented as a mean to
preserve the \textit{gauge symmetry} of perturbative \textit{quantum}
gauge theories \cite{tHooft,Bollini,Breitenlohner,Collins}. Our basic
problem here is to respect the gauge symmetry associated with the
\textit{diffeomorphism invariance} of the \textit{classical} general
relativistic description of interacting point masses.} However, it was
also mentioned at the time \cite{Damour:1982wm} that the
generalization of Riesz analytic continuation beyond the 2.5PN level
did not look straightforward because of the appearance of poles,
proportional to $A^{-1}$, at the 3PN level (when using harmonic
coordinates).

Recently, Damour, Jaranowski and Sch\"afer \cite{Damour:2001bu} showed
how to use dimensional regularization within the ADM canonical
formalism. They found that the reduced Hamiltonian describing the
dynamics of two point masses in space-time dimension $D \equiv d+1$
was \textit{finite} (no pole part) as $d \rightarrow 3$. They also
found that the unique 3PN Hamiltonian defined by the analytic
continuation of $d$ towards 3 had two properties: (i) the
velocity-dependent terms had the unique structure compatible with
global Poincar\'e invariance,\footnote{Thus the ``kinetic ambiguity''
parameter $\omega_k$, originally introduced in the ADM approach
\cite{Jaranowski:1998ky,Jaranowski:1999ye}, takes the unique value
$\omega_k = \frac{41}{24}$. This value was obtained in
\cite{Blanchet:2000nv} using the result for the binary energy function
in the case of circular orbits, as calculated in the
harmonic-coordinates formalism, and also directly from the requirement
of Poincar\'e invariance in the ADM formalism \cite{Damour:2000kk}.}
and (ii) the velocity-independent (``static'') terms led to an
unambiguous determination of the unknown ADM parameter $\omega_s$,
namely
\begin{equation}
\omega_s^\text{dim. reg. ADM}=0\,.
\label{omegas}
\end{equation}
The fact that the dimensionally regularized 3PN ADM Hamiltonian ends
up being globally Poincar\'e invariant is a confirmation of the
consistency of dimensional regularization, because this symmetry is
not at all manifest within the ADM approach which uses a
space-plus-time split from the start. By contrast, the global
Poincar\'e symmetry is manifest in harmonic coordinates, and indeed
the 3PN harmonic-coordinates equations of motion derived in
\cite{Blanchet:2000nv,Blanchet:2000ub} were found to be manifestly
Poincar\'e invariant.

\subsection{Method and main results}
\label{Method}
In the present paper, we shall show how to implement dimensional
regularization (henceforth often abbreviated as ``dim. reg.'' or even
``dr'') in the computation of the equations of motion in harmonic
coordinates, \textit{i.e.}, following the same iterative post-Newtonian
formalism as in
Refs.~\cite{Blanchet:1998vx,Blanchet:2000nv,Blanchet:2000ub}. Similarly
to the ADM calculation of Ref. \cite{Damour:2001bu}, our strategy will
essentially consist of computing the \textit{difference} between the
$d$-dimensional result and the 3-dimensional one
\cite{Blanchet:2000nv,Blanchet:2000ub} corresponding to Hadamard
regularization. This difference is computed in the form of a Laurent
expansion in $\varepsilon \equiv d-3$, where $d$ denotes the spatial
dimension. The main reason for computing the $\varepsilon$-expansion of
the difference is that it depends only on the singular behavior of
various metric coefficients in the vicinity of the point particles, so
that the functions involved in the delicate divergent integrals can all
be computed in $d$ dimensions in the form of local expansions in powers
of $r_1$ or $r_2$ (where $r_a \equiv \vert\mathbf{x} - \mathbf{y}_a
\vert$; $\mathbf{y}_a$, $a=1,2$, denoting the locations of the two
point masses). Dimensional regularization as we use it here can then be
seen as a powerful argument for completing the 3-dimensional
Hadamard-regularization results of
\cite{Blanchet:2000nv,Blanchet:2000ub} and fixing the value of the
unknown parameter. We leave to future work the task of an exact
calculation of the $d$-dimensional equations of motion, instead of the
calculation of the first few terms in a Laurent expansion in
$\varepsilon$ around $d=3$, as done here. The first step towards such a
calculation is taken in Appendix \ref{Littleg}, where we give the
explicit expression of the basic quadratically non-linear Green
function $\textsl{g}(\mathbf{x} , \mathbf{y}_1 , \mathbf{y}_2)$ in $d$
dimensions.

The detailed way of computing the difference between dim. reg. and
Hadamard's reg. will turn out to be significantly more intricate than
in the ADM case. This added complexity has several sources. A first
source of complexity is that the harmonic-gauge $d$-dimensional
calculation will be seen to contain (as anticipated long ago
\cite{Damour:1982wm}) poles, proportional to $(d-3)^{-1}$, by contrast
to the ADM calculation which is finite as $d \rightarrow 3$. A second
source of complexity is that the end results \cite{Blanchet:2000ub}
for the 3-dimensional 3PN equations of motion have been derived using
systematically an \textit{extended} version of the Hadamard
regularization method, incorporating both a generalized theory of
singular pseudo-functions and their associated (generalized)
distributional derivatives \cite{Blanchet:2000nu}, and an improved
definition of the finite part as $\mathbf{x} \rightarrow
\mathbf{y}_1$, say $[F]_1$, of a singular function $F(\mathbf{x} ,
\mathbf{y}_1 , \mathbf{y}_2)$, designed so as to respect the global
Poincar\'e symmetry of the problem \cite{Blanchet:2000cw}. We shall
then find it technically convenient to subtract the various
contributions to the end results of \cite{Blanchet:2000ub} which arose
because of the specific use of the extended Hadamard regularization
methods of \cite{Blanchet:2000nu,Blanchet:2000cw} before considering
the difference with the $d$-dimensional result. A third source of
added complexity (with respect to the ADM case\,\footnote{The specific
form of the 3PN ADM Hamiltonian $H$ derived in \cite{Damour:1999cr}
and used (in its $d$-dimensional generalization) in
\cite{Damour:2001bu} was written, on purpose, in a way which does not
involve any hidden distributional terms (the only delta-function
contributions it contains being explicit contact terms $F(\mathbf{x})
\, \delta^{(3)}_a$). This allowed one to estimate the difference
between the $d$-dimensional Hamiltonian $H^{(d)}$ and the
Hadamard-regularized 3-dimensional one Hr$[H^{(3)}]$ without having to
worry about distributional derivatives. However, as a check on the
consistency of dim. reg., the authors of \cite{Damour:2001bu} did
perform another calculation of $H$ based on a starting form of the
Hamiltonian which involved hidden distributional terms, with the same
final result.}) comes from the presence in the harmonic-gauge
integrals we shall evaluate of ``hidden-distributional'' terms in the
integrands. By hidden distributional terms we mean terms proportional
to the second spatial derivatives of the Poisson kernel $\Delta^{-1}
\, \delta^{(d)}_a \propto r_a^{2-d}$, or to the fourth spatial
derivatives of the iterated Poisson kernel $\Delta^{-2} \,
\delta^{(d)}_a \propto r_a^{4-d}$. Such terms, $\partial_{ij} \,
r_a^{2-d}$ or $\partial_{ijkl} \, r_a^{4-d}$, considered as Schwartz
distributional derivatives \cite{Schwartz}, contain pieces
proportional to the delta function $\delta^{(d)}_a$, which need to be
treated with care. The generalized distributional derivative defined
in \cite{Blanchet:2000nu}, and used to compute the end results of
\cite{Blanchet:2000ub}, led to an improved way, compared to the normal
Schwartz distributional derivative, of evaluating contributions coming
{}from the product of a singular function and a derivative of the
type $\partial_{ij} \, r_a^{-1}$ or $\partial_{ijkl} \, r_a$, and
more generally of any derivatives of singular functions in a certain
class. We shall find it convenient to subtract these additional
non-Schwartzian contributions to the 3PN equations of motion before
applying dimensional regularization. However, we shall note at the end
that dim. reg. automatically incorporates all of these non-Schwartzian
contributions.

A fourth, but minor, source of complexity concerns the dependence of
the end results of \cite{Blanchet:2000ub} for the 3PN acceleration of
the first particle (label $a=1$), say $\mathbf{a}_1^\text{BF}$, on two
arbitrary length scales $r'_1$ and $r'_2$, and on the ``ambiguity''
parameter $\lambda$. Explicitly, we define
\begin{equation}
\mathbf{a}_1^\text{BF} \bigl[\lambda;r_1',r_2'\bigr] \equiv
\text{R.H.S. of Eq.~(7.16) in Ref. \cite{Blanchet:2000ub}}\,.
\label{a1BF}
\end{equation}
Here the acceleration is considered as a function of the two masses
$m_1$ and $m_2$, the relative distance $\mathbf{y}_1-\mathbf{y}_2\equiv
r_{12}\mathbf{n}_{12}$ (where $\mathbf{n}_{12}$ is the unit vector
directed from particle 2 to particle 1), the two coordinate velocities
$\mathbf{v}_1$ and $\mathbf{v}_2$, and also, as emphasized in
(\ref{a1BF}), the parameter $\lambda$ as well as two regularization
length scales $r_1'$ and $r_2'$. The latter length scales enter the
equations of motion at the 3PN level through the logarithms $\ln
(r_{12}/r_1')$ and $\ln (r_{12}/r_2')$. They come from the
regularization as the field point $\mathbf{x}'$ tends to $\mathbf{y}_1$
or $\mathbf{y}_2$ of Poisson-type integrals (see Section
\ref{HadamardPoisson} below). The length scales $r_1'$, $r_2'$ are
``pure gauge'' in the sense that they can be removed by the effect
induced on the world-lines of a coordinate transformation of the bulk
metric \cite{Blanchet:2000ub}. On the other hand, the dimensionless
parameter $\lambda$ entering the final result (\ref{a1BF}) corresponds
to genuine physical effects. It was introduced by requiring that the
3PN equations of motion admit a conserved energy (and more generally be
derivable from a Lagrangian). This extra requirement imposed
\textit{two relations} between the two length scales $r'_1$, $r'_2$ and
two other length scales $s_1$, $s_2$ entering originally into the
formalism, namely the constants $s_1$ and $s_2$ parametrizing the
Hadamard partie finie of an integral as defined by Eq. (\ref{Pf})
below. These relations were found to be of the form
\begin{equation}
\ln\Bigl(\frac{r_2'}{s_2}\Bigr)=\frac{159}{308}+\lambda
\frac{m_1+m_2}{m_2}~~\text{and $1\leftrightarrow 2$}\,,
\label{lnr2s2}
\end{equation}
where the so introduced \textit{single} dimensionless parameter
$\lambda$ has been proved to be a purely numerical coefficient
(independent of the two masses). When estimating the difference
between dim. reg. and Hadamard reg. it will be convenient to insert
Eq. (\ref{lnr2s2}) into (\ref{a1BF}) and to reexpress the acceleration
of particle 1 in terms of the \textit{original} regularization length
scales entering the Hadamard regularization of $\mathbf{a}_1$, which
were in fact $r'_1$ and $s_2$. Thus we can consider alternatively
\begin{equation}
\mathbf{a}_1^\text{BF} [r'_1 , s_2] \equiv \mathbf{a}_1^\text{BF}
\bigl[\lambda ; r_1', r'_2 (s_2 , \lambda)\bigr]~~\text{and
$1\leftrightarrow 2$}\,,
\label{Foncta1BF}
\end{equation}
where the regularization constants are subject to the constraints
(\ref{lnr2s2}) [we will then check that the $\lambda$-dependence on the
R.H.S. of (\ref{Foncta1BF}) disappears when using Eq.~(\ref{lnr2s2})
to replace $r'_2$ as a function of $s_2$ and $\lambda$].

Our strategy will consist of \textit{two steps}. The \textit{first
step} consists of subtracting all the extra contributions to
Eq.~(\ref{a1BF}), or equivalently Eq.~(\ref{Foncta1BF}), which were
specific consequences of the extended Hadamard regularization defined
in \cite{Blanchet:2000nu,Blanchet:2000cw}. As we shall detail below,
there are \textit{seven} such extra contributions $\delta^A
\mathbf{a}_1$, $A = 1, \cdots , 7$. Essentially, subtracting these
contributions boils down to estimating the value of $\mathbf{a}_1$ that
would be obtained by using a ``pure'' Hadamard regularization, together
with Schwartz distributional derivatives. Such a ``pure
Hadamard-Schwartz'' (pHS) acceleration was in fact essentially the
result of the first stage of the calculation of $\mathbf{a}_1$, as
reported in the (unpublished) thesis \cite{FayeThesis}. It is given by
\begin{equation}
\mathbf{a}_1^\text{pHS}\bigl[r_1',s_2\bigr]=\mathbf{a}_1^\text{BF}
[r'_1 , s_2]-\sum_{A=1}^7\delta^A\mathbf{a}_1\,.
\label{accpH}
\end{equation}
The \textit{second step} of our method consists of evaluating the
Laurent expansion, in powers of $\varepsilon = d-3$, of the
\textit{difference} between the $d$-dimensional and the pure
Hadamard-Schwartz (3-dimensional) computations of the acceleration
$\mathbf{a}_1$. We shall see that this difference makes a contribution
only when a term generates a \textit{pole} $\sim 1/\varepsilon$, in
which case dim. reg. adds an extra contribution, made of the pole and
the finite part associated with the pole [we consistently neglect all
terms $\mathcal{O}(\varepsilon)$]. One must then be especially wary of
combinations of terms whose pole parts finally cancel (``cancelled
poles'') but whose dimensionally regularized finite parts generally do
not, and must be evaluated with care. We denote the above defined
difference
\begin{equation}
\mathcal{D}\mathbf{a}_1=\mathcal{D}\mathbf{a}_1
\bigl[\varepsilon,\ell_0;r_1',s_2\bigr]\equiv\mathcal{D}\mathbf{a}_1
\bigl[\varepsilon,\ell_0;\lambda ; r_1', r'_2\bigr]\,.
\label{deltaacc}
\end{equation}
It depends both on the Hadamard regularization scales $r_1'$ and $s_2$
(or equivalently on $\lambda$ and $r_1'$, $r_2'$) and on the
regularizing parameters of dimensional regularization, namely
$\varepsilon$ and the characteristic length $\ell_0$ associated with
dim. reg. and introduced in Eq.~(\ref{l0}) below. We shall explain in
detail below the techniques we have used to compute
$\mathcal{D}\mathbf{a}_1$ (see Section \ref{Difference}). Finally, our
main result will be the explicit computation of the
$\varepsilon$-expansion of the dim. reg. acceleration as
\begin{equation}
\mathbf{a}_1^\text{dr} [\varepsilon , \ell_0] = \mathbf{a}_1^\text{pHS}
[r'_1 , s_2] + \mathcal{D}\mathbf{a}_1 [\varepsilon , \ell_0 ; r'_1 ,
s_2]\,.
\label{a1DimReg}
\end{equation}
With this result in hands, we shall prove (in Section
\ref{Renormalise}) two theorems.

\vskip 1pc
\noindent
\textbf{Theorem 1}\quad
\textit{The pole part $\propto 1/\varepsilon$ of the
dimensionally-regularized acceleration (\ref{a1DimReg}), as well as of
the metric field $g_{\mu\nu}(x)$ outside the particles, can be
re-absorbed (\textit{i.e.}, renormalized away) into some shifts of the
two ``bare'' world-lines: $\mathbf{y}_a \rightarrow
\mathbf{y}_a+\bm{\xi}_a$, with, say, $\bm{\xi}_a \propto 1/\varepsilon$
(``minimal subtraction''; MS), so that the result, expressed in terms
of the ``dressed'' quantities, is finite when $\varepsilon\rightarrow
0$.}

\vskip 1pc
\noindent
The situation in harmonic coordinates is to be contrasted with the
calculation in ADM-type coordinates within the Hamiltonian formalism
\cite{Damour:2001bu}, where it was shown that all pole parts directly
cancel out in the total 3PN Hamiltonian (no shifts of the world-lines
were needed). The central result of the paper is then as follows.

\vskip 1pc
\noindent
\textbf{Theorem 2}\quad
\textit{The ``renormalized'' (finite) dimensionally-regularized
acceleration is physically equivalent to the
extended-Hadamard-regularized acceleration (end result of
Ref.~\cite{Blanchet:2000ub}), in the sense that there exist some shift
vectors $\bm{\xi}_1 (\varepsilon , \ell_0 ; r'_1)$ and $\bm{\xi}_2
(\varepsilon , \ell_0 ; r'_2)$, such that
\begin{equation}
\mathbf{a}_1^\mathrm{BF} [\lambda, r'_1 , r'_2] =
\lim_{\varepsilon\rightarrow
0} \, \bigl[\mathbf{a}_1^\mathrm{dr} [\varepsilon , \ell_0] +
\delta_{\bm{\xi} (\varepsilon , \ell_0 ; r'_1 , r'_2)} \,
\mathbf{a}_1 \bigr]
\label{eta}
\end{equation}
(where $\delta_{\bm{\xi}} \, \mathbf{a}_1$ denotes the effect of the
shifts on the acceleration\,\footnote{When working at the level of the
equations of motion (not considering the metric outside the
world-lines), the effect of shifts can be seen as being induced by a
coordinate transformation of the bulk metric as in
Ref.~\cite{Blanchet:2000ub} (we comment on this point in Section
\ref{Shift} below).}), if and only if the heretofore unknown parameter
$\lambda$ entering the harmonic-coordinates equations of motion takes
the value
\begin{equation}
\lambda^\mathrm{dim. ~reg. ~harmonic} = -\frac{1987}{3080}\,.
\label{lambda}
\end{equation}}

\noindent
The precise shifts $\bm{\xi}_a (\varepsilon)$ needed in Theorem~2
involve not only a pole contribution $\propto 1/\varepsilon$, which
defines the ``minimal'' (MS) shifts considered in Theorem~1, but also a
finite contribution when $\varepsilon\rightarrow 0$. Their explicit
expressions read:
\begin{equation}
\bm{\xi}_1=\frac{11}{3}\frac{G_N^2\,m_1^2}{c^6}\left[
\frac{1}{\varepsilon}-2\ln\left(
\frac{r'_1\overline{q}^{1/2}}{\ell_0}\right)
-\frac{327}{1540}\right] \mathbf{a}_{N1}~~\text{and}~~1\leftrightarrow
2\,,
\end{equation}
where $G_N$ is the usual Newton's constant [see Eq. (\ref{l0}) below],
$\mathbf{a}_{N1}$ denotes the acceleration of the particle 1 (in $d$
dimensions) at the Newtonian level, and $\overline{q}\equiv 4\pi e^C$
depends on the Euler constant $C=0.577\cdots$. [The detailed proofs
of Theorems~1 and 2 will consist of our investigations
expounded in the successive sections of the paper, and will be
completed at the end of Sections~\ref{Shift} and
\ref{LambdaDetermined} respectively, taking also into account the
results of Section~\ref{Kinetic}.]

Notice that an alternative way of presenting our central result is to
say that, in fact, each choice of specific renormalization prescription
(within dim. reg.), such as ``minimal subtraction'' as assumed in
Theorem~1 for conceptual simplicity,\footnote{However, for
technical simplicity we shall prefer in Section \ref{LambdaDetermined}
below to use a modified minimal subtraction that we shall denote
$\bm{\xi}_{\overline{\text{MS}}}$.} leads to renormalized equations of
motion which depend only on the dim. reg. characteristic length scale
$\ell_0$ through the logarithm $\ln (r_{12}/\ell_0)$, and that any of
these renormalized equations of motion are physically equivalent to the
final results of \cite{Blanchet:2000ub}. In particular, this means, as
we shall see below, that each choice of renormalization prescription
within dim. reg. determines the two regularization length scales
$r'_1$, $r'_2$ entering Eq. (\ref{a1BF}). Of course, what is important
is not the particular values these constants can take in a particular
renormalization scheme [indeed $r'_1$ and $r'_2$ are simply ``gauge''
constants which can anyway be removed by a coordinate transformation],
but the fact that the different renormalization prescriptions yield
equations of motion falling into the ``parametric'' class
(\textit{i.e.}, parametrized by $r'_1$ and $r'_2$) of equations of
motion obtained in \cite{Blanchet:2000ub}.

An alternative way to phrase the result (\ref{eta})-(\ref{lambda}), is
to combine Eqs.~(\ref{accpH}) and (\ref{a1DimReg}) in order to arrive
at
\begin{equation}
\lim_{\varepsilon\rightarrow 0} \, \Bigl[\mathcal{D}\mathbf{a}_1
\bigl[\varepsilon,\ell_0; \hbox{$-\frac{1987}{3080}$} ;
r_1', r'_2\bigr]
+ \delta_{\bm{\xi} (\varepsilon , \ell_0 ; r'_1 , r'_2)} \,
\mathbf{a}_1 \Bigr] = \sum_{A=1}^7\delta^A\mathbf{a}_1\,.
\label{equiveta}
\end{equation}
Under this form one sees that the sum of the additional terms
$\delta^A\mathbf{a}_1$ differs by a mere shift, \textit{when and only
when} $\lambda$ takes
the value (\ref{lambda}), from the specific contribution
$\mathcal{D}\mathbf{a}_1$ we shall evaluate in this paper, which comes
directly from dimensional regularization. Therefore one can say that,
when $\lambda = -\frac{1987}{3080}$, the extended-Hadamard
regularization \cite{Blanchet:2000nu,Blanchet:2000cw} is in fact
(physically) equivalent to dimensional regularization. However the
extended-Hadamard regularization is incomplete, both because it is
unable to determine $\lambda$, and also because it necessitates some
``external'' requirements such as the imposition of the link
(\ref{lnr2s2}) in order to ensure the existence of a conserved energy
--- and in fact of the ten first integrals linked to the Poincar\'e
group. By contrast dim. reg. succeeds automatically (without extra
inputs) in guaranteeing the existence of the ten conserved
integrals of the Poincar\'e group, as already found in
Ref.~\cite{Damour:2001bu}.

In view of the necessary link (\ref{lambdaomegas}) provided by the
equivalence between the ADM-Hamiltonian and the harmonic-coordinates
equations of motion, our result (\ref{lambda}) is in perfect agreement
with the previous result (\ref{omegas}) obtained in
\cite{Damour:2001bu}.\footnote{One may wonder why the value of
$\lambda$ is a complicated rational fraction while $\omega_s$ is so
simple. This is because $\omega_s$ was introduced precisely to measure
the amount of ambiguities of certain integrals, and that the ADM
Hamiltonian reported in \cite{Damour:1999cr} was put in a minimally
ambiguous form, already in three dimensions, for which an \textit{a
posteriori} look at the ``ambiguities'' discussed in the Appendix A of
\cite{Damour:1999cr} already showed that $\omega_s = 0$. By contrast,
$\lambda$ has been introduced as the only possible unknown constant in
the link between the four arbitrary scales $r'_1 , r'_2 , s_1 , s_2$
(which has \textit{a priori} nothing to do with ambiguities of
integrals), in a framework where the use of the extended Hadamard
regularization makes in fact the calculation to be unambiguous.} Our
result is also in agreement with the recent finding of Itoh and
Futamase \cite{itoh1,itoh2} (see also \cite{Itoh:2001np}), who derived
the 3PN equations of motion in harmonic gauge using a
``surface-integral'' approach, aimed at describing \textit{extended}
relativistic compact binary systems in the strong-field point particle
limit. The surface-integral approach of Refs.~\cite{itoh1,itoh2} is
interesting because, like the matching method used at 2.5PN order in
\cite{Damour:1982wm}, it is based on the physical notion of extended
compact bodies. In this respect, we recall that the matching method
used in \cite{Damour:1982wm} showed that the internal structure (Love
numbers) of the constituent bodies would start influencing the
equations of motion of (non-spinning) compact bodies only at the 5PN
level. This \textit{effacement property} strongly suggests that it is
possible to model, in a physically preferred manner, two compact bodies
as being two point-like particles, described by two masses and two
world-lines, up to the 4.5PN level included. It remains, however, to
prove that the dimensional regularization of delta-function sources
does yield the physically unique equations of motion of two compact
bodies up to the 4.5 PN order. The work \cite{Damour:1982wm} proved it
at the 2.5 PN level, and the agreement of the present results with
those of \cite{itoh1,itoh2} indicates that this is also true at the 3PN
level.

Besides the independent confirmation of the value of $\omega_s$ or
$\lambda$, let us also mention that our work provides a confirmation
of the \textit{consistency} of dim. reg., because our explicit
calculations [which involved combinations of hundreds of Laurent
expansions of the form $a_{-1} \, \varepsilon^{-1} + a_0 + {\cal O}
(\varepsilon)$] are entirely different from the ones of
\cite{Damour:2001bu}: We use harmonic coordinates (instead of ADM-type
ones), we work at the level of the equations of motion (instead of the
Hamiltonian), we use a different form of Einstein's field equations
and we solve them by a different iteration scheme.

Finally, from a practical point of view our confirmation of the value
of $\omega_s$ or$\lambda$ allows one to use the full 3PN accuracy in
the analytical computation of the dynamics of the last orbits of binary
orbits \cite{Damour:2000we,Blanchet:2001id}. It remains, however, the
task of computing, using dimensional regularization, the parameter
$\theta$ entering the 3.5PN gravitational energy flux
\cite{Blanchet:2001aw,Blanchet:2001ax} to be able to have full 3.5PN
accuracy in the computation of the gravitational waveforms emitted by
inspiralling compact binaries (see, \textit{e.g.}, \cite{Damour:2002vi}
and references therein).

\vskip1pc
The organization of the paper is as follows.
In Section~\ref{FieldEq} we derive our basic 3PN solution of the field
equations for general fluid sources in $d$ spatial dimensions, using
$d$-dimensional generalizations of the elementary potentials
introduced in Ref.~\cite{Blanchet:2000ub}. Section~\ref{Hadamard}
collects all the additional terms included in \cite{Blanchet:2000ub}
which are due specifically to the extended Hadamard regularization, and
derives the pure Hadamard-Schwartz (pHS) contribution to the equations
of motion. The differences between the dimensional and pHS
regularizations for all the potentials and their gradients are computed
in Section~\ref{Difference}. Then the dim. reg. equations of motion are
obtained in Section~\ref{Results}, where we comment also on their
interpretation in terms of space-time diagrams.
Section~\ref{Renormalise} is devoted to the renormalization of the dim.
reg. equations by means of suitable shifts of the particles'
world-lines, and to the equivalence with the end results of
\cite{Blanchet:2000ub} when Eq. (\ref{lambda}) holds. At this stage,
the proofs of Theorems~1 and 2 stated above are finally
completed.

We end the paper with some conclusions (Section~\ref{Conclusions}) and
three appendices. Appendix~\ref{ExpFieldEq} provides further material
on the $d$-dimensional metric and geodesic equation,
Appendix~\ref{Formulae} gives a compendium of useful formulae for
working in $d$ dimensions, and Appendix~\ref{Littleg} generalizes the
well-known quadratic-order elementary kernel $\textsl{g}^{(d=3)}
(\mathbf{x}) = \ln (r_1 + r_2 + r_{12})$ to $d$ dimensions. The latter
calculation of the $d$-dimensional kernel $\textsl{g}^{(d)}$ is
not directly employed in the present paper, but represents a first
step in obtaining the equations of motion in any dimension $d$ (not
necessarily of the form $3+\varepsilon$).

\section{Field equations in \lowercase{$d+1$} space-time dimensions}
\label{FieldEq}
This section is devoted to the field equations of general relativity
in $d+1$ space-time dimensions, and to the geodesic equation describing
the motion of point particles. We use the sign conventions of
Ref.~\cite{MTW}, and in particular our metric signature is mostly $+$.
Space-time indices are denoted by greek letters, and spatial indices by
latin letters ($i,j,\cdots$ run from 1 to $d$). A summation is
understood for any pair of repeated indices. We work in the harmonic
gauge, which is such that
\begin{equation}
\Gamma^\lambda \equiv g^{\alpha\beta}\Gamma^\lambda_{\alpha\beta} = 0
\,.
\label{harmGauge}
\end{equation}
As usual, $g^{\alpha\beta}$ denotes the inverse metric and
$\Gamma^\lambda_{\alpha\beta}$ the Christoffel symbols. Using this
gauge condition, one can easily prove that the Ricci tensor reads
\textit{in any dimension}
\begin{eqnarray}
2 R_{\mu\nu}^\text{harm}&=&-g^{\alpha\beta} g_{\mu\nu,\alpha\beta}
+g^{\alpha\beta} g^{\gamma\delta}
\biggl(g_{\mu\alpha,\gamma}\,g_{\nu\beta,\delta}
-g_{\mu\alpha,\gamma}\,g_{\nu\delta,\beta}\nonumber\\
&&+g_{\mu\alpha,\gamma}\,g_{\beta\delta,\nu}
+g_{\nu\alpha,\gamma}\,g_{\beta\delta,\mu}
-\frac{1}{2}\,g_{\alpha\gamma,\mu}\,g_{\beta\delta,\nu}\biggr),
\label{RicciHarm}
\end{eqnarray}
where a comma denotes partial derivation. Note that the spatial
dimension $d$ does not appear explicitly in this expression, whereas
some $d$-dependent coefficients do appear when expressing the Ricci
tensor in terms of the so-called ``gothic'' metric
$\mathfrak{g}^{\mu\nu} \equiv \sqrt{-g}\, g^{\mu\nu}$ [see
Eq.~(\ref{RicciGoth}) in Appendix~\ref{ExpFieldEq} below].

In any dimension, the Einstein field equations read
\begin{equation}
R^{\mu\nu} -\frac{1}{2}g^{\mu\nu}R = \frac{8\pi G}{c^4}\, T^{\mu\nu}\,,
\label{Einstein}
\end{equation}
where $T^{\mu\nu}$ denotes the matter stress-energy tensor, given by
the functional derivative $\sqrt{-g}\, T^{\mu\nu}\equiv 2 c\, \delta
S_m/\delta g_{\mu\nu}$ of the matter action $S_m$ with respect to the
metric tensor. \textit{By definition}, $G$ denotes the constant
involved in Eq.~(\ref{Einstein}), which shows that its dimension is
such that
\begin{equation}
G = G_N \ell_0^{d-3}\,,
\label{l0}
\end{equation}
where $G_N$ is the usual Newton constant (in 3 spatial dimensions)
and $\ell_0$ is an arbitrary length scale. This scale will be involved
in our dimensionally regularized results below, but we will finally
show that the physical observables do not depend on it. As is well
known, the combination of Eq.~(\ref{Einstein}) with its trace allows us
to rewrite it as
\begin{equation}
R_{\mu\nu} = \frac{8\pi G}{c^4} \left(T_{\mu\nu}
-\frac{1}{d-1}\,g_{\mu\nu} T^\lambda_\lambda\right),
\label{Ricci}
\end{equation}
in which the spatial dimension $d$ now appears explicitly.

We wish to expand in powers of $1/c$ the field equations resulting
{}from~(\ref{RicciHarm}) and~(\ref{Ricci}). The basic idea is to
introduce a sequence of ``elementary potentials'', $V$, $V_i$, $\hat
W_{ij}$, \dots which allow one to parametrize conveniently the
successive post-Minkowskian contributions to the metric $g_{\mu\nu}
(x)$. For instance, at the first post-Minkowskian order it is
convenient to parametrize the metric as
\begin{equation}
g_{00} \equiv -1 +2V/c^2 +{\mathcal{O}}(G^2)\,,\quad
g_{0i} \equiv -4 V_i/c^3 +{\mathcal{O}}(G^2)\,,
\label{gLowest}
\end{equation}
where the so-introduced elementary potentials $V$ and $V_i$ satisfy
equations of the form
\begin{equation}
\Box V = -4 \pi G \sigma\,,\quad
\Box V_i = -4 \pi G \sigma_i\ ,
\label{VLowest}
\end{equation}
where $\Box \equiv \partial_i^2 -(1/c^2)\partial_t^2$ denotes the flat
d'Alembertian and where, \textit{by definition}, the sources $\sigma$
and $\sigma_i$ are linear combinations of the contravariant components
$T^{\mu\nu}$ of the stress-energy tensor of the matter. Let us
underline that the factor $-4\pi G$ in these equations is a
\textit{choice}. We could of course introduce here a functional
dependence on the spatial dimension $d$, for instance by replacing the
factor $4\pi$ by the surface of the unit $(d-1)$-dimensional sphere
[see Eq.~(\ref{Omegad1}) in Appendix~\ref{Formulae}], but this would
only complicate the intermediate expressions without changing our final
result. The matter sources $\sigma$ and $\sigma_i$ \textit{defined} by
Eqs.~(\ref{gLowest}), (\ref{VLowest}) read (in $d$ spatial dimensions):
\begin{equation}
\sigma \equiv \frac{2}{d-1}\ \frac{(d-2)T^{00}+T^{ii}}{c^2}\,,\qquad
\sigma_i \equiv \frac{T^{0i}}{c}\ ,\qquad
\sigma_{ij} \equiv T^{ij}\,.
\label{sigma}
\end{equation}
The definition for $\sigma_{ij}$ has been added for future use. Note
that $\sigma_i$ and $\sigma_{ij}$ take the same forms as usual in 3
dimensions (see Eqs.~(3.9) of Ref.~\cite{Blanchet:2000ub}), but that
the definition of $\sigma$ involves an explicit dependence on $d$.
Conversely, the first and third of these equations allow us to express
$T^{00}$ in terms of the above matter sources: $T^{00} = \left(
\frac{d-1}{2}\, \sigma c^2 -\sigma_{ii}\right)/(d-2)$. A simple
consequence of the expression of $\sigma$ is that the $d$-dimensional
\textit{Newtonian potential} generated by a mass $m_a$ located at
$\mathbf{y}_a$ reads explicitly
\begin{equation}
U_a (\mathbf{x}) = V_a (\mathbf{x}) + {\cal O} \left( \frac{1}{c^2}
\right) = 2\,
\frac{d-2}{d-1} \, \tilde k \, \frac{G \, m_a}{r_a^{d-2}} + {\cal O}
\left( \frac{1}{c^2} \right) \, ,
\label{Newton}
\end{equation}
where the factor $2(d-2)/(d-1)$ comes from $\sigma$ (\textit{i.e.},
{}from Einstein's equations), while the factor $\tilde k = \Gamma
\bigl(\frac{d-2}{2} \bigr) / \pi^{\frac{d-2}{2}}$ comes from the
expression of the Green function of the Laplacian in $d$ dimensions
(see Eq.~(\ref{Green}) below and Appendix~\ref{Formulae}).

We give below the simplest forms of the metric and of the potential
equations that we could obtain. We will explain afterwards which rules
we followed to simplify them. Let us first define the useful
combination
\begin{equation}
\mathcal{V} \equiv
V -\frac{2}{c^2} \left(\frac{d-3}{d-2}\right) K
+\frac{4 \hat X}{c^4} +\frac{16 \hat T}{c^6}\,.
\label{calV}
\end{equation}
Then the metric components can be written in a rather compact form:
\begin{subequations}
\label{metric}
\begin{eqnarray}
g_{00}&=&-e^{-2\mathcal{V}/c^2}\left(1 -\frac{8 V_i V_i}{c^6}
-\frac{32 \hat R_i V_i}{c^8}\right)
+{\mathcal{O}}\left(\frac{1}{c^{10}}\right),
\label{g00}\\
g_{0i}&=&-e^{-\frac{(d-3)\mathcal{V}}{(d-2)c^2}}
\left\{\frac{4 V_i}{c^3}\left[
1+\frac{1}{2}\left(\frac{d-1}{d-2}\ \frac{V}{c^2}\right)^2\right]
+\frac{8\hat R_i}{c^5}
+\frac{16}{c^7}\left[\hat Y_i+\frac{1}{2}\hat
W_{ij}V_j\right]\right\}
+{\mathcal{O}}\left(\frac{1}{c^9}\right),\nonumber\\
\label{g0i}\\
g_{ij}&=&e^{\frac{2\mathcal{V}}{(d-2)c^2}}\left\{\delta_{ij}
+\frac{4}{c^4} \hat W_{ij}
+\frac{16}{c^6} \left[\hat Z_{ij}-V_i V_j
+\frac{1}{2(d-2)}\, \delta_{ij} V_k V_k
\right]\right\}
+{\mathcal{O}}\left(\frac{1}{c^8}\right).
\label{gij}
\end{eqnarray}
\end{subequations}
The various elementary potentials $V$, $V_i$, $K$, $\hat W_{ij}$, $\hat
R_i$, $\hat X$, $\hat Z_{ij}$, $\hat Y_i$ and $\hat T$ introduced in
these definitions are $d$-dimensional analogues of those used in
Eqs.~(3.24) of
Ref.~\cite{Blanchet:2000ub}. Actually, an extra potential is needed for
$d\neq 3$, and it has been denoted $K$ in Eq.~(\ref{calV}) above. We
give in Appendix~\ref{ExpFieldEq} the explicit expansion of this metric
in powers of $1/c$, as well as its inverse $g^{\mu\nu}$ and its
determinant $g$, which can be useful for future works. Note that the
first post-Newtonian order of the spatial metric, $g_{ij} =
\delta_{ij}\bigl[1+\frac{2V}{(d-2)c^2}\bigr] +{\mathcal{O}}(1/ c^4)$,
explicitly depends on $d$ contrary to our choice~(\ref{gLowest}) for
$g_{00}$. This dissymmetry between $g_{00}$ and $g_{ij}$ is imposed by
the field equations~(\ref{Ricci}).

The successive post-Newtonian truncations of the field
equations~(\ref{RicciHarm})-(\ref{Ricci}) give us the sources for these
various potentials. The equations for $\Box V$ and $\Box V_i$ have
already been written in Eqs.~(\ref{VLowest}) above. We
get for the remaining potentials:
{\allowdisplaybreaks
\begin{subequations}
\label{potentialEq}
\begin{eqnarray}
\Box K&=&-4\pi G \sigma V\,,
\label{dalK}\\
\Box\hat W_{ij}&=&-4\pi G\left(\sigma_{ij}
-\delta_{ij}\,\frac{\sigma_{kk}}{d-2}\right)
-\frac{1}{2}\left(\frac{d-1}{d-2}\right)\partial_i V \partial_j V\,,
\label{dalWij}\\
\Box\hat R_i&=&-\frac{4\pi G}{d-2}\left(\frac{5-d}{2}\, V \sigma_i
-\frac{d-1}{2}\, V_i\, \sigma\right)\nonumber\\
&&-\frac{d-1}{d-2}\,\partial_k V\partial_i V_k
-\frac{d(d-1)}{4(d-2)^2}\,\partial_t V \partial_i V\,,
\label{dalRi}\\
\Box\hat X&=&-4\pi G\left[\frac{V\sigma_{ii}}{d-2}
+2\left(\frac{d-3}{d-1}\right)\sigma_i V_i
+\left(\frac{d-3}{d-2}\right)^2
\sigma\left(\frac{V^2}{2} +K\right)\right]\nonumber\\
&&+\hat W_{ij}\, \partial_{ij}V
+2 V_i\,\partial_t\partial_i V
+\frac{1}{2}\left(\frac{d-1}{d-2}\right) V \partial^2_t V
\nonumber\\
&&+\frac{d(d-1)}{4(d-2)^2}\left(\partial_t V\right)^2
-2 \partial_i V_j\, \partial_j V_i\ ,
\label{dalX}\\
\Box\hat Z_{ij}&=&-\frac{4\pi G}{d-2}\, V\left(\sigma_{ij}
-\delta_{ij}\,\frac{\sigma_{kk}}{d-2}\right)
-\frac{d-1}{d-2}\, \partial_t V_{(i}\, \partial_{j)}V
+\partial_i V_k\, \partial_j V_k\nonumber\\
&&+\partial_k V_i\, \partial_k V_j
-2 \partial_k V_{(i}\, \partial_{j)}V_k
-\frac{\delta_{ij}}{d-2}\, \partial_k V_m
\left(\partial_k V_m -\partial_m V_k\right)\nonumber\\
&&-\frac{d(d-1)}{8(d-2)^3}\, \delta_{ij}\left(\partial_t V\right)^2
+\frac{(d-1)(d-3)}{2(d-2)^2}\, \partial_{(i} V\partial_{j)} K\,,
\label{dalZij}\\
\Box\hat Y_i&=&-4\pi G
\biggl[-\frac{1}{2}\left(\frac{d-1}{d-2}\right)\sigma\hat R_i
-\frac{(5-d)(d-1)}{4(d-2)^2}\, \sigma V V_i
+\frac{1}{2}\, \sigma_k\hat W_{ik}
+\frac{1}{2}\sigma_{ik} V_k\nonumber\\
&&\hphantom{-4\pi G \biggl[}+\frac{1}{2(d-2)}\, \sigma_{kk}V_i
-\frac{d-3}{(d-2)^2}\, \sigma_i \left(V^2 +\frac{5-d}{2}\,
K\right)\biggr]
\nonumber\\
&&+\hat W_{kl}\, \partial_{kl} V_i
-\frac{1}{2}\left(\frac{d-1}{d-2}\right)
\partial_t\hat W_{ik}\, \partial_k V
+\partial_i\hat W_{kl}\, \partial_k V_l
-\partial_k\hat W_{il}\, \partial_l V_k
\nonumber\\
&&-\frac{d-1}{d-2}\, \partial_k V \partial_i \hat R_k
-\frac{d(d-1)}{4 (d-2)^2}\, V_k\, \partial_i V \partial_k V
-\frac{d(d-1)^2}{8 (d-2)^3}\, V\partial_t V\partial_i V
\nonumber\\
&&-\frac{1}{2}\left(\frac{d-1}{d-2}\right)^2 V \partial_k V
\partial_k V_i
+\frac{1}{2}\left(\frac{d-1}{d-2}\right) V \partial^2_tV_i
+2 V_k\, \partial_k\partial_t V_i
\nonumber\\
&&+\frac{(d-1)(d-3)}{(d-2)^2}\, \partial_k K \partial_i V_k
+\frac{d(d-1)(d-3)}{4(d-2)^3}
\left(\partial_t V\partial_i K +\partial_i V\partial_t K\right),
\label{dalYi}\\
\Box\hat T&=&-4\pi G\biggl[\frac{1}{2(d-1)}\, \sigma_{ij} \hat W_{ij}
+\frac{5-d}{4(d-2)^2}\, V^2\sigma_{ii}
+\frac{1}{d-2}\, \sigma V_i V_i
-\frac{1}{2}\left(\frac{d-3}{d-2}\right)\sigma\hat X\nonumber\\
&&\hphantom{-4\pi G\biggl[}-\frac{1}{12}\left(\frac{d-3}{d-2}\right)^3
\sigma V^3 -\frac{1}{2}\left(\frac{d-3}{d-2}\right)^3 \sigma V K
+\frac{(5-d)(d-3)}{2(d-1)(d-2)}\, \sigma_i V_i V\nonumber\\
&&\hphantom{-4\pi G\biggl[}+\frac{d-3}{d-1}\, \sigma_i\hat R_i
-\frac{d-3}{2(d-2)^2}\, \sigma_{ii} K\biggr]
+\hat Z_{ij}\, \partial_{ij}V
+\hat R_i\, \partial_t\partial_i V\nonumber\\
&&-2 \partial_i V_j\, \partial_j\hat R_i
-\partial_i V_j\, \partial_t\hat W_{ij}
+\frac{1}{2}\left(\frac{d-1}{d-2}\right) V V_i\, \partial_t\partial_i V
+\frac{d-1}{d-2}\, V_i\, \partial_j V_i\, \partial_j V\nonumber\\
&&+\frac{d(d-1)}{4(d-2)^2}\, V_i\, \partial_t V\, \partial_i V
+\frac{1}{8}\left(\frac{d-1}{d-2}\right)^2 V^2\partial^2_t V
+\frac{d(d-1)^2}{8(d-2)^3}\, V\left(\partial_t V\right)^2\nonumber\\
&&-\frac{1}{2}\left(\partial_t V_i\right)^2
-\frac{(d-1)(d-3)}{4(d-2)^2}\, V \partial^2_t K
-\frac{d(d-1)(d-3)}{4(d-2)^3}\, \partial_t V\, \partial_t K\nonumber\\
&&-\frac{(d-1)(d-3)}{4(d-2)^2}\, K \partial^2_t V
-\frac{d-3}{d-2}\, V_i\, \partial_t\partial_i K
-\frac{1}{2}\left(\frac{d-3}{d-2}\right)\hat W_{ij}\,
\partial_{ij} K\,.
\label{dalT}
\end{eqnarray}
\end{subequations}}

\noindent In Eq.~(\ref{dalZij}), parentheses around indices mean their
symmetrization, \textit{i.e.}, $a_{(ij)} \equiv \frac{1}{2} \, (a_{ij}
+ a_{ji})$. For $d=3$, the above set of equations~(\ref{potentialEq})
reduces to Eqs.~(3.26) and (3.27) of Ref.~\cite{Blanchet:2000ub}. The
order of the terms and their writing has been chosen to be as close as
possible to this reference.

The harmonic gauge conditions~(\ref{harmGauge}) impose the following
differential identities between the potentials:
{\allowdisplaybreaks
\begin{subequations}
\label{GaugeIdentities}
\begin{eqnarray}
g^{\mu\nu}\Gamma^0_{\mu\nu} = 0&\Rightarrow&\partial_t\Biggl\{
\frac{1}{2}\left(\frac{d-1}{d-2}\right) V
+\frac{1}{2 c^2}\left[\hat W
+\left(\frac{d-1}{d-2}\right)^2 V^2
-\frac{2(d-1)(d-3)}{(d-2)^2}\, K\right]\nonumber\\
&&\hphantom{\partial_t\Biggl\{}+\frac{2}{c^4}
\left(\frac{d-1}{d-2}\right)\biggl[\hat X
+\frac{d-2}{d-1}\,\hat Z
-\frac{d-3}{d-1}\,V_k V_k
+\frac{1}{2}\,V\hat W\nonumber\\
&&\hphantom{\partial_t\Biggl\{+\frac{2}{c^4}
\left(\frac{d-1}{d-2}\right)\biggl[}+\frac{(d-1)^2}{6(d-2)}\, V^3
-\frac{(d-1)(d-3)}{(d-2)^2}\, V K
\biggr]
\Biggr\}\nonumber\\
&&+\partial_i\Biggl\{V_i
+\frac{2}{c^2}\left[\hat R_i
+\frac{1}{2}\left(\frac{d-1}{d-2}\right) V V_i\right]\nonumber\\
&&\hphantom{+\partial_i\Biggl\{}+\frac{4}{c^4}\Biggl[\hat Y_i
-\frac{1}{2}\, \hat W_{ij} V_j
+\frac{1}{2}\, \hat W V_i
+\frac{1}{2}\left(\frac{d-1}{d-2}\right) V \hat R_i
+\frac{1}{4}\left(\frac{d-1}{d-2}\right)^2 V^2 V_i\nonumber\\
&&\hphantom{+\partial_i\Biggl\{+\frac{4}{c^4}\Biggl[}-
\frac{(d-1)(d-3)}{2 (d-2)^2}\, K V_i
\Biggr]
\Biggr\} = {\mathcal{O}}\left(\frac{1}{c^6}\right),
\label{diVi}\\
g^{\mu\nu}\Gamma^i_{\mu\nu} = 0&\Rightarrow&\partial_t\left\{
V_i +\frac{2}{c^2}\left[
\hat R_i +\frac{1}{2}\left(\frac{d-1}{d-2}\right)V V_i
\right]\right\}\nonumber\\
&&+\partial_j\left\{
\hat W_{ij} -\frac{1}{2}\, \hat W \delta_{ij}
+\frac{4}{c^2}\left[\hat Z_{ij}
-\frac{1}{2}\, \hat Z \delta_{ij}\right]
\right\} = {\mathcal{O}}\left(\frac{1}{c^4}\right),
\label{djWij}
\end{eqnarray}
\end{subequations}}$\!\!$
where $\hat W \equiv \hat W_{kk}$ and $\hat Z \equiv \hat Z_{kk}$
denote the traces of potentials $\hat W_{ij}$ and $\hat Z_{ij}$. For
$d=3$, these identities reduce to Eqs.~(3.28) of
Ref.~\cite{Blanchet:2000ub}. In this paper we shall check (see
Sections~\ref{IteratedEinstein} and \ref{Poles}) that all the
dimensionally-regularized potentials we use obey, at the indicated
accuracy, the differential identities (\ref{GaugeIdentities})
equivalent to the harmonic gauge conditions.

In order to simplify as much as possible the above
equations~(\ref{potentialEq}) for the potentials, we used the following
rules:
\begin{enumerate}
\item[(i)]We used the harmonic gauge condition~(\ref{diVi}) to replace
everywhere $\partial_i V_i$ in terms of $\partial_t V$ and higher
post-Newtonian order terms, and the gauge condition~(\ref{djWij}) to
replace $\partial_j\hat W_{ij}$ in terms of $\partial_i \hat W$ and
$\partial_t V_i$ [our knowledge of the higher order terms
${\mathcal{O}}(1/c^2)$ in Eq.~(\ref{djWij}) was actually not necessary
for the simplification of Eqs.~(\ref{potentialEq})]. We also used the
lowest order terms of Eqs.~(\ref{GaugeIdentities}) to simplify their
own higher order contributions.
\item[(ii)]If the source of a potential $P$ contained a double
(contracted) gradient of the form $\Box P = \partial_k A \partial_k B
+(\text{other terms})$, where $A$ and $B$ were two lower-order
potentials, we got rid of the double gradient by defining another
potential $P' \equiv P -\frac{1}{2} AB$. We could then write its
equation as $\Box P' = -\frac{1}{2}(\Box A)B -\frac{1}{2}A(\Box B)
+\frac{1}{c^2}\, \partial_t A \partial_t B +(\text{other terms})$, in
which $\Box A$ and $\Box B$ were replaced by their own explicit
sources. The contribution proportional to $1/c^2$ was then transferred
into the source of a higher order potential.
\item[(iii)]At order ${\mathcal{O}}(1/c^6)$,
equation~(\ref{RicciHarm})-(\ref{Ricci}) for $R_{00}$ (\textit{i.e.},
for $\Box g_{00}$) contains the term $\hat W_{ij}\,\partial_{ij} V$,
that we introduced in the source of potential $\hat X$,
Eq.~(\ref{dalX}). In all other equations involving the same source
$\hat W_{ij}\,\partial_{ij} V$, we used $\Box \hat X$ to eliminate
it, instead of reintroducing it in the sources of other potentials.
This is the reason why $\hat X$ is involved in the spatial metric
$g_{ij}$ too at order ${\mathcal{O}}(1/c^6)$ [\textit{via} the
exponential of $\mathcal{V}$ in Eq.~(\ref{gij})], and why $V\hat X$
appears again in $g_{00}$ at order ${\mathcal{O}}(1/c^8)$. See the
expanded form of the metric~(\ref{metricExp}) in
Appendix~\ref{ExpFieldEq}.
\item[(iv)]In the equation for $R_{00}$ at order
${\mathcal{O}}(1/c^8)$, we \textit{chose} to eliminate a source
proportional to $V\partial_i V_j\,\partial_i V_j$, by including an
all-integrated term $V V_i V_i$ in the definition of $g_{00}$,
Eq.~(\ref{g00}). On the other hand, we could not eliminate at the same
time the source term proportional to $V_i\, \partial_j V_i\, \partial_j
V$ in $\Box g_{00}$, although it involves a double (contracted)
gradient too. This is the reason why such a term appears in
Eq.~(\ref{dalT}).
\end{enumerate}
The above simplification rules have been applied systematically with a
single exception. Indeed, Eq.~(\ref{dalZij}) for $\Box\hat Z_{ij}$
involves double (contracted) gradients $\partial_k V_i\, \partial_k
V_j$ and $\delta_{ij}\, \partial_k V_m\, \partial_k V_m$. Therefore,
the application of rule~(ii) would have yielded another potential
\begin{equation}
\widetilde Z_{ij}\equiv \hat Z_{ij}
-\frac{1}{2} V_i V_j +\frac{1}{2(d-2)}\, \delta_{ij} V_k V_k\ ,
\label{Ztilde}
\end{equation}
such that no double gradient appears in its source (but extra compact
sources $\sigma_{(i}V_{j)}$ and $\delta_{ij}\sigma_k V_k$ would have
been involved). Although this modified potential $\widetilde Z_{ij}$
actually simplifies slightly some equations (but not all of them), we
have chosen to use $\hat Z_{ij}$ which is the direct $d$-dimensional
analogue of the potential written in Eq.~(3.27c) of
Ref.~\cite{Blanchet:2000ub}. Indeed, as explained in the following
sections, the 3-dimensional results of this reference will be necessary
for our $d$-dimensional calculations, and it is more convenient to keep
the same notation.

Notice also that after the above simplifications, the resulting metric
involves only potentials which are at most cubically non-linear (like
for the term $\hat W_{ij}\,\partial_{ij} V$ in the potential $\hat X$
--- using the terminology of Section \ref{DimRegStat} below). There is
no need to introduce any quartically non-linear elementary potential
because it turns out that it is possible to ``integrate directly'' all
of them (at the 3PN level) in terms of other potentials. The only
quartic contributions are the terms composed of $V^4$ and $V \hat X$ in
the metric component $g_{00}$ [see Eq. (\ref{g00Exp}) in Appendix
\ref{ExpFieldEq}]. The fact that there are no intrinsically quartic
potentials at the 3PN order made the closed-form calculation in
\cite{Blanchet:2000nv,Blanchet:2000ub} possible. We shall comment more
on this interesting fact in Section \ref{IteratedEinstein}.

Let us now apply the general potential parametrization of the metric
defined above to the specific case of (monopolar) point particles,
\textit{i.e.}, to the action
\begin{equation}
S = \int \frac{d^{d+1} \, x}{c} \, \sqrt{-g} \, \frac{c^4}{16 \pi G} \,
R(g) - \sum_a m_a \, c \int \sqrt{-g_{\mu\nu} (y_a^{\lambda}) \,
dy_a^{\mu} \, dy_a^{\nu}}\ .
\label{action}
\end{equation}
The stress-energy tensor $T^{\mu\nu} (x) \equiv [2c / \sqrt{-g(x)}] \,
\delta \, S_\text{matter} / \delta \, g_{\mu\nu} (x)$ deduced from this
action reads
\begin{equation}
T^{\mu\nu} (x) = \sum_a m_a \, c^2 \int ds_a \, \frac{dy_a^{\mu}}{ds_a}
\, \frac{dy_a^{\nu}}{ds_a} \, [-g(y_a)]^{-\frac{1}{2}} \, \delta^{(d+1)}
(x^{\lambda} - y_a^{\lambda} (s_a))\,,
\label{Tmunu}
\end{equation}
where $ds_a \equiv \sqrt{-g_{\mu\nu} \, (y_a^{\lambda}) \, dy_a^{\mu}
\, dy_a^{\nu}}$ is ($c$ times) the proper time along the world-line of
the $a^\text{th}$ particle and where $\delta^{(d+1)}$ is the Dirac
density in $d+1$ dimensions ($\int d^{d+1} x \, \delta^{(d+1)} (x) =
1$). Here, we take advantage of the fact (emphasized in
\cite{Damour:2001bu}) that dim. reg. respects the basic properties of
the algebraic and differential calculus: associativity, commutativity
and distributivity of point-wise addition and multiplication, Leibniz's
rule, Schwarz's rule $(\partial_{\mu} \, \partial_{\nu} \, f =
\partial_{\nu} \, \partial_{\mu} \, f)$, integration by parts, etc.
In addition, the post-Newtonian expansion of $g_{\mu\nu} (x)$ yields
``$d$-dimensional functions'' which are formally as smooth as wished
(by taking the real part of $d$ small enough) in the vicinity of the
world-lines: see for instance Eq. (\ref{Newton}). This allows us to
work with self-gravitating point particles essentially as if they were
\textit{test} particles. For instance, we can use $F [g_{\mu\nu} (x)]
\, \delta^{(d)} \, (\mathbf{x}-\mathbf{y}_a) = F [g_{\mu\nu} (y_a)] \,
\delta^{(d)} \, (\mathbf{x}-\mathbf{y}_a)$. In particular, the
$y_a$-evaluated determinant factor $[-g(y_a)]^{-\frac{1}{2}}$ in
(\ref{Tmunu}) came from the field-point dependent factor
$[-g(x)]^{-\frac{1}{2}}$ in the definition of $T^{\mu\nu} (x)$.
Similarly, the usual derivation of the equations of motion of a test
particle formally generalizes to the case of self-gravitating point
particles in $d$ dimensions. One then finds that the equations of
motion of point particles can equivalently be written as
\begin{equation}
\nabla_\nu \, T^{\mu\nu} (x) = 0\,,
\label{Geod1}
\end{equation}
or as the usual geodesic equations. The latter can be written either
in covariant form $u_a^{\nu} \, \nabla_\nu \, u_a^{\mu} = 0$
($u_a^{\mu} \equiv dy_a^{\mu} / ds_a$), \textit{i.e.}, explicitly
\begin{equation}
\frac{d^2 \, y_a^{\lambda}}{ds_a^2} + \Gamma_{\mu\nu}^{\lambda} \,
[g(y_a) , \partial g(y_a)] \, \frac{dy_a^{\mu}}{ds_a} \,
\frac{dy_a^{\nu}}{ds_a} = 0 \,,
\label{Geod2}
\end{equation}
where $\Gamma_{\mu\nu}^{\lambda} = \frac{1}{2} \, g^{\lambda \sigma}
(\partial_{\mu} \, g_{\nu\sigma} + \partial_{\nu} \, g_{\mu\sigma} -
\partial_{\sigma} \, g_{\mu\nu})$ as usual, or in the explicit form
corresponding to using the coordinate time $t = y_a^0 / c$ as parameter
along the world-lines, which is easily derived from the covariant
expression with a lower index, $u_a^{\nu} \, \nabla_\nu \, u^a_{\mu} =
0 \Leftrightarrow d(g_{\mu\rho} u_a^\rho)/ds_a =
\frac{1}{2}\partial_\mu g_{\nu\rho}\, u_a^\nu u_a^\rho$. Like in 3
dimensions, cf. Eqs.~(3.32)-(3.33) of Ref.~\cite{Blanchet:2000ub}, it
can thus be put in the form
\begin{equation}
\frac{d P^i}{dt} = F^i\,,
\label{eqGeod}
\end{equation}
where
\begin{equation}
P^i \equiv \frac{g_{i\mu} v^\mu}{\sqrt{-g_{\rho\sigma}
v^\rho v^\sigma/c^2}}\ ,\quad
F^i \equiv\frac{1}{2}\,
\frac{\partial_ig_{\mu\nu} v^\mu v^\nu}{\sqrt{-g_{\rho\sigma}
v^\rho v^\sigma/c^2}}\ ,
\label{PiFi}
\end{equation}
$v^\mu \equiv dx^\mu/dt = (c, \mathbf{v})$ denoting the coordinate
velocity. Let us emphasize again that in $d$ dimensions, all the
non-linear functions of $g_{\mu\nu} (y_a)$ and $\partial_{\lambda} \,
g_{\mu\nu} (y_a)$ that will enter our calculation of
(\ref{eqGeod})-(\ref{PiFi}) can be treated as in the $\mathbf{x}
\rightarrow \mathbf{y}_a$ evaluation of smooth functions of
$\mathbf{x}$. For instance, denoting for simplicity $f \equiv
2(d-2)/(d-1)$, the Newtonian approximation, say
$U^{(d)}(\mathbf{x})\equiv U(\mathbf{x})$, of the basic scalar
potential $V(\mathbf{x})$, reads, in the vicinity of $\mathbf{x} =
\mathbf{y}_1$,
\begin{equation}
U (\mathbf{x}) = f \, \tilde k \, G \, m_1 \, r_1^{2-d} +
U_2 (\mathbf{x})\,,
\label{U}
\end{equation}
where $U_2 (\mathbf{x}) = f \, \tilde k \, G \, m_2 \, r_2^{2-d}$ is
(in any $d$) an indefinitely differentiable function of $\mathbf{x}$
near $\mathbf{y}_1$. Analytically continuing $d$ to sufficiently
``low'' (and even with negative real part, if needed) values, we see
not only that $[U(\mathbf{x})]_{\mathbf{x} = \mathbf{y}_1} = U_2
(\mathbf{y}_1)$, but that $[U^n (\mathbf{x})]_{\mathbf{x} =
\mathbf{y}_1} = (U_2 (\mathbf{y}_1))^n$, and, \textit{e.g.}, $[U^p
(\mathbf{x}) \, \partial_i \, U(\mathbf{x})]_{\mathbf{x} =
\mathbf{y}_1} = (U_2 (\mathbf{y}_1))^p \, \partial_i \, U_2
(\mathbf{y}_1)$, etc.

Although the expressions (\ref{PiFi}) do not depend explicitly on the
dimension $d$, the metric~(\ref{metric}) does depend on it, and
therefore the post-Newtonian expansions of Eqs.~(\ref{PiFi}) involve
many $d$-dependent coefficients. We give their full expressions in
Appendix~\ref{ExpFieldEq}, Eqs.~(\ref{ExpPi})-(\ref{ExpFi}), but we
quote below only their Newtonian orders and the very few terms which
will contribute to the poles $\propto 1/(d-3)$ in our dimensionally
regularized calculations:
\begin{equation}
P^i= v^i +\cdots
-\frac{8}{c^4} \hat R_i
-\frac{16}{c^6}\hat Y_i
+{\mathcal{O}}\left(\frac{1}{c^{8}}\right),\quad
F^i=\partial_i V +\cdots
+\frac{4}{c^4}\partial_i\hat X
+\frac{16}{c^6}\partial_i\hat T
+{\mathcal{O}}\left(\frac{1}{c^{8}}\right).
\label{SmallExpPiFi}
\end{equation}
The acceleration $\mathbf{a} \equiv d\mathbf{v}/dt$ can thus be
written as
\begin{eqnarray}
a^i&=&F^i - \frac{d\left(P^i-v^i\right)}{dt} \label{SmallAccel}\\
&=&\partial_i V +\frac{1}{c^2}\Bigl[\cdots\Bigr]
+\frac{4}{c^4}\biggl[\partial_i\hat X + 2\frac{d\hat
R_i}{dt}+\cdots\biggr] +\frac{16}{c^6}\biggl[\partial_i\hat T +
\frac{d\hat Y_i}{dt}+\cdots\biggr]
+{\mathcal{O}}\left(\frac{1}{c^{8}}\right).\nonumber
\end{eqnarray}
In Section \ref{IteratedEinstein} we shall give flesh to the formal
expressions written above by explaining by which algorithm one can
compute, with the required accuracy, the explicit $d$-dimensional
expansions near $\mathbf{x} = \mathbf{y}_1$ [analogous to the simple
case (\ref{U})] of the various elementary potentials entering
Eq.~(\ref{SmallAccel}), and notably of the crucial ones $\hat X$, $\hat
R_i$ (to be computed with 1PN accuracy) and $\hat T$, $\hat Y_i$ (to be
computed at Newtonian order only).

\section{Hadamard self-field regularizations in 3 dimensions}
\label{Hadamard}
The main aim of this Section is to complete the \textit{first step} of
the strategy outlined in the Introduction, \textit{i.e.} to collect a
complete list of the additional contributions to the equations of
motion which are specific consequences of the use of the
\textit{extended} Hadamard regularization methods defined in
\cite{Blanchet:2000nu,Blanchet:2000cw}. However, to do that we need to
start by recalling some material concerning the Hadamard regularization
in 3 dimensions, and by contrasting it with dimensional regularization.
Such material is needed for understanding our computation based on the
``difference'' in Section \ref{Difference}. We shall start by recalling
the definition of the ``ordinary'' Hadamard regularization and complete
it by defining what we shall call the ``pure'' Hadamard regularization.
Then we shall recall the main new features of the \textit{extended}
Hadamard regularization defined in
\cite{Blanchet:2000nu,Blanchet:2000cw}, and collect the additional
contributions to the equations of motion which are specific
consequences of the use of the extended Hadamard regularization (there
are seven such additional contributions).

\subsection{Ordinary and ``pure'' Hadamard regularizations}
\label{PureHadamard}
The phrase ``Hadamard regularization'' covers two distinct concepts:
(i) the regularization of the ``limit''
$\lim_{\mathbf{x}\rightarrow\mathbf{y}_1} \, F(\mathbf{x} ;
\mathbf{y}_1 , \mathbf{y}_2)$ where $F(\mathbf{x} ; \mathbf{y}_1 ,
\mathbf{y}_2)$ belongs to a class ${\cal F}$ of singular functions
(generated by the iteration of Einstein's equations), and (ii)
the regularization of the 3-dimensional integral $\int d^3 \mathbf{x}
\, F (\mathbf{x} ; \mathbf{y}_1 , \mathbf{y}_2)$ of some function
$F \in {\cal F}$. The class of functions $\mathcal{F}$ consists of all
functions $F(\mathbf{x})$ on $\mathbb{R}^3$ that are smooth except at
$\mathbf{y}_1$ and $\mathbf{y}_2$, around which they admit Laurent-type
expansions in powers of $r_1$ or $r_2$ (see Section II.A of
\cite{Blanchet:2000nu} for the precise definition of $\mathcal{F}$).
When $r_1\equiv \vert\mathbf{x}-\mathbf{y}_1\vert\rightarrow 0$
(\textit{i.e.}, around singularity 1) we have, $\forall
N\in\mathbb{N}$,
\begin{equation}
F(\mathbf{x})=\sum_{p_0\leq p\leq N}r_1^p
\mathop{f}_1{}_p(\mathbf{n}_1)+o(r_1^N)\,,
\label{Fx}
\end{equation}
where the Landau $o$-symbol takes its usual meaning, and the
${}_1f_p(\mathbf{n}_1)$'s denote the coefficients of the various
powers of $r_1$, which are functions of the positions and velocities
of the particles, and of the unit direction $\mathbf{n}_1\equiv
(\mathbf{x}-\mathbf{y}_1)/r_1$ of approach to singularity 1. The
powers of $r_1$ are relative integers, $p\in \mathbb{Z}$, bounded from
below by some typically negative $p_0$ depending on the $F$ in
question.

The Hadamard ``\textit{partie finie}'' of the singular function $F$ at
the location of the singular point 1 (first meaning of Hadamard
regularization) is defined as the angular average of the zeroth-order
coefficient in the expansion~(\ref{Fx}). It is denoted $(F)_1$, so
that
\begin{equation}
(F)_1 \equiv \bigl<\mathop{f}_1{}_0\bigr>\equiv\int
\frac{d\Omega(\mathbf{n}_1)}{4\pi}\mathop{f}_1{}_0(\mathbf{n}_1)\,,
\label{F1}
\end{equation}
where $d\Omega(\mathbf{n}_1)$ denotes the usual surface element on
the 2-dimensional sphere centered on 1. We shall employ systematically
a bracket notation $\bigl<\bigr>$ for the angular average of a function
of the angles (either $\mathbf{n}_1$ or
$\mathbf{n}_2$).\footnote{Since this will always be clear from the
context, we do not specify on the brackets if the angular integration
should be performed around point 1 or 2. We do not indicate either if
the integration sphere is two-dimensional or $(d-1)$-dimensional (as
we shall see later there can be no confusion about this).} A
distinctive feature of the Hadamard partie finie (\ref{F1}) is its
``non-distributivity'' in the sense that
\begin{equation}
(FG)_1\not=(F)_1(G)_1~~\text{in general for $F,G\in\mathcal{F}$.}
\label{nondistr}
\end{equation}
The non-distributivity represents a crucial departure away from the
simple algebraic properties of the analog of $(F)_1$ in
dim. reg. which is merely $F^{(d)}(\mathbf{y}_1)$. It is an
interesting fact that in a post-Newtonian expansion the
non-distributivity starts playing a role only at the 3PN order
(because the functions there become singular enough). Up to the 2PN
order one can show that $(FG)_1=(F)_1(G)_1$ for all the functions
involved in the equations of motion in harmonic coordinates
\cite{Blanchet:1998vx}. Several of the problems of the Hadamard
self-field regularization (in the ``ordinary'' sense) when applied at
the 3PN level (\textit{e.g.} the occurrence of the unknown constant
$\lambda$) are related to the latter non-distributivity.

The second notion of Hadamard \textit{partie finie} (denoted
$\text{Pf}$ in the following) is to give a meaning to the generally
divergent integral $\int d^3\mathbf{x}\,F(\mathbf{x})$. In this work we
shall have to consider only the ultra-violet (UV) divergencies of the
integrals, \textit{i.e.}, at the locations of the two local
singularities $\mathbf{y}_1$ and $\mathbf{y}_2$. All functions involved
at the 3PN order are such that there are no infra-red (IR) divergencies
when $\vert\mathbf{x}\vert\rightarrow \infty$ (this is true not only in
3 dimensions but also for any dimension $d=3+\varepsilon$ in a
neighborhood of $d=3$). The Hadamard partie finie of the (UV)
divergencies is then defined as
\begin{eqnarray}
\text{Pf}_{s_1,s_2}\int
d^3\mathbf{x}\,F(\mathbf{x})&\equiv&\lim_{s\rightarrow
0}\biggl\{\int_{\mathbb{R}^3\setminus B_1(s)\cup
B_2(s)}d^3\mathbf{x}\,F(\mathbf{x})\nonumber\\
&&+4\pi\sum_{p+3<0}\frac{s^{p+3}}{p+3}\bigl<\mathop{f}_1{}_p\bigr>
+4\pi\ln \left(
\frac{s}{s_1}\right)\bigl<\mathop{f}_1{}_{-3}\bigr>+1\leftrightarrow
2\biggr\}.
\label{Pf}
\end{eqnarray}
The description of this formula in words is as follows. One first
excises two \textit{spherical} balls $B_1(s)$ and $B_2(s)$ surrounding
the two singularities (each one having the same radius $s$), and one
computes the integral on the volume external to these balls,
\textit{i.e.}, $\mathbb{R}^3\setminus B_1(s)\cup B_2(s)$ --- first term
in Eq. (\ref{Pf}). That integral tends to infinity when $s\rightarrow
0$, but we can subtract from it its purely divergent part, which is
given by the additional terms in (\ref{Pf}) (which obviously are to be
duplicated when there are 2 singularities, \textit{cf.} the symbol
$1\leftrightarrow 2$). The limit $s\rightarrow 0$ then exists (by
definition) and defines Hadamard's partie finie.

Notice the crucial dependence of the partie finie on two constants
$s_1$ and $s_2$ entering the log-terms. These constants have the
dimension of length. We shall say that $s_1$ is the regularization
length scale associated to the Hadamard regularization of the
divergencies near $\mathbf{x} = \mathbf{y}_1$ (similarly for $s_2$).
Note also that the Hadamard partie finie does not depend (modulo
changing the values of $s_1$ and $s_2$) on the \textit{shape} of the
regularization volumes $B_1$ and $B_2$, above chosen as simple
spherical balls (see the discussion in Ref.~\cite{Blanchet:2000nu}).

An important consequence of the definition (\ref{Pf}) is that, in
general, the integral of a gradient $\partial_iF$ is not zero, because
the surface integrals surrounding the singularities become infinite
when the surface areas tend to zero, and may possess a finite part. We
find (see Eq.~(3.4) in \cite{Blanchet:2000nu})
\begin{equation}
\text{Pf}\int d^3\mathbf{x}\,\partial_iF(\mathbf{x})=-4\pi\bigl<n_1^i
\mathop{f}_1{}_{-2}\bigr>+1\leftrightarrow 2\,.
\label{Pfdiv}
\end{equation}
For a general $F\in \mathcal{F}$ the R.H.S. is typically non-zero. This
fact shows that the application of the ordinary Hadamard regularization
in the post-Newtonian iteration has to be supplemented by a notion of
distributional derivatives, in order to ensure that the integrals of
gradients are zero like in the case of regular functions. Notice that
the constants $s_1$ and $s_2$ disappear from the result (\ref{Pfdiv}).
[We shall also see the need, within dim. reg., to consider some
derivatives in the sense of distribution theory.]

Let us apply the definition (\ref{Pf}) to the integral of a
compact-support or ``contact'' term, \textit{i.e.}, made of the product
of some $F$ and a Dirac delta-function at the point 1. Let us formally
assume that\,\footnote{Actually this assumption should be viewed as the
\textit{definition} of a new object we can call
$\text{Pf}\delta^{(3)}(\mathbf{x}-\mathbf{y}_1)$ and which takes the
property (\ref{Pfdelta}). This is exactly what we do in the context of
the extended Hadamard regularization.}
\begin{equation}
\text{Pf}\int d^3\mathbf{x}\,F(\mathbf{x})\delta^{(3)}
(\mathbf{x}-\mathbf{y}_1)=(F)_1\,,
\label{Pfdelta}
\end{equation}
which is the most natural way, within Hadamard's regularization, to
give a sense to such integral. Now the problem with that definition is
that if we want to dispose of some \textit{local} meaning (at any field
point $\mathbf{x}$) for the product of $F$ with the delta-function,
then as a consequence of the non-distributivity we cannot simply equate
$F(\mathbf{x})\delta^{(3)}(\mathbf{x}-\mathbf{y}_1)$ with
$(F)_1\delta^{(3)}(\mathbf{x}-\mathbf{y}_1)$, \textit{i.e.}:
\begin{equation}
F(\mathbf{x})\delta^{(3)}(\mathbf{x}-\mathbf{y}_1)
\not=(F)_1\delta^{(3)}(
\mathbf{x}-\mathbf{y}_1)~~\text{in general for $F\in\mathcal{F}$.}
\label{deltanondistr}
\end{equation}
Indeed, if it were true that $F\delta^{(3)}_1=(F)_1\delta^{(3)}_1$
[for simplicity we denote
$\delta^{(3)}_1\equiv\delta^{(3)}(\mathbf{x}-\mathbf{y}_1)$], then
multiplying by any $G$ we would have
$FG\delta^{(3)}_1 = (F)_1G\delta^{(3)}_1$, and by integrating over
$\mathbb{R}^3$ following the rule (\ref{Pfdelta}) this would yield
$(FG)_1=(F)_1(G)_1$ in contradiction with the violation of
distributivity (\ref{nondistr}). Therefore, both the violation of
distributivity (\ref{nondistr}) and the consequence
(\ref{deltanondistr}) are unescapable in the ordinary Hadamard
regularization.

The previous situation should be contrasted with the $d$-dimensional
case for which the distributivity is always satisfied, as we have
simply
\begin{equation}
(F^{(d)} \, G^{(d)}) (\mathbf{y}_1) = F^{(d)} (\mathbf{y}_1) \,
G^{(d)} (\mathbf{y}_1)\,,
\label{distrd}
\end{equation}
and
\begin{equation}
F^{(d)} (\mathbf{x}) \, \delta^{(d)} (\mathbf{x} - \mathbf{y}_1) =
F^{(d)} (\mathbf{y}_1) \, \delta^{(d)} (\mathbf{x} - \mathbf{y}_1)\,.
\label{contactd}
\end{equation}
Finally, taking the $d \rightarrow 3$ limit, we see that the dim. reg.
way of regularizing a three-dimensional ``contact term'', \textit{i.e.}
a term like $F(\mathbf{x}) \, \delta^{(3)} (\mathbf{x} -
\mathbf{y}_1)$, is by considering it as the $d\rightarrow 3$ limit of
its $d$-dimensional analogue (\ref{contactd}). That is,
\begin{equation}
\text{dim. reg.} \left[ F (\mathbf{x}) \, \delta^{(3)} (\mathbf{x}
- \mathbf{y}_1) \right] \equiv
\left( \lim_{d\rightarrow 3} F^{(d)} (\mathbf{y}_1) \right)
\delta^{(3)} (\mathbf{x} - \mathbf{y}_1)\,,
\label{purecontact}
\end{equation}
where $F^{(d)}$ is the $d$-dimensional version of
$F$, as obtained by solving Einstein's equations in $d$ dimensions
(using the method explained in Section~\ref{IteratedEinstein} below).
There are no poles
in the calculation of the ``contact'' terms in any of the potentials
at the 3PN order so the limit $d\rightarrow 3$ in
Eq. (\ref{purecontact}) always exists. Once again the
dim. reg. prescription (\ref{purecontact}) owns all the good features
one wishes, notably the distributivity as we have emphasized in
Eqs. (\ref{distrd})-(\ref{contactd}).

In the following it will be convenient, in order to compare the
present dim.-reg. calculation with the Hadamard-based work
\cite{Blanchet:2000ub}, to introduce the terminology \textit{pure
Hadamard} regularization to refer to the following ``minimal'' version
of the Hadamard regularization: (a) an integral $\int d^3 \mathbf{x}
\, F(\mathbf{x})$, where $F$ is made of some product of derivatives of
the non-linear potentials $V$, $V_i$, $\cdots$, is regularized by the
ordinary Hadamard partie finie prescription (\ref{Pf}), without
bringing in any distributional contributions (see below for the
treatment of these); (b) the regularization of a product of potentials
$V$, $V_i$, $\hat{W}_{ij}$, $\cdots$ (and their gradients) is assumed
to be distributive, which means that the value at the singular point
$\mathbf{y}_1$ of some polynomial in $V$, $V_i$, $\hat{W}_{ij}$,
$\cdots$ and their gradients, say ${\cal F} [ V, V_i, \hat{W}_{ij},
\partial_i V, \cdots]$, is given by the replacement rule
\begin{equation}
\bigl({\cal F} [ V, V_i, \hat{W}_{ij}, \partial_i V,
\cdots]\bigr)_1\longrightarrow {\cal F} [ (V)_1, (V_i)_1,
(\hat{W}_{ij})_1, (\partial_i V)_1, \cdots]\,;\label{calF1}
\end{equation}
and (c) a contact term, \textit{i.e.} of the form
$F(\mathbf{x})\delta^{(3)}(\mathbf{x}-\mathbf{y}_1)$, appearing in the
calculation of the \textit{sources} of the non-linear potentials, is
regularized by using the rule
\begin{equation}
{\cal F} [ V, V_i, \hat{W}_{ij}, \cdots]
\delta^{(3)}(\mathbf{x}-\mathbf{y}_1)\longrightarrow {\cal F} [ (V)_1,
(V_i)_1, (\hat{W}_{ij})_1, \cdots]\delta^{(3)}(
\mathbf{x}-\mathbf{y}_1)\,,\label{calFdelta}
\end{equation}
(there are no gradients of potentials in the contact terms). The rules
(\ref{calF1})-(\ref{calFdelta}) of the pure-Hadamard regularization
are formally equivalent to assuming the replacement rules
$(FG)_1\longrightarrow (F)_1(G)_1$ together with
$F(\mathbf{x})\delta^{(3)}(\mathbf{x}-\mathbf{y}_1)\longrightarrow
(F)_1 \delta^{(3)}( \mathbf{x}-\mathbf{y}_1)$, in the case where $F$
and $G$ are made of products of our elementary potentials and their
gradients.\footnote{Thus we shall write
$(V\hat{W}_{ij})_1\longrightarrow (V)_1(\hat{W}_{ij})_1$ or
$(V^3\partial_iV)_1\longrightarrow [(V)_1]^3(\partial_iV)_1$, but not,
for instance, $(V)_1\longrightarrow [\bigl(\sqrt{V}\bigr)_1]^2$.}
The rules of the pure Hadamard regularization are, however, well
defined, and are not submitted (by their very definition) to the
consequences of the \textit{ordinary} Hadamard regularization
(\ref{nondistr}) and (\ref{deltanondistr}). Note also that, as done in
previous computations of the 3PN ADM-Hamiltonian
\cite{Jaranowski:1998ky,Jaranowski:1999ye} and the 3PN binary's energy
flux \cite{Blanchet:2001aw}, one can formally use
(\ref{calF1})-(\ref{calFdelta}) at the price of adding a limited number
of arbitrary parameters (considered as unknown).

The definition (\ref{calFdelta}) of pure-Hadamard regularization for
contact terms is useful because we have checked that, when using the
dim. reg. prescription (\ref{purecontact}) (in the limit where
$d\rightarrow 3$), all the contact terms in the sources of the
non-linear potentials $V$, $V_i$, $\hat W_{ij}$, $\cdots$ needed at the
3PN order, \textit{agree} with the result of the pure-Hadamard
regularization. [Of course, we would not need to introduce a notion of
pure-Hadamard regularization in a direct calculation of the equations
of motion in $d$ dimensions, \textit{i.e.}, not based on the
``difference'' between Hadamard and dim. reg., because in such a pure
dim. reg. approach the contact terms would be treated unambiguously
{}from the start using Eq.~(\ref{purecontact}).] On the other hand,
when computing the value at the singular point of the potentials for
insertion into the geodesic equations, we do find some departure
between the dim.-reg. calculation and the (ordinary or extended)
Hadamard one. Let us illustrate these differences by means of the
simplest example which does enter our 3PN calculation, namely the
regularization of $(U)^3\partial_iU$ where $U$ is the Newtonian
potential. In $d$ dimensions $U^{(d)}(\mathbf{x})$ is given by
Eq.~(\ref{U}) [we add here a superscript $(d)$ to indicate the
$d$-dimensionality of a potential and pose $U\equiv U^{(3)}$].
Therefore in dim. reg. the result is simply
\begin{equation}
\lim_{d\rightarrow
3}\Bigl([U^{(d)}(\mathbf{y}_1)]^3\partial_iU^{(d)}(\mathbf{y}_1)\Bigr)
=[U_2(\mathbf{y}_1)]^3
\partial_iU_2(\mathbf{y}_1)~~~~\text{(dim. reg.)}\,,
\label{U4dr}\end{equation}
where $U_2(\mathbf{y}_1)=G m_2/r_{12}$ is the value at point 1 of the
potential of the other particle.
The result (\ref{U4dr}) is the same as when using the pure Hadamard
regularization. Indeed, we find first that
$(U)_1=U_2(\mathbf{y}_1)$ and
$(\partial_iU)_1=\partial_iU_2(\mathbf{y}_1)$, and then, by using the
definition (\ref{calF1}),
\begin{equation}
(U^3\partial_iU)_1\stackrel{\text{def}}{\longrightarrow}
[(U)_1]^3(\partial_iU)_1
=[U_2(\mathbf{y}_1)]^3\partial_iU_2
(\mathbf{y}_1)~~~~\text{(pure-Hadamard)}\,.
\label{U4purehad}\end{equation}
On the other hand, the
latter results contrast with the application of the ordinary Hadamard
regularization for which we find
\begin{equation}
(U^3\partial_iU)_1=[U_2(\mathbf{y}_1)]^3\partial_i
U_2(\mathbf{y}_1)+\frac{6}{5}[U_1(\mathbf{y}_2)]^2
U_2(\mathbf{y}_1)\partial_iU_2
(\mathbf{y}_1)~~~~\text{(ordinary-Hadamard)}\,.
\label{U4had}\end{equation}
The first term is in fact the ``pure-Hadamard'' result which is in
agreement with the dim. reg. one. The second term is an example of the
non-distributivity of the ordinary Hadamard
regularization,\footnote{In the ADM-Hamiltonian the analogue of this
example is the regularization of $U^4$, which gives automatically
$[U_2(\mathbf{y}_1)]^4$ in dim. reg. and also (by definition) in the
pure-Hadamard reg., while
$$(U^4)_1=[U_2(\mathbf{y}_1)]^4+2[U_1(\mathbf{y}_2)]^2[U_2
(\mathbf{y}_1)]^2\,.$$
The latter example represents in fact the only
source of ambiguity present in (the static part of) the
ADM-Hamiltonian formalism \cite{Damour:2001bu}.} which is also
systematically taken into account in the extended-Hadamard
regularization that we shall describe in
Section~\ref{ImprovedHadamard}.

\subsection{Ordinary Hadamard regularization of three-dimensional
Poisson integrals}
\label{HadamardPoisson}
Let us give some reminders of the way we apply the considerations of
Section \ref{PureHadamard} to the computation of Hadamard-regularized
potentials having the form of Poisson or Poisson-like integrals. Let
us first discuss the prescription one has taken in $d=3$ to define the
``value at $\mathbf{x}' = \mathbf{y}_1$'' of a (singular) Poisson
potential $P(\mathbf{x}')$. In $d=3$, the Poisson integral
$P(\mathbf{x}')$, at some field point $\mathbf{x}' \in \mathbb{R}^3$,
of some singular source function $F(\mathbf{x})$ in the class ${\cal
F}$ is defined in the sense of the partie-finie integral (\ref{Pf}),
namely
\begin{equation}
P(\mathbf{x}')=-\frac{1}{4\pi}\text{Pf}_{s_1,s_2}
\int\frac{d^3\mathbf{x}}{\vert\mathbf{x}-
\mathbf{x}'\vert}F(\mathbf{x})\,,
\label{Px}
\end{equation}
where $s_1$ and $s_2$ are the two constants introduced in
Eq.~(\ref{Pf}). At first sight we could think that a good choice for
defining the pure Hadamard value $[P(\mathbf{x}')]_{\mathbf{x}' =
\mathbf{y}_1}$ is simply to replace $\mathbf{x}'=\mathbf{y}_1$ in
(\ref{Px}), \textit{i.e.},
\begin{equation}
P(\mathbf{y}_1) \equiv -\frac{1}{4\pi}\text{Pf}_{s_1,s_2}
\int\frac{d^3\mathbf{x}}{r_1}F(\mathbf{x})\,.
\label{Py1}
\end{equation}
However, the work on the 3PN equations of motion
\cite{Blanchet:2000nv,Blanchet:2000ub} suggested that the definition
(\ref{Py1}) is not acceptable: it did not seem to be able to yield
equations of motion compatible with basic physical properties such as
energy conservation.

The choice adopted in \cite{Blanchet:2000nv,Blanchet:2000ub} is to
define the regularized ``value at $\mathbf{x}' = \mathbf{y}_1$'' of
the function $P(\mathbf{x}')$ by taking the Hadamard partie finie in
the singular limit $\mathbf{x}' \rightarrow \mathbf{y}_1$. Notice
first that $P(\mathbf{x}')$ does not belong (in general) to the class
$\mathcal{F}$ because the Poisson integral will generate some
\textit{logarithms} of $r_1'$ in its expansion when $r_1'\rightarrow
0$. Thus, we shall have, rather than an expansion of type (\ref{Fx}),
\begin{equation}
P(\mathbf{x}')=\sum_{p_0\leq p\leq N}{r_1'}^p
\left[\mathop{g}_1{}_p(\mathbf{n}_1')+
\mathop{h}_1{}_p(\mathbf{n}_1')\ln r_1'\right]+o({r_1'}^N)\,,
\label{ExpandPx}
\end{equation}
where the coefficients ${}_1g_p$ and ${}_1h_p$ depend on the angles
$\mathbf{n}_1'$, and also on the constants $s_1$ and $s_2$, in such a
way that when combining together the terms in (\ref{ExpandPx}) the
constant $r_1'$ always appears in ``adimensionalized'' form like in
$\ln (r_1'/s_1)$. Then we define the Hadamard partie finie at point 1
exactly in the same way as in Eq.~(\ref{F1}), except that we now
include a contribution linked to the (divergent) logarithm of $r_1'$,
which is possibly present into the zeroth-order power of $r_1'$. More
precisely, we define
\begin{equation}
(P)_1 \equiv \bigl<\mathop{g}_1{}_0\bigr>+
\bigl<\mathop{h}_1{}_0\bigr>\ln r_1'\,,
\label{P1def}
\end{equation}
where we introduced a \textit{new regularization length scale} denoted
$r'_1$, which can be seen as some ``small'' but finite cut-off length
scale [so that $\ln r'_1$ in Eq.~(\ref{P1def}) is a finite, but
``large'' cut-off dependent contribution]. We shall see later that the
dependence on $r'_1$ disappears (as it should) when adding to $(P)_1$
the difference $\mathcal{D} P(1) \equiv P^{(d)} (\mathbf{y}_1) -
(P)_1$. To compute the partie finie one must apply the definition
(\ref{P1def}) to the Poisson integral (\ref{Px}), which involves
evaluating correctly the angular integration therein. The result,
proved in Theorem 3 of \cite{Blanchet:2000nu}, is
\begin{equation}
(P)_1=-\frac{1}{4\pi}\text{Pf}_{s_1,s_2}
\int\frac{d^3\mathbf{x}}{r_1}F(\mathbf{x})+\left[\ln\left(
\frac{r_1'}{s_1}\right)-1\right]\bigl<\mathop{f}_1{}_{-2}\bigr>\,.
\label{P1}
\end{equation}
We recover in the first term the value of the potential at the point 1:
$P(\mathbf{y}_1)$, given by Eq.~(\ref{Py1}). The supplementary term
makes the partie finie to differ from the ``na\"{\i}ve'' guess
$P(\mathbf{y}_1)$ in a way which was found to play a significant role
in the computations of \cite{Blanchet:2000nv,Blanchet:2000ub}. The
apparent dependence of the result (\ref{P1}) on the scale $s_1$ is
illusory. The $s_1$-dependence of the R.H.S. of Eq.~(\ref{P1}) cancels
between the first and the second term, so the result depends only on
the constants $r_1'$ and $s_2$, and we have in fact the following
simpler rewriting of (\ref{P1}),
\begin{equation}
(P)_1 \equiv -\frac{1}{4\pi}\text{Pf}_{r_1',s_2}
\int\frac{d^3\mathbf{x}}{r_1}
F(\mathbf{x})-\bigl<\mathop{f}_1{}_{-2}\bigr>\, .
\label{P1'}
\end{equation}
Similarly the regularization performed at point 2 will depend on
$r_2'$ and $s_1$, so that the binary's point-particle dynamics in
Hadamard's regularization depends on four (\textit{a priori}
independent) length scales $r_1'$, $s_2$ and $r_2'$, $s_1$. The
explicit expression of the result (\ref{P1'}) is readily obtained from
the definition of the partie-finie integral (\ref{Pf}). We find (see
the details in Ref.~\cite{Blanchet:2000nu})

\begin{eqnarray}
(P)_1=\lim_{s\rightarrow 0}\Biggl\{&-&\frac{1}{4\pi}
\int_{\mathbb{R}^3\setminus
B_1(s)\cup B_2(s)}\frac{d^3\mathbf{x}}{r_1}F(\mathbf{x})\nonumber\\
&-&\sum_{p+2<0}\frac{s^{p+2}}{p+2}\bigl<\mathop{f}_1{}_p\bigr>-\left[
\ln\left(
\frac{s}{r_1'}\right)+1\right]
\bigl<\mathop{f}_1{}_{-2}\bigr>\nonumber\\
&-&\sum_{\ell=0}^{+\infty}\frac{(-)^\ell}{\ell!}\partial_L
\left(\frac{1}{r_{12}}\right)\left[\sum_{p+\ell+3<0
}\frac{s^{p+\ell+3}}{p+\ell+3}\bigl<n_2^L\mathop{f}_2{}_p\bigr>
+\ln\left(\frac{s}{s_2}\right)\bigl<n_2^L
\mathop{f}_2{}_{-\ell-3}\bigr>\right]\Biggr\}.\nonumber\\
\label{P1result}
\end{eqnarray}
Note that the terms corresponding to singularity 2 involve the
multipolar expansion around the point $\mathbf{y}_2$ of the factor
$1/r_1=1/\vert\mathbf{x}-\mathbf{y}_1\vert$ present into the
integrand.\footnote{We write the multipole expansion in the form
$$\frac{1}{r_1}=\sum_{\ell=0}^{+\infty}\frac{(-)^\ell}{\ell!}\partial_L
\left(\frac{1}{r_{12}}\right)r_2^\ell n_2^L\,,$$ employing our usual
notation where capital letters denote multi-indices: $L\equiv
i_1i_2\cdots i_\ell$, and, for instance, $n_2^L\equiv n_2^{i_1}\cdots
n_2^{i_\ell}$. The expansion is symmetric-trace-free (STF) because
$\delta_{i_\ell i_{\ell-1}}\partial_L(1/r_{12})=\partial_{L-2}\Delta
(1/r_{12})=0$. Here $\partial_L(1/r_{12})$ is a short-hand for $\ell$
partial derivatives $\partial /\partial y_{12}^i$ of $1/r_{12}\equiv
1/\vert\mathbf{y}_1-\mathbf{y}_2\vert$. The multipole expansion in $d$
dimensions (also STF) is given by Eq.~(\ref{r1ofr2}) below.}

Because we work at the level of the equations of motion, many of the
terms we shall need in this paper are in the form of the
\textit{gradient} of a Poisson-like potential. For the gradient we
have a formula analogous to (\ref{P1'}) and given by Eq.~(5.17a) of
\cite{Blanchet:2000nu}, namely
\begin{eqnarray}
(\partial_iP)_1&=&-\frac{1}{4\pi} \text{Pf}_{s_1,s_2}\int d^3
\mathbf{x}\frac{n_1^i}{r_1^2}F(\mathbf{x})
+\ln\left(\frac{r_1'}{s_1}\right)\bigl<n_1^i
\mathop{f}_1{}_{-1}\bigr>
\nonumber\\
&=&-\frac{1}{4\pi} \text{Pf}_{r_1',s_2}\int d^3
\mathbf{x}\frac{n_1^i}{r_1^2}F(\mathbf{x})\,,
\label{diP1}
\end{eqnarray}
where we have taken into account (in the rewriting of the second line)
the always correct fact that the constant $s_1$ cancels out and gets
``replaced'' by $r_1'$. Notice that in (\ref{diP1}) there is no
additional term to the partie finie integral similar to the last term
in (\ref{P1'}). The corresponding explicit expression is
\begin{eqnarray}
(\partial_iP)_1=\lim_{s\rightarrow
0}\Biggl\{&-&\frac{1}{4\pi}\int_{\mathbb{R}^3\setminus B_1(s)\cup
B_2(s)}d^3\mathbf{x}\,\frac{n_1^i}{r_1^2}F(\mathbf{x})\nonumber\\
&-&\sum_{p+1<0}\frac{s^{p+1}}{p+1}\bigl<n_1^i\mathop{f}_1{}_p\bigr>-
\ln\left(
\frac{s}{r_1'}\right)\bigl<n_1^i\mathop{f}_1{}_{-1}\bigr>\nonumber\\
&-&\sum_{\ell=0}^{+\infty}\frac{(-)^\ell}{\ell!}\partial_{iL}
\left(\frac{1}{r_{12}}\right)\left[\sum_{
p+\ell+3<0}\frac{s^{p+\ell+3}}{p+\ell+3}\bigl<n_2^L
\mathop{f}_2{}_p\bigr>+\ln\left(\frac{s}{s_2}
\right)\bigl<n_2^L\mathop{f}_2{}_{-\ell-3}\bigr>\right]\Biggr\}.
\nonumber\\
\label{diP1result}
\end{eqnarray}

Finally we must also treat the more general case of potentials in the
form of retarded integrals [see Eqs.~(\ref{potentialEq})], but because
we shall have to consider (in Section \ref{Diff}) only the
\textit{difference} between the dimensional and Hadamard
regularizations, it will turn out that in fact the first-order
retardation (1PN relative order) is sufficient for this
purpose. Actually, in this paper we are not interested in
radiation-reaction effects, so we shall use the symmetric
(half-retarded plus half-advanced) integral. At the 1PN order we thus
have to evaluate
\begin{equation}
R(\mathbf{x}')=P(\mathbf{x}')+
\frac{1}{2c^2}Q(\mathbf{x}')+\mathcal{O}\left(\frac{1}{c^4}\right),
\label{RPQ}
\end{equation}
where $P(\mathbf{x}')$ is given by (\ref{Px}), and where
$Q(\mathbf{x}')$ denotes (two times) the double or ``twice-iterated''
Poisson integral of the second-time derivative, still endowed with a
prescription of taking the Hadamard partie finie, namely
\begin{equation}
Q(\mathbf{x}')=-\frac{1}{4\pi}\text{Pf}_{s_1,s_2}\int
d^3\mathbf{x}\,\vert\mathbf{x}-\mathbf{x}'
\vert\partial_t^2F(\mathbf{x})\,.
\label{Qx'}
\end{equation}
In the case of $Q(\mathbf{x}')$ the results concerning the partie
finie at point 1 were given by Eqs.~(5.16) and (5.17b) of
\cite{Blanchet:2000nu},
\begin{subequations}\label{Q1}\begin{eqnarray}
(Q)_1&=&-\frac{1}{4\pi}\text{Pf}_{r_1',s_2}\int d^3\mathbf{x}\,r_1
\partial_t^2F(\mathbf{x})+\frac{1}{2}\bigl<\mathop{k}_1{}_{-4}\bigr>
\,,\\
(\partial_iQ)_1&=&\frac{1}{4\pi}\text{Pf}_{r_1',s_2}\int
d^3\mathbf{x}\,n_1^i \partial_t^2F(\mathbf{x})+\frac{1}{2}\bigl<
n_1^i\mathop{k}_1{}_{-3}\bigr>\,,
\end{eqnarray}\end{subequations}
where the ${}_1k_p$'s denote the analogues of the coefficients
${}_1f_p$, parametrizing the expansion of $F$ when $r_1'\rightarrow 0$,
but corresponding to the double time-derivative $\partial_t^2F$ instead
of $F$. [In the following we shall not need the explicit forms of the
results (\ref{Q1}).]

Let us clarify an important point concerning the treatment of the
repeated time derivative $\partial_t^2F(\mathbf{x})$ in
Eqs.~(\ref{Q1}). As we are talking here about Hadamard-regularized
integrals (which excise small balls around both $\mathbf{y}_1$ and
$\mathbf{y}_2$), the value of $\partial_t^2 F(\mathbf{x})$ can be
simply taken in the sense of ordinary functions, \textit{i.e.},
without including eventual ``distributional'' contributions
proportional to $\delta (\mathbf{x} - \mathbf{y}_1)$ or $\delta
(\mathbf{x} - \mathbf{y}_2)$ and their derivatives. However, we know
that such terms are necessary for the consistency of the
calculation. This is why we must also include somewhere in our
formalism the \textit{difference} between the evaluation of these
distributional terms in $d$ dimensions, and the specific
distributional contributions issued from the generalized framework
used in \cite{Blanchet:2000ub}. This difference will be included in
Section \ref{additional} below, among the complete list of additional
contributions specifically related to the use of the extended
regularization approach we shall now describe.

\subsection{Extended-Hadamard regularization}
\label{ImprovedHadamard}
The ``\textit{extended-Hadamard}'' regularization, proposed in
Refs.~\cite{Blanchet:2000nu,Blanchet:2000cw}, tackles the
particular properties of the ordinary Hadamard regularization, notably
the non-distributivity of Eqs. (\ref{nondistr}) and
(\ref{deltanondistr}), and the fact that the integral of a gradient is
not zero [Eq. (\ref{Pfdiv})]. These properties are implemented within
a theory of pseudo-functions, \textit{viz} linear forms defined on the
set of singular functions $\mathcal{F}$. The use of pseudo-functions
in this context enables one to give a precise meaning to the object
$F\delta(\mathbf{x}-\mathbf{y}_1)$ needed in the computation of the
contact terms, and which is otherwise ill-defined in distribution
theory. Furthermore the use of some generalized versions of
distributional derivatives permits a systematic treatment of integrals
and a natural implementation of the property that the integral of a
gradient is always zero. In this paper we shall content ourselves with
recalling the principle of the extended-Hadamard regularization,
and with presenting its ``ready-to-use'' consequences.

To any $F\in \mathcal{F}$ we associate the ``partie finie''
pseudo-function $\text{Pf}F$, which is the linear form on
$\mathcal{F}$ defined by the duality bracket
\begin{equation}
<\text{Pf}F, G > \equiv \text{Pf}\int d^3\mathbf{x} \,
F(\mathbf{x})G(\mathbf{x})\,,
\label{PfF}
\end{equation}
which means that the action of $\text{Pf}F$ on any $G\in\mathcal{F}$
is the partie-finie integral, as given by (\ref{Pf}), of the ordinary
product. The pseudo-function $\text{Pf}F$ reduces to a distribution in
the ordinary sense of Schwartz \cite{Schwartz} when restricted to the
usual set $\mathcal{D}$ of smooth functions with compact support on
$\mathbb{R}^3$. The product of pseudo-functions coincides, by
definition, with the ordinary point-wise product, namely
$\text{Pf}F.\text{Pf}G \equiv \text{Pf}(FG)$. In the class of
pseudo-functions constructed in Ref. \cite{Blanchet:2000nu}, the
``Dirac-delta'' pseudo-function $\text{Pf}\delta_1$ is defined by
\begin{equation}
< \text{Pf}\delta_1, F > \equiv \text{Pf} \int d^3
\mathbf{x}~\delta_1(\mathbf{x})F(\mathbf{x}) \equiv (F)_1\ ,
\label{Pfdelta1}
\end{equation}
where $(F)_1$ denotes Hadamard's partie finie (\ref{F1}). This
definition, which obviously yields a natural extension of the Dirac
function $\delta_1(\mathbf{x})\equiv
\delta^{(3)}(\mathbf{x}-\mathbf{y}_1)$ in the context of Hadamard's
regularization, leads also to new objects which have no equivalent in
distribution theory, the most important one being the pseudo-function
$\text{Pf}(F\delta_1)$ which played a crucial role in
\cite{Blanchet:2000nv,Blanchet:2000ub} for the calculation of the
compact-support parts of potentials as well as the purely
distributional parts of derivatives. It is given by
\begin{equation}
< \text{Pf}(F\delta_1),G > \equiv (FG)_1\ ,
\label{PfFdelta1}
\end{equation}
where one should be reminded that it is in general not allowed to
replace the R.H.S. by the product of regularizations:
$(FG)_1\not=(F)_1(G)_1$.

In the actual computation \cite{Blanchet:2000nv,Blanchet:2000ub} the
pseudo-function $\text{Pf}(F\delta_1)$ acts always on smooth functions
with compact support ($\in\mathcal{D}$), in which case it reduces to a
distribution in the ordinary sense, which was shown to admit the
``intrinsic'' form
\begin{equation}
\text{Pf}(F\delta_1)=\sum_{\ell=0}^{+\infty} \frac{(-)^\ell}{\ell
!}\bigl<n_1^L
\mathop{f}_1{}_{-\ell}\bigr>\partial_L\delta_1~~~~\text{(when
restricted to $\mathcal{D}$)}\,.
\label{PfFdelta}
\end{equation}
Here $L\equiv i_1\cdots i_\ell$ denotes a multi-index composed of
$\ell$ multipolar indices $i_1, \cdots, i_\ell$,
$\partial_L\equiv\partial_{i_1}\cdots \partial_{i_\ell}$ means a
product of $\ell$ partial derivatives $\partial_i=\partial/\partial
x^i$, and $n_1^L\equiv n_1^{i_1}\cdots n_1^{i_\ell}$ a product of
$\ell$ unit vectors (we do not write the $\ell$ summation symbols,
{}from 1 to 3, over the indices composing $L$). Notice that the sum in
Eq. (\ref{PfFdelta}) is finite because $F$ admits some maximal order of
divergency when $r_1\rightarrow 0$. Now we discover that the
``monopole'' term in the latter multipolar sum, having $\ell=0$, is
nothing but $(F)_1\delta_1$ which is exactly the result we would get
following the pure-Hadamard regularization rule (\ref{calFdelta}).
[Indeed, as we are considering here only the contact terms entering the
source terms for the 3PN-level nonlinear potentials, the ``ordinary''
Hadamard regularization $(F)_1$ coincides with the ``pure'' Hadamard
regularization (\ref{calFdelta}).] The sum of the other terms then
define what we can call some non-distributive contributions because
their appearance is the direct consequence of the violation of
distributivity (\ref{nondistr}). Thus,
\begin{equation}
\text{Pf}(F\delta_1)=(F)_1 \delta_1~+~\text{``non-distributivity''
contributions}\,.
\label{PfFdelta'}
\end{equation}
In the work \cite{Blanchet:2000ub} care has been taken of all such
non-distributivity terms. Consider for instance the Poisson integral
of a compact-support term $\text{Pf}(F\delta_1)$ (say, proportional to
the matter source densities $\sigma$, $\sigma_i$ or
$\sigma_{ij}$). Using (\ref{PfFdelta}) the Poisson integral
reads\,\footnote{To apply (\ref{PfFdelta}) we assume that
$\mathbf{x}'$ is distinct from the 2 singularities $\mathbf{y}_1$
and $\mathbf{y}_2$; see \cite{Blanchet:2000nu} for more details.}
\begin{equation}
\int\frac{d^3\mathbf{x}}{\vert\mathbf{x}-
\mathbf{x}'\vert}\text{Pf}(F\delta_1)
=\sum_{\ell=0}^{+\infty}\frac{(-)^\ell}{\ell !}\bigl<n_1^L
\mathop{f}_1{}_{-\ell}\bigr>\partial_L\left(\frac{1}{r_1'}\right)\, ,
\label{PoissonFdelta}
\end{equation}
[where $\partial_L\left(1/r_1'\right)$ should be better written
$\partial_L'\left(1/r_1'\right)$]. Evaluating now the partie finie
(\ref{F1}) at both singular points [\textit{i.e.}, when
$r_1'\rightarrow 0$ and $r_2'\rightarrow 0$] we obtain
\begin{subequations}\label{PoissonFdeltaab}\begin{eqnarray}
\left(\int\frac{d^3\mathbf{x}}{\vert\mathbf{x}-\mathbf{x}'\vert}
\text{Pf}(F\delta_1)\right)_1&=&0\,,\label{PoissonFdelta1}\\
\left(\int\frac{d^3\mathbf{x}}{\vert\mathbf{x}-\mathbf{x}'\vert}
\text{Pf}(F\delta_1)\right)_2&=& \sum_{\ell=0}^{+\infty} \frac{1}{\ell
!}\bigl<n_1^L \mathop{f}_1{}_{-\ell}\bigr>
\partial_L\left(\frac{1}{r_{12}}\right)
=\int\frac{d^3\mathbf{x}}{r_2}\text{Pf}(F\delta_1).
\label{PoissonFdelta2}
\end{eqnarray}\end{subequations}
The result (\ref{PoissonFdelta1}) is in agreement with the
pure-Hadamard regularization; however Eq.~(\ref{PoissonFdelta2}) does
involve some extra terms with respect to the pure-Hadamard calculation,
since the latter is easily seen to simply yield $(F)_1/r_{12}$, which
is nothing but the ``monopolar'' term $\ell=0$ of the multipolar sum in
the R.H.S. of (\ref{PoissonFdelta2}). Therefore we decompose
(\ref{PoissonFdelta2}) as
\begin{equation}
\left(\text{Pf}\int
\frac{d^3\mathbf{x}}{\vert\mathbf{x}-\mathbf{x}'\vert}
F\delta_1\right)_2=\frac{(F)_1}{r_{12}}~+~\text{``non-distributivity''
contributions}\,.
\label{PoissonFdelta2nondistr}
\end{equation}

The second ingredient of the extended-Hadamard regularization concerns
the treatment of partial derivatives in some (extended) distributional
sense. Essentially, one requires \cite{Blanchet:2000nu} that the
derivative reduces to the ordinary derivative in the case of regular
functions, and is such that one can integrate by parts any integrals.
The latter property (valid for the spatial derivative) translates into
\begin{equation}
< \partial_i(\text{Pf}F),G > = - < \partial_i(\text{Pf}G),F >\,.
\label{intparts}
\end{equation}
This rule contains the standard definition of the distributional
derivative \cite{Schwartz} as a particular case. It implies the
important property that the integral of a divergence is zero. Let us
pose
\begin{equation}
\partial_i (\text{Pf} F) = \text{Pf}(\partial_i F) + \text{D}_i[F]\,,
\label{derivedistr}
\end{equation}
where $\text{Pf}(\partial_i F)$ denotes the derivative of $F$ viewed as
an ``ordinary'' pseudo-function, and $\text{D}_i[F]$ represents the
purely distributional part of the spatial derivative (with support
concentrated on $\mathbf{y}_1$ or $\mathbf{y}_2$).

Looking for explicit solutions of the basic
relation (\ref{intparts}) we have found~\cite{Blanchet:2000nu}, with
the help of Eq.~(\ref{Pfdiv}):
\begin{equation}
\text{D}_i[F] = 4\pi \, \text{Pf} \Biggl( n_1^i \biggl[ \frac{1}{2} \,
r_1 \, \mathop{f}_1{}_{-1}+\sum_{k\geq 0} \frac{1}{r_1^k} \,
\mathop{f}_1{}_{-2-k}\biggl] \delta_1 \Biggr) + 1 \leftrightarrow 2
\,.
\label{distrpart}
\end{equation}
Notice that $\text{D}_i[F]$ depends only on the \textit{singular}
coefficients of $F$ (coefficients of negative powers of $r_1$ in the
expansion of $F$). The derivative operator defined by
Eqs.~(\ref{derivedistr})-(\ref{distrpart}) does not represent the
unique solution of (\ref{intparts}), but it has been checked during
the calculation \cite{Blanchet:2000ub} that using another possible
solution results in \textit{physically equivalent} equations of motion
at 3PN order (\textit{i.e.}, reducing to each other by a gauge
transformation). Concerning multiple derivatives we dispose of the
general formula
\begin{equation}
\text{D}_{L}[F]=\sum_{k=1}^\ell\partial_{i_1\dots i_{k-1}}
\text{D}_{i_{k}}[\partial_{i_{k+1}\dots i_\ell}F]\,,
\label{multdistr}
\end{equation}
giving the distributional term associated with the $\ell$-th spatial
derivative, $\text{D}_L[F]\equiv\partial_L\text{Pf}F
-\text{Pf}\partial_L F$ (where $L=i_1i_2\cdots i_\ell$), in terms of
the single derivative $\text{D}_i[F]$. As an example, to treat the
second-derivative of the Newtonian potential, $\partial_{ij}U$ where
$U=G m_1/r_1+G m_2/r_2$, one uses
\begin{equation}
\text{D}_{ij}\left[\frac{1}{r_1}\right]=-\frac{4\pi}{3}\,
\text{Pf}\left(\delta^{ij}+\frac{15}{2}\hat{n}_1^{ij}\right)\delta_1
\ ,\label{ex}
\end{equation}
where $\hat{n}_1^{ij}\equiv
n_1^{ij}-\frac{1}{3}\delta_{ij}$. Therefore the extended
distributional derivative differs in general from the usual Schwartz's
derivative [\textit{cf.} the second term in (\ref{ex})]. [This is
unavoidable if one wants to respect the basic rule of integration by
parts (\ref{intparts}) for general functions in the class
$\mathcal{F}$.] Notice also that we do find a distributional term in
the case of the first derivative: $\text{D}_i\left[1/r_1\right]=2\pi\,
\text{Pf}\left(r_1 n_1^i \delta_1\right)$. We recall also (for future
use) the case of the partial time-derivative, $\partial_t (\text{Pf}
F) = \text{Pf}(\partial_t F) + \text{D}_t[F]$, whose distributional
term is given by (following Ref.~\cite{Blanchet:2000nu})
\begin{equation}
\text{D}_t[F]=v_1^i\mathop{\text{D}}_1{}_i[F]
+v_2^i\mathop{\text{D}}_2{}_i[F]\,,
\label{DtF}
\end{equation}
in terms of the partial derivatives with respect to the
\textit{source} points $\mathbf{y}_1$ and $\mathbf{y}_2$, namely
${}_1\text{D}_i[F]$ and ${}_2\text{D}_i[F]$. The explicit expression
reads
\begin{equation}
\text{D}_t[F] = -4\pi \, \text{Pf} \Biggl( (n_1v_1)\biggl[ \frac{1}{2}
\, r_1 \, \mathop{f}_1{}_{-1}+\sum_{k\geq 0} \frac{1}{r_1^k} \,
\mathop{f}_1{}_{-2-k}\biggl] \delta_1 \Biggr) + 1 \leftrightarrow 2\,,
\label{distrtpart}
\end{equation}
where $(n_1v_1)$ denotes the ordinary scalar product [notice the
over-all sign difference with respect to
Eq.~(\ref{distrpart})]. Multiple time-derivatives can be treated
accordingly to Eq.~(\ref{multdistr}). For instance,
\begin{equation}
\text{D}_{tt}[F] = \text{D}_t[\partial_tF]+\partial_t\text{D}_t[F]\,.
\label{Dtt}
\end{equation}
Following the regularization \cite{Blanchet:2000nu} all the
distributional terms [of type $\text{Pf}(F\delta_1)$ and
$\text{Pf}(F\delta_2)$] issued from the latter distributional
derivatives are to be treated when computing the potentials according
to the extended contact term definitions of
Eqs.~(\ref{PoissonFdelta1})-(\ref{PoissonFdelta2}).

Finally let us turn to the extension of the Hadamard regularization
(introduced in \cite{Blanchet:2000cw}) concerning the definition of a
new operation of regularization, denoted $[F]_1$, consisting of
performing the Hadamard regularization $(F)_1$ within the spatial
hypersurface that is geometrically orthogonal (in a Minkowskian sense)
to the four-velocity of the particle 1. The regularization $[F]_1$
differs from $(F)_1$ by a series of relativistic corrections
calculated in \cite{Blanchet:2000cw}. Together with the other
improvements of the extended-Hadamard regularization, it resulted in
equations of motion in harmonic-coordinates which are manifestly
Lorentz invariant at the 3PN order \cite{Blanchet:2000nv,
Blanchet:2000ub}. Here we give a formula, sufficient for the present
purpose, for expressing $[F]_1$ in terms of the basic regularization
$(F)_1$, defined by (\ref{F1}), at the 1PN order:
\begin{equation}
[F]_1=\biggl(F+\frac{1}{c^2}(\mathbf{r}_1.\mathbf{v}_1)
\Bigl[\partial_tF+\frac{1}{2}v^i_1\partial_i
F\Bigr]\biggr)_1+\mathcal{O}\left(\frac{1}{c^4}\right).
\label{lorentzian}
\end{equation}
The first term is simply $(F)_1$, while the other terms define a set
of relativistic corrections required for ensuring the Lorentz
invariance of the final equations in Hadamard's regularization. Hence,
we decompose (\ref{lorentzian}) into
\begin{equation}
[F]_1=(F)_1~+~\text{``Lorentz'' contributions}\,.
\label{lorentzian'}
\end{equation}

\subsection{Contributions due to the extended-Hadamard regularization}
\label{additional}

After the reminders of the last subsections, we are now in position to
explain the origin of all the contributions [included in the final
result (\ref{a1BF})] which were due to the specific use of the
extended-Hadamard regularization. Actually, we shall list here the
contributions due to the use of the full prescriptions of
\cite{Blanchet:2000nu,Blanchet:2000cw} with respect to those that would
follow from using what we shall call a ``\textit{pure
Hadamard-Schwartz}'' (pHS) regularization. By this we mean: (1)
treating the contact terms of all the non-linear potentials $V$, $V_i$,
$\hat{W}_{ij}$, $\cdots$ as in (\ref{calFdelta}) [we have checked that
for all the potentials involved this is equivalent to
(\ref{purecontact})]; (2) treating the distributional part of an
integrand such as $F_{ij} \, \partial_{ij} \, U$ in the normal
Schwartz's distributional way, for instance\,\footnote{Notice that the
distributional term differs from the extended-Hadamard prescription
(\ref{ex}).}
\begin{equation}
\partial_{ij} \, \left(\frac{1}{r_1}\right) = \frac{3 \, n_1^i \,
n_1^j - \delta^{ij}}{r_1^3} - \frac{4\pi}{3} \, \delta_{ij} \,
\delta^{(3)} (\mathbf{x} - \mathbf{y}_1)\,,
\label{schwder}
\end{equation}
and evaluating the contact term generated by the delta function in the
``pure Hadamard'' way (\ref{calFdelta}); (3) regularize any
three-dimensional integral by the ordinary Hadamard prescription
(\ref{Pf}); and, finally, (4) using systematically, in the
last stage of the calculation where one replaces the metric into the
geodesic equations, the pure Hadamard replacement rule appearing in
(\ref{calF1}) [for instance, we write
$(V^3\partial_iV)_1\longrightarrow [(V)_1]^3(\partial_iV)_1$, creating
therefore a net difference with respect to the ordinary and/or
extended Hadamard regularizations for which
$(V^3\partial_iV)_1\not=[(V)_1]^3(\partial_iV)_1$].

The usefulness of the definition of such a pHS regularization is to
``localize'' the additional contributions brought by dim. reg. to the
occurrence of poles $\propto 1/\varepsilon$ (or ``cancelled'' poles) in
$d$ dimensions.

Our complete list of additional contributions contains seven items.
First of all there are four ``non-distributivity'' contributions of the
type given by Eq.~(\ref{PoissonFdelta2nondistr}):
\begin{enumerate}
\item[(i)] The so-called ``self'' terms, for which the delta-function
in $\text{Pf}(F\delta_1)$ comes from the purely distributional part of
the distributional derivative given by (\ref{distrpart}). The self
terms were derived in Eq. (6.20) in \cite{Blanchet:2000ub}; they read
explicitly
\begin{equation}
\delta^\text{self}a_1^i=\frac{151}{9}\frac{G^4m_1m_2^3}
{c^6r_{12}^5}n_{12}^i+\frac{G^3m_2^3}
{c^6}\left[-\frac{1}{2}(n_{12}v_2)^2n_{12}^i+\frac{1}{10}v_2^2
n_{12}^i+\frac{1}{5}(n_{12}v_2)v_2^i\right],
\label{deltaself0}
\end{equation}
where $(n_{12}v_2)$ denotes the usual scalar product between
$\mathbf{n}_{12}$ and $\mathbf{v}_2$, and where
$v_2^2=\mathbf{v}_2^2$. The expression (\ref{deltaself0}) can be
rewritten in a simpler way as (where we denote for simplicity
$v_2^{jk}\equiv v_2^jv_2^k$)
\begin{equation}
\delta^\text{self}a_1^i=\frac{151}{
9}\frac{G^4m_1m_2^3}{c^6r_{12}^5}n_{12}^i+\frac{1}{30}\frac{G^3m_2^3}
{c^6}v_2^{jk}\partial_{ijk}\left(\frac{1}{r_{12}}\right).
\label{deltaself}
\end{equation}
\item[(ii)] The so-called ``Leibniz'' terms, which are additional
contributions due to the extended distributional derivative, taking
into account the violation of the Leibniz rule when performing some
simplifications of the non-linear potentials at the 3PN order (see the
explanations in Section III B in \cite{Blanchet:2000ub}). The Leibniz
terms were written in Eq.~(6.19) in \cite{Blanchet:2000ub}, and read
\begin{equation}
\delta^\text{Leibniz}a_1^i=-\frac{88}{9}\frac{G^4m_1m_2^3}
{c^6r_{12}^5}n_{12}^i-\frac{1}{6}\frac{
G^3m_2^3}{c^6}v_2^{jk}\partial_{ijk}
\left(\frac{1}{r_{12}}\right).
\label{deltaLeibniz}
\end{equation}
We emphasize that the contributions (\ref{deltaself}) and
(\ref{deltaLeibniz}) represent some additive effects of the use of the
distributional derivative introduced in Ref.~\cite{Blanchet:2000nu},
when compared to the effect of the Schwartz derivative in the pHS
regularization. Note that both Eqs.~(\ref{deltaself}) and
(\ref{deltaLeibniz}) depend on the choice of distributional
derivative, and we have given them here in the case of the
``particular'' derivative\,\footnote{Another derivative was introduced
and discussed in \cite{Blanchet:2000nu} where it is called the
``correct'' one, but it yields physically equivalent 3PN equations of
motion.} defined by (\ref{distrpart}).
\item[(iii)] A special non-distributivity in the compact-support
potential $V$ when it is computed at the 3PN order. In this case the
$\text{Pf}(F\delta_1)$ comes simply from the compact-support
point-particle source $\sigma$ of the potential. Eq. (4.17) of
\cite{Blanchet:2000ub} gives for that term
\begin{equation}
\delta^{V}a_1^i=5\frac{G^4m_1m_2^3}{ c^6r_{12}^5}n_{12}^i\,.
\label{deltaV}
\end{equation}
\item[(iv)] A contribution coming from the compact-support part of the
potential ${\hat T}$ parametrizing the metric at the 3PN order, and
derived at the end of Section IV A in \cite{Blanchet:2000ub}:
\begin{equation}
\delta^{\hat{T}}a_1^i=\frac{1}{15}\frac{G^3m_2^3}
{c^6}v_2^{jk}\partial_{ijk}\left(\frac{1}{r_{12}}\right).
\label{deltaT}
\end{equation}
\end{enumerate}
Besides the non-distributivity of the type
(\ref{PoissonFdelta2nondistr}), we have also the more ``direct''
non-distributivity due to the fact that the pure Hadamard prescription
for the regularization of the value of an expression ``at
$\mathbf{y}_1$'', Eq.~(\ref{calF1}) , differs from the ordinary and/or
extended Hadamard ones [see for instance Eq.~(\ref{U4had})]. It plays
a role only in the last stage of the computation of the 3PN equations
of motion, once we substitute all the potentials computed at the right
PN order into the geodesic equations. We thus have
\begin{enumerate}
\item[(v)] A ``direct'' non-distributivity contribution, which can be
called non-distributivity in the equations of motion (EOM), and given
by Eq.~(6.34) in \cite{Blanchet:2000ub},
\begin{equation}
\delta^\text{EOM}a_1^i=\frac{G^4m_1^2m_2}{
c^6r_{12}^5}\left[\frac{779}{ 210}m_1-\frac{97}{
210}m_2\right]n_{12}^i-\frac{779}{420}\frac{G^3m_1^2m_2}{c^6}\,
v_{12}^{jk}\,\partial_{ijk}\left(\frac{1}{r_{12}}\right),
\label{deltaEOM}
\end{equation}
where $v_{12}^{jk}=v_{12}^jv_{12}^k$ and
$v_{12}^j=v_1^j-v_2^j$. This term involves some combinations of
masses different from those in
Eqs.~(\ref{deltaself})-(\ref{deltaT}). Note that because
$(FG)_1\not=(F)_1(G)_1$ the non-distributivity in the EOM depends on
which prescription has been chosen for the stress-energy tensor of
point-particles. Eq.~(\ref{deltaEOM}) corresponds to the particular
prescription advocated in Section V of \cite{Blanchet:2000cw}. However
it was checked in \cite{Blanchet:2000ub} that different prescriptions
yield physically equivalent equations of motion.
\end{enumerate}
The next correction brought about by the extended-Hadamard
regularization is the one due to the regularization $[F]_1$, performed
in the Lorentzian rest frame of the particle. In practice the effect
of such ``Lorentzian'' regularization boils down to applying
Eq.~(\ref{lorentzian}). It turned out that the only new contribution
of this type came from the regularization of the potential $\hat{X}$
at the 1PN order [and also when deriving the result for
$\delta^\text{EOM}a_1^i$, which is Galilean-invariant, in
Eq.~(\ref{deltaEOM})], leading to
\begin{enumerate}
\item[(vi)] The so-called ``Lorentz'' contribution to the
acceleration, given by Eq.~(5.35) in \cite{Blanchet:2000ub} as
\begin{equation}
\delta^\text{Lorentz}a_1^i=\frac{G^3m_1^2m_2}
{c^6}\left[-\frac{9}{70}v_1^{jk}+\frac{1}{5}v_1^jv_2^k\right]
\partial_{ijk}\left(\frac{1}{r_{12}}\right).
\label{deltaLorentz}
\end{equation}
This term has been crucial for ensuring the Lorentz invariance of the
final 3PN equations of motion in
\cite{Blanchet:2000nv,Blanchet:2000ub}.
\end{enumerate}
Finally, one must also take care of one additional contribution (with
respect to the ``pHS'' definitions) due to the non-Schwartzian way of
treating distributional derivatives. We have already mentioned two
contributions coming from this origin: (i) and (ii) above. Actually,
there is a third one with the same origin and which comes from our
computation (see Section \ref{Difference} below) of the ``difference''
between the dimensional and Hadamard regularizations of retarded
potentials, namely the crucial potentials $\hat X$ and $\hat R_i$ which
must both be expanded to 1PN fractional accuracy. More precisely, this
contribution is due to the repeated time-derivative operator
$\partial_t^2$ coming when expanding the time-symmetric Green function
of the d'Alembertian as $\Box^{-1} = \Delta^{-1} +
c^{-2}\Delta^{-2}\partial_t^2 + {\cal O} (c^{-4})$. We shall explicitly
exhibit in Eqs.~(\ref{DRdiRresult}) below the way these derivatives
enter our calculation of the difference. For technical reasons the
time-derivative $\partial_t^2$ must be kept \textit{inside} the
integrals, so it has to be considered in a distributional sense, and we
have therefore to take into account the different ways of treating the
distributional derivatives in both regularizations. In the
extended-Hadamard regularization the distributional terms are given by
$\text{D}_{tt}[F]$ which is shown in Eqs. (\ref{DtF})-(\ref{Dtt}), and
when they enter the source of some Poisson-type integral they are
evaluated according to Eqs.~(\ref{PoissonFdeltaab}). On the other hand,
in dim. reg. one uses the ordinary Schwartz derivative (in $d$
dimensions) which is described in Section \ref{DistrDeriv}. In this
case the double time-derivative $\partial_t^2$ is computed with the
help of the Gel'fand-Shilov formulae (\ref{gelfand})-(\ref{gelfandt})
below. When examining the difference between the contact terms in
$\text{D}_{tt}[F]$ and those issued from $\lim_{d\rightarrow
3}\partial_t^2 F^{(d)}$, we find that only the source for the 1PN
potential $\hat X$ (or rather for the combination $4 \partial_i \hat X
/ c^4$ which enters the equations of motion) contributes. This gives:
\begin{enumerate}
\item[(vii)] The following ``time-derivative'' contribution to the
acceleration,
\begin{equation}
\delta^\text{time-derivative}a_1^i=-\frac{2}{15}\frac{G^4m_1m_2^3}
{c^6r_{12}^5}n_{12}^i + \frac{4}{35}\frac{G^3m_2^3}{c^6}\,
v_2^{jk}\,\partial_{ijk}\left(\frac{1}{r_{12}}\right).
\label{deltatime}
\end{equation}
This term was part of the final result of
\cite{Blanchet:2000ub}. However it is not mentioned in
\cite{Blanchet:2000ub} because this reference never tried to compare
the results of the extended distributional derivative with those given
by the ordinary Schwartz derivative in 3 dimensions, except in those
cases, \textit{items} (i) and (ii) above, for which the Schwartz
derivative yielded in fact some ill-defined (formally infinite)
expressions in 3 dimensions. [The latter expressions turn out
to be rigorously zero when computed in dimensional regularization.]
\end{enumerate}
In summary, there are in all \textit{seven} different terms,
(i)--(vii), which are specifically due to the extended version of the
Hadamard regularization. The ``pure-Hadamard-Schwartz'' equations of
motion are then obtained from the end result of
\cite{Blanchet:2000ub}, \textit{i.e.}, $\mathbf{a}_1^\text{BF}$ given
by Eq.~(7.16) of \cite{Blanchet:2000ub}, by subtracting these
terms. Therefore we define (see also Section \ref{PureHadStat} below)
\begin{equation}
\mathbf{a}_1^\text{pHS}\equiv\mathbf{a}_1^\text{BF}
-\left(\delta^\text{self}\mathbf{a}_1+\delta^\text{Leibniz}\mathbf{a}_1
+\delta^V\mathbf{a}_1+\delta^{\hat{T}}\mathbf{a}_1
+\delta^\text{EOM}\mathbf{a}_1+\delta^\text{Lorentz}\mathbf{a}_1
+\delta^\text{time-derivative}\mathbf{a}_1\right) ,
\label{defpH}
\end{equation}
and \textit{idem} with $1\leftrightarrow 2$ for the other particle.

\section{Dimensional versus Hadamard regularizations}
\label{Difference}
In this Section we come to the core of our technique for evaluating
the difference between the $d$-dimensional equations of motion and
their pure-Hadamard-Schwartz expressions, defined above and given in
practice by Eq.~(\ref{defpH}).

\subsection{Iteration of Einstein's equations in $d$ dimensions}
\label{IteratedEinstein}
Let us start by indicating how we solved (with sufficient accuracy)
Einstein's field equations in $d$ dimensions. One writes the
post-Minkowskian expansion of Einstein's equations in the guise of
explicit formulae for the elementary potentials $V , V_i , \cdots ,
\hat T$, as given in Section~\ref{FieldEq}. Note that it is
crucial to take into account the explicit $d$-dependence of the
coefficients entering these equations. The first step of the formalism
is to get sufficiently accurate explicit expressions for the basic
linear potentials $V$ and $V_i$. As we do not need to consider here
radiation reaction effects (which do not mix with the UV divergencies
arising at the 3PN level) it is enough to solve Eqs.~(\ref{VLowest}) by
means of the PN expansion of the time-symmetric Green function. For
instance, we have
\begin{equation}
V = -4 \pi G \, \Box_\text{sym}^{-1} \sigma = -4 \pi G \left(
\Delta^{-1} \sigma + \frac{1}{c^2} \, \Delta^{-2} \, \partial_t^2 \,
\sigma + \frac{1}{c^4} \, \Delta^{-3} \, \partial_t^4 \, \sigma +
\frac{1}{c^6} \, \Delta^{-4} \, \partial_t^6 \, \sigma
\right)+\mathcal{O}\left(\frac{1}{c^8}\right).
\label{Gsigma}
\end{equation}
{}From Eq.~(\ref{Tmunu}) we see that the source $\sigma$ reads
\begin{equation}
\sigma (\mathbf{x} , t) = \tilde\mu_1 (t) \, \delta^{(d)} [\mathbf{x}
- \mathbf{y}_1 (t)] + 1 \leftrightarrow 2\,,
\end{equation}
where
\begin{equation}
\tilde\mu_1 (t) = \frac{2}{d-1}\,\frac{m_1 \, c}{\sqrt{-g_{\rho \sigma}
(\mathbf{y}_1, t) \, v_1^{\rho} \, v_1^{\sigma}}} \, \frac{(d-2) +
\mathbf{v}_1^2 / c^2}{\sqrt{-g (\mathbf{y}_1 , t)}}\,.
\end{equation}
Note the presence of many ``contact'' evaluations of field quantities
in $\sigma$. Such terms are unambiguously defined in dimensional
regularization. They are computed by successive iterations
(\textit{e.g.} to get $\tilde\mu_1 (t)$ to 1PN fractional accuracy we
need to have already computed $g_{\mu\nu}$ to order ${\cal O}
(c^{-2})$ included). Those evaluations do not give rise to pole terms
in $\sigma$, up to the 3PN accuracy. Hence, as we said above, we can
consider that their $d \rightarrow 3$ limits define a certain
(three-dimensional) way of estimating contact terms, that we have
checked to be in full agreement with the ``pure-Hadamard''
prescription defined in the previous Section.

Coming now to the spatial dependence of the scalar potential $V$ we get
{}from Eq.~(\ref{Gsigma})
\begin{equation}
V(\mathbf{x},t) = G \, \tilde\mu_1 (t) \, u_1 + \frac{G}{c^2} \,
\partial_t^2 \left[\tilde\mu_1 (t) \, v_1\right] + \cdots + 1
\leftrightarrow 2\,,
\label{Vexpand}
\end{equation}
where we introduced the following elementary solutions $u_1 \equiv
\Delta^{-1} \bigl(-4 \pi \delta_1^{(d)}\bigr)$, $v_1 \equiv \Delta^{-1}
\, u_1$, etc., whose explicit forms are
\begin{subequations}\label{u1v1}\begin{eqnarray}
u_1&=&\tilde{k}\,r_1^{2-d}\,,\label{u1}\\
v_1&=&\frac{\tilde{k}\,r_1^{4-d}}{2(4-d)}\,,
\label{v1}\end{eqnarray}\end{subequations}
where $\tilde{k}$ is related to the usual Eulerian $\Gamma$-function
by\,\footnote{The constant $\tilde{k}$ adopted here is related by
$\tilde{k}=4\pi k$ to the constant $k$ chosen in
\cite{Damour:2001bu}. Our present choice is motivated by the
easy-to-remember fact that $\lim_{d\rightarrow 3}\tilde{k}=1$.}
\begin{equation}
\tilde{k}=\frac{\Gamma\left(\frac{d-2}{2}\right)}{\pi^{
\frac{d-2}{2}}}\,.
\label{ktilde}
\end{equation}
Inserting the explicit expression (\ref{Vexpand}) of $V$ into, say,
the non-linear terms in the R.H.S. of Eq.~(\ref{dalWij}), yields a
d'Alembert equation for the non-linear potential $\hat W_{ij}$ with a
``source function'' which is the sum of some contact terms
$S(\mathbf{x}) \, \delta^{(d)} (\mathbf{x} - \mathbf{y}_a)$ and of an
extended non-linear source $F^{(d)} (\mathbf{x})$ which belongs to the
$d$-dimensional analogue of the class ${\cal F}$, say ${\cal
F}^{(d)}$. More precisely, at each stage of the iteration we find
inhomogeneous wave equations of the type
\begin{equation}
\Box \, W^{(d)} (\mathbf{x}) = F^{(d)} (\mathbf{x}) + \sum_a S_a
(\mathbf{x}) \, \delta^{(d)} (\mathbf{x} - \mathbf{y}_a)\,,
\end{equation}
where the extended source function $F^{(d)} (\mathbf{x})$ is regular
everywhere except at the points $\mathbf{y}_1$ and $\mathbf{y}_2$, in
the vicinity of which it admits an expansion of the general form
($\forall N\in\mathbb{N}$)
\begin{equation}
F^{(d)}(\mathbf{x})=\sum_{\substack{p_0\leq p\leq N\\
q_0\leq q\leq
q_1}}r_1^{p+q\varepsilon}\mathop{f}_1{}_{p,q}^{
(\varepsilon)}(\mathbf{n}_1)+o(r_1^N)\,,
\label{Fd}
\end{equation}
where $p$ and $q$ are relative integers ($p,q\in\mathbb{Z}$), whose
values are limited by some $p_0$, $q_0$ and $q_1$ as indicated. The
expansion (\ref{Fd}) differs from the corresponding expansion in 3
dimensions, as given in (\ref{Fx}), by the appearance of integer powers
of $r_1^\varepsilon$ where $\varepsilon \equiv d-3$. The coefficients
${}_1f_{p,q}^{(\varepsilon)}$ depend on the unit vector $\mathbf{n}_1$
in $d$ dimensions, on the positions and coordinate velocities of the
particles, and also on the characteristic length scale $\ell_0$ of
dimensional regularization. Because $F^{(d)}\rightarrow F$ when
$d\rightarrow 3$ we necessarily have the constraint ($\forall p\geq
p_0$)
\begin{equation}
\sum_{q_0\leq q\leq q_1}\mathop{f}_1{}_{p,q}^{(0)}=\mathop{f}_1{}_p
\label{f1p}\,.
\end{equation}
The iteration continues by inverting the wave operator by means of the
time-symmetric expansion (\ref{Gsigma}). The basic terms of this
expansion which will turn out to be crucial for our 3PN calculation
based on the \textit{difference} are in fact the first two
terms. Focussing on the terms generated by the extended source
$F^{(d)} (\mathbf{x})$ (rather than the simpler contact terms) we can
write the $d$-dimensional analogue of (\ref{RPQ}) as
\begin{equation}
R^{(d)}(\mathbf{x}') \equiv \Box_\text{sym}^{-1} \bigl[F^{(d)}
(\mathbf{x})\bigr] = P^{(d)}(\mathbf{x}') + \frac{1}{2c^2} Q^{(d)}
(\mathbf{x}') + \mathcal{O}\left(\frac{1}{c^4}\right),
\label{Rd}
\end{equation}
where the $d$-dimensional Poisson integral of $F^{(d)}$ reads
\begin{equation}
P^{(d)}(\mathbf{x}')=\Delta^{-1}\bigl[F^{(d)}(\mathbf{x})\bigr]\equiv
-\frac{\tilde{k}}{4\pi}
\int\frac{d^d\mathbf{x}}{\vert\mathbf{x}-\mathbf{x}'\vert^{d-2}}
F^{(d)}(\mathbf{x})\,.
\label{Pdx}
\end{equation}
We have used the fact, already mentioned above, that the
$d$-dimensional elementary solution of the Laplacian reads
\begin{equation}
\Delta\left(\tilde{k}\vert\mathbf{x}-\mathbf{x}'\vert^{2-d}\right)=
-4\pi\delta^{(d)}(\mathbf{x}-\mathbf{x}')\,,
\label{Green}
\end{equation}
(see Appendix \ref{Formulae} for a proof of Eq.~(\ref{Green}) and for
other useful formulae valid in $d$ dim.), while the 1PN term is given
by
\begin{equation}
Q^{(d)}(\mathbf{x}')=
2\Delta^{-2}\bigl[\partial_t^2F^{(d)}(\mathbf{x})\bigr]=
-\frac{\tilde{k}}{4\pi(4-d)} \int
d^d\mathbf{x}\,\vert\mathbf{x}-\mathbf{x}'\vert^{4-d}
\partial_t^2F^{(d)}(\mathbf{x})\,.
\label{Qdx}
\end{equation}
Note the important point that in $d$ dimensions, as in 3 dimensions,
the time-derivative operator $\partial_t^2$ present in the integrand
of (\ref{Qdx}) is to be considered in the sense of distributions (see
further discussion in Section \ref{DistrDeriv} below).

An important technical aspect of the $d$-dimensional PN iteration of
the elementary potentials $V , \cdots , \hat T$ is the existence of
the generalization (\ref{Green}) of the usual Green's function for the
Laplace equation, as well as of its higher PN analogues $\Delta^{-n}
\, \delta^{(d)}$, allowing one to explicitly compute the spatial
dependence of the \textit{linear} potentials $V$, $V_i$ and $K$ for
instance. However, starting with $\hat{W}_{ij}$ we need to
Poisson-integrate \textit{non-linear} sources, such as $\Delta^{-1}
(\partial_i \, U \, \partial_j \, U)$. In three dimensions, these
non-linear contributions are reducible to the knowledge of the basic
non-linear potential $\textsl{g}$, such that $\Delta \textsl{g} =
r_1^{-1} \, r_2^{-1}$. We have succeeded in explicitly computing the
$d$-dimensional analogue of the $\textsl{g}$ potential, namely
\begin{equation}
\textsl{g}^{(d)} (\mathbf{x}) \equiv \Delta^{-1} (r_1^{2-d} \,
r_2^{2-d})\,.
\label{gd}
\end{equation}
Our result is reported in Appendix \ref{Littleg}. As indicated there,
if we wished to explicitly compute some of the higher PN potentials
needed to write the closed form of the non-linear sources relevant to
the 3PN equations of motion, we should extend the calculation of the
potential $\textsl{g}^{(d)}$ to the potentials $f^{(d)}$ and
$f^{(d)}_{12}$ of Appendix \ref{Littleg}.

Luckily, it is not needed to use a closed-form expression for any of
the non-linear potentials. Indeed, similarly to what was used long ago
\cite{Damour:1982wm} when discussing the iteration generated by
Riesz-type sources, Eq.~(\ref{TRiesz}), one can control the UV
\textit{singular} part of $\Box^{-1} F(\mathbf{x})$ from the knowledge
of the UV singular part of its non-linear source
$F(\mathbf{x})$.\footnote{We have checked that for all the non-compact
(extended) potentials involved in this calculation, there are no IR
divergencies, \textit{i.e.}, the integrals converge at infinity
$|\mathbf{x}|\rightarrow\infty$ for any small enough value of
$\varepsilon = d-3$.} More precisely, in the vicinity say of
$\mathbf{y}_1$, at each iteration stage we can decompose the source in
$F(\mathbf{x}) = \text{Sing}_F (\mathbf{x}) + \text{Reg}_F
(\mathbf{x})$ where the \textit{singular} part $\text{Sing}_F
(\mathbf{x})$ (with respect to $\mathbf{y}_1$) is a sum of terms of the
form Eq.~(\ref{Fd}), which are not (in the limit $d \rightarrow 3$)
smooth functions of $\mathbf{x} - \mathbf{y}_1$, and where the
\textit{regular} part $\text{Reg}_F (\mathbf{x})$ is a smooth
$(C^{\infty})$ function of $\mathbf{x} - \mathbf{y}_1$. [The simplest
example of this decomposition is Eq.~(\ref{U}) with, near point
$\mathbf{y}_1$, $\text{Sing}_U (\mathbf{x}) = f \, \tilde k \, G \, m_1
\, r_1^{2-d}$ and $\text{Reg}_U (\mathbf{x}) = U_2 (\mathbf{x})$.] If,
for concreteness, we then consider $P(\mathbf{x}) \equiv \Delta^{-1} \,
F (\mathbf{x})$, the above decomposition entails a corresponding
decomposition of $P(\mathbf{x})$, and it is easy to see that $\Delta \,
\text{Sing}_P (\mathbf{x}) = \text{Sing}_F (\mathbf{x})$. From this
result, we can uniquely determine $\text{Sing}_P (\mathbf{x})$ from
$\text{Sing}_F (\mathbf{x})$ using, \textit{e.g.}, the formula
(\ref{intform}) in Appendix~\ref{Formulae}. This local procedure does
not allow one to compute the regular part of the Poisson potential in
$d$ dimensions. Fortunately, thanks to particular simplifications that
occur in the structure of Einstein's field equations, the knowledge of
$\text{Reg}_P(\mathbf{x})$ in $d$ dimensions for the complicated
non-linear sources is not needed. Indeed, one can see on our explicit
solution of Einstein's field equations at 3PN given in Section
\ref{FieldEq} that there are no ``quartically non-linear'' source terms
of the form, say, $\partial_i V \, \partial_j V \, \hat{W}_{ij}$ or
$\partial_i \hat{W}_{jk} \, \partial_j \hat{W}_{ki}$ for $g_{00}$ at
the 3PN order (see Fig.~\ref{fig5} below).

As explained in Section \ref{DimRegStat} below, a nice way to
understand the origin of the poles $\propto (d-3)^{-1}$ appearing in
the 3PN equations of motion is to use a diagrammatic representation. A
pole can arise in $\mathbf{a}_1$ only when three propagator lines
(including the extra one coming from $\Box^{-1}$ when solving $\Box \,
g_{\mu \nu} =$ non-linear source) can all shrink towards the first
world-line. If terms of the type above (\textit{e.g.} $\partial_i V \,
\partial_j V \, \hat{W}_{ij}$) were present in the source one could
have a diagram where the three shrinking propagators come from
$\Box^{-1}$, $\partial_i V_1$ and $\partial_j V_1$. Then
$\text{Reg}^{(d)} \, [\hat{W}_{ij} (\mathbf{x})]$ would remain as an
external attachment to this diagram (and would then fork into two
``feet'' on the second world-line). In view of the pole $\propto
\varepsilon^{-1}$ (with $\varepsilon \equiv d-3$) arising from the
triplet of shrinking propagators, one would need to know
$\text{Reg}^{(d)} [\hat{W}_{ij} (\mathbf{x})]$ up to $\varepsilon$
accuracy, \textit{i.e.}, $\text{Reg}^{(d)} [\hat{W}_{ij} (\mathbf{x})]
= \text{Reg}^{(3)} [\hat{W}_{ij} (\mathbf{x})] + \varepsilon \,
\hat{W}'_{ij} (\mathbf{x}) + {\cal O} (\varepsilon^2)$ [in which
$\hat{W}'_{ij} (\mathbf{x})$ is defined by this expansion]. If such a
term had been present we would have needed to use the full
$d$-dimensional, globally determined $\textsl{g}$-potential given in
Appendix \ref{Littleg} to determine $\hat{W}'_{ij}$, which would have
entered the final, renormalized equations of motion. However, because
all such terms are absent at the 3PN order, the only external
attachments to the dangerous shrinking diagrams are simple lines, such
for instance as the lines ending on $\mathbf{y}_2 (t)$ in
Figs.~\ref{fig2}d, \ref{fig3}b or \ref{fig4}b presented below. Such
lines do need to be evaluated to accuracy $\varepsilon$, but this is
easy because they represent linear potentials such as $V$ or $V_i$
which are known in dimension $d$ \textit{via} Eq.~(\ref{Vexpand}).

In conclusion, the algorithm we use to solve, with sufficient accuracy,
Einstein's equations in $d$ dimensions consists of: (1) starting from
the fully $d$-dimensional expressions for the linear potentials $V$,
$V_i$ (and more generally for the parts of the non-linear potentials
with delta-function sources); (2) determining the local expansions,
near $\mathbf{y}_a$, of the singular parts of the non-linear potentials
by inverting $\Delta \text{Sing}_P^{(d)} (\mathbf{x}) =
\text{Sing}_F^{(d)} (\mathbf{x})$ \textit{via} formulae (\ref{intform})
of Appendix~\ref{Formulae}; (3) completing $P^{(d)} (\mathbf{x})$ by
adding to $\text{Sing}_P^{(d)} (\mathbf{x})$ the limit when $d
\rightarrow 3$ of $\text{Reg}_P^{(d)} (\mathbf{x})$, namely
$\text{Reg}_P^{(3)} (\mathbf{x})$ which is known from the previous work
on the 3PN equations of motion in 3-dimensions \cite{Blanchet:2000ub}.
Note that we denote by $\text{Reg}_P^{(3)} (\mathbf{x})$ a formal
$d$-dimensional function, $\mathbf{x} \in \mathbb{R}^d$, the explicit
expression of which in terms of $r_1$, $\mathbf{n}_1$, etc. coincides
with its 3-dimensional counterpart. For instance
$\text{Reg}_\textsl{g}^{(3)} (\mathbf{x})$ denotes the usual regular
part of $\textsl{g}^{(3)} (\mathbf{x})$, obtained by subtracting from
$\textsl{g}^{(3)} (\mathbf{x}) \equiv \ln (r_1 + r_2 + r_{12})$ the
two three-dimensional locally singular expansions of
$\Delta_{(3)}^{-1} (r_1^{-1} \, r_2^{-1})$ around $\mathbf{y}_1$ and
$\mathbf{y}_2$ as given by the $d \rightarrow 3$ limit of
Eq.~(\ref{eqD9}) and its $1 \leftrightarrow 2$ analog. After this
double subtraction, $\text{Reg}_\textsl{g}^{(3)} (\mathbf{x})$ is
considered as a function in $\mathbb{R}^d$, and we can use as
approximation to $\textsl{g}^{(d)} (\mathbf{x})$ the explicit
expression $\textsl{g}_\text{loc 1}^{(d)} (\mathbf{x}) +
\textsl{g}_\text{loc 2}^{(d)} (\mathbf{x}) +
\text{Reg}_\textsl{g}^{(3)} (\mathbf{x})$. More generally, in our
calculations we use as approximation to $P^{(d)}$ [which symbolizes
here the non-linear potentials $\hat{R}_i$ and $\hat{Z}_{ij}$ at
Newtonian order, and $\hat{W}_{ij}$ at the 1PN order] the expression
$\text{Sing}_P^{(d)} (\mathbf{x})+\text{Reg}_P^{(3)} (\mathbf{x})$.
Evidently, the subtraction of the singular part needs to be performed
only up to some finite order in $r_1^N$ and $r_2^N$. We have checked
the choice we made of $N$ in each calculation by doing two separate
calculations for the values $N$ and $N+1$, and checking that the
corresponding final results are the same. We performed also direct
checks of the independence of the final results on the precise
$d$-dimensional extensions of the ``regular'' part of the non-linear
potentials, such as $\text{Reg}_P^{(d)} (\mathbf{x})=\text{Reg}_P^{(3)}
(\mathbf{x})+\varepsilon P'(\mathbf{x})+\mathcal{O}(\varepsilon^2)$
[in which $P'(\mathbf{x})$ is defined by this expansion].
We systematically added in all our non-linear potentials $\hat{R}_i$,
$\cdots$, $\hat{W}_{ij}$ some smooth contributions to
$\text{Reg}_P^{(3)} (\mathbf{x})$ vanishing with $\varepsilon$,
\textit{i.e.}, some substitutes for the actual $P'(\mathbf{x})$. These
``substitutes'' were determined in such a way that (i)~they are
homogeneous solutions of the d'Alembertian equation at the required
post-Newtonian order, (ii)~the differential identities obeyed by the
potentials in $d$ dimensions, Eqs.~(\ref{diVi})-(\ref{djWij}), are
indeed satisfied up to the order $\varepsilon$, and with the required
precision $N$ in powers\footnote{Therefore, our verification that the
potentials we need do satisfy the harmonicity conditions
(\ref{GaugeIdentities}) has been done only in the vicinity of the two
particles.} of $r_1$ or $r_2$. And we checked that our final results
are totally insensitive to the introduction of such substitutes for the
function $P'(\mathbf{x})$.

Finally, when evaluating the equations of motion, as given by
Eq.~(\ref{SmallAccel}), we must evaluate the value at $\mathbf{x}' =
\mathbf{y}_1$ of many terms given either by Poisson integrals of the
form (\ref{Pdx}) or their 1PN generalizations (\ref{Qdx}). This is
quite easy to do in dim. reg., because the nice properties of analytic
continuation allow simply to get $[P^{(d)} (\mathbf{x}')]_{\mathbf{x}'
= \mathbf{y}_1}$ (say) by replacing $\mathbf{x}'$ by $\mathbf{y}_1$ in
the explicit integral form (\ref{Pdx}). Finally, we simply have for
the values at $\mathbf{x}' = \mathbf{y}_1$ of the potentials,
\begin{subequations}
\label{PQd}
\begin{eqnarray}
P^{(d)}(\mathbf{y}_1)&=&-\frac{\tilde{k}}{4\pi}
\int\frac{d^d\mathbf{x}}{r_1^{d-2}}F^{(d)}(\mathbf{x})\,,\label{Pd}\\
Q^{(d)}(\mathbf{y}_1)&=&-\frac{\tilde{k}}{4\pi(4-d)} \int
d^d\mathbf{x}\,r_1^{4-d}\partial_t^2F^{(d)}(\mathbf{x})\,,
\label{Qd}
\end{eqnarray}
\end{subequations}
as well as for their spatial gradients,
\begin{subequations}
\label{dPQd}\begin{eqnarray}
\partial_iP^{(d)}(\mathbf{y}_1)&=&-\frac{\tilde{k}(d-2)}{4\pi} \int
d^d\mathbf{x}\,\frac{n_1^i}{r_1^{d-1}}F^{(d)}(\mathbf{x})\,,
\label{dPd}\\
\partial_iQ^{(d)}(\mathbf{y}_1)&=&\frac{\tilde{k}}{4\pi} \int
d^d\mathbf{x}\,n_1^i r_1^{3-d}\partial_t^2F^{(d)}(\mathbf{x})\,.
\label{dQd}
\end{eqnarray}
\end{subequations}
As said above, the main technical step of our strategy will then
consist of computing the \textit{difference} between such
$d$-dimensional Poisson-type potentials (\ref{PQd}) or (\ref{dPQd}),
and their ``pure Hadamard-Schwartz'' 3-dimensional counterparts, which
were already obtained in Section \ref{HadamardPoisson}.

\subsection{Difference between the dimensional and Hadamard
regularizations}
\label{Diff}

We denote the difference between the prescriptions of dimensional and
``pure Hadamard-Schwartz'' regularizations by means of the script
letter $\mathcal{D}$. Given the results $(P)_1$ and
$P^{(d)}(\mathbf{y}_1)$ of the two regularizations [respectively
obtained in Eqs.~(\ref{P1'}) and (\ref{Pd})] we pose
\begin{equation}
\mathcal{D}P(1)\equiv P^{(d)}(\mathbf{y}_1)-(P)_1\,.
\label{DP1}
\end{equation}
That is, $\mathcal{D}P(1)$ is what we shall have to \textit{add} to
the pure Hadamard-Schwartz result (\ref{defpH}) in order to get the
correct $d$-dimensional result. Note that, in this paper, we shall
only compute the first two terms, $a_{-1} \, \varepsilon^{-1} + a_0 +
\mathcal{O} (\varepsilon)$, of the Laurent expansion of
$\mathcal{D}P(1)$ when $\varepsilon \rightarrow 0$. This is the
information we shall need to fix the value of the parameter
$\lambda$. We leave to future work an eventual computation of the
$d$-dimensional equations of motion as an exact function of the
complex number $d$.

Similarly to the evaluation of the difference $\mathcal{D} H \equiv
H^{(d)} - \text{Hadamard} [H^{(3)}]$ in Ref.~\cite{Damour:2001bu}, the
difference (\ref{DP1}) can be obtained by splitting the
$d$-dimensional integral (\ref{Pd}) into three volumes, two spherical
balls $B_1^{(d)}(s)$ and $B_2^{(d)}(s)$ of radius $s$ and centered on
the two singularities, and the external volume ${\mathbb R}^d\setminus
B_1^{(d)}(s)\cup B_2^{(d)}(s)$. When $d\rightarrow 3$ (with fixed
$s$), $B_1^{(d)}(s)$ and $B_2^{(d)}(s)$ tend to the regularization
volumes $B_1(s)$ and $B_2(s)$ we introduced in
Eq.~(\ref{P1result}). Consider first, for a given value $s>0$, the
external integral, over ${\mathbb R}^d\setminus B_1^{(d)}(s)\cup
B_2^{(d)}(s)$. [If wished, two balls with different radii could be
used, with the same result.] Since the integrand is regular on this
domain, it is clear that the external integral reduces in the limit
$\varepsilon\rightarrow 0$ to the one in 3 dimensions that is part of
the Hadamard regularization (\ref{P1result}). So we can write (for any
$s>0$)
\begin{equation}
-\frac{\tilde{k}}{4\pi}\int_{{\mathbb R}^d\setminus B_1^{(d)}(s)\cup
B_2^{(d)}(s)}\frac{d^d\mathbf{x}}{r_1^{d-2}}F^{(d)}(\mathbf{x})=
-\frac{1}{4\pi}\int_{\mathbb{R}^3\setminus B_1(s)\cup
B_2(s)}\frac{d^3\mathbf{x}}{r_1}F(\mathbf{x})+\mathcal{O}(\varepsilon)
\,,
\label{IntExt}
\end{equation}
and we see that when computing the difference $\mathcal{D}P(1)$ the
exterior contributions will cancel out modulo
$\mathcal{O}(\varepsilon)$. Thus we obtain, after this preliminary
step [following Eq.~(\ref{P1result})],
\begin{eqnarray}
\mathcal{D}P(1)&=&\lim_{s\rightarrow
0}\Biggl\{-\frac{\tilde{k}}{4\pi}\int_{B_1^{(d)}(s)}
\frac{d^d\mathbf{x}}{r_1^{d-2}}F^{(d)}(\mathbf{x})-
\frac{\tilde{k}}{4\pi}\int_{B_2^{(d)}(s)}
\frac{d^d\mathbf{x}}{r_1^{d-2}}F^{(d)}
(\mathbf{x})\nonumber\\
&&+\sum_{p\leq
-3}\frac{s^{p+2}}{p+2}\bigl<\mathop{f}_1{}_p\bigr>+\left[\ln\left(
\frac{s}{r_1'}\right)+1\right]
\bigl<\mathop{f}_1{}_{-2}\bigr>\nonumber\\
&&+\sum_{\ell\geq 0}\frac{(-)^\ell}{\ell!}\partial_L
\left(\frac{1}{r_{12}}\right)\left[\sum_{p\leq
-\ell-4}\frac{s^{p+\ell+3}}{p+\ell+3}\bigl<n_2^L
\mathop{f}_2{}_p\bigr>+\ln\left(\frac{s}{s_2}
\right)\bigl<n_2^L\mathop{f}_2{}_{-\ell-3}\bigr>\right]\Biggr\}
\nonumber\\
&&+\mathcal{O}(\varepsilon)\,.
\label{DP1result}
\end{eqnarray}
See Section~IV of Ref.~\cite{Damour:2001bu} for a careful
justification of the formal interversions of limits $s \rightarrow 0$
and $\varepsilon \rightarrow 0$ that we shall do here. The point is
that in order to obtain the difference $\mathcal{D}P(1)$ we do not
need the expression of $F^{(d)}$ for an arbitrary source point
$\mathbf{x}\in\mathbb{R}^d$ but only in the vicinity of the two
singularities: indeed the two local integrals over $B_1^{(d)}(s)$ and
$B_2^{(d)}(s)$ in Eq. (\ref{DP1result}) can be computed by replacing
$F^{(d)}$ by its expansions when $r_1\rightarrow 0$ and
$r_2\rightarrow 0$ respectively. We substitute the $r_1$-expansion
Eq.~(\ref{Fd}) into the local integral over $B_1^{(d)}(s)$, and
integrate that expansion term by term. This readily leads to
\begin{equation}
-\frac{\tilde{k}}{4\pi}\int_{B_1^{(d)}(s)}\frac{d^d\mathbf{x}}
{r_1^{d-2}}F^{(d)}(\mathbf{x})=-\frac{1}{1+\varepsilon}\sum_{p,q}
\frac{s^{p+2+q\varepsilon}}{p+2+q\varepsilon}\bigl<
\mathop{f}_1{}_{p,q}^{(\varepsilon)}\bigr>\,,
\label{Int1}
\end{equation}
where we still use the bracket notation to denote the angular average,
but now performed in $d$ dimensions, \textit{i.e.},
\begin{equation}
\bigl<\mathop{f}_1{}_{p,q}^{(\varepsilon)}\bigr>
\equiv\int\frac{d\Omega_{d-1}(\mathbf{n}_1)}{\Omega_{d-1}}
\mathop{f}_1{}_{p,q}^{(\varepsilon)}(\mathbf{n}_1)\,.
\label{IntAngul}
\end{equation}
Here $d\Omega_{d-1}$is the solid angle element around the direction
$\mathbf{n}_1$, and
$\Omega_{d-1}=2\pi^{\frac{d}{2}}/\Gamma\left(\frac{d}{2}\right)$ is
the volume of the unit sphere with $d-1$ dimensions (see Appendix
\ref{Formulae} for more discussion). To derive (\ref{Int1}) we used
the following relation linking $\tilde{k}$ and $\Omega_{d-1}$,
\begin{equation}
\tilde{k}=\frac{4\pi}{(d-2)\Omega_{d-1}}\,.
\label{unitsphere}
\end{equation}
Concerning the other local integral, over $B_2^{(d)}(s)$, things are a
little bit more involved because we need to perform a multipolar
re-expansion of the factor $r_1^{2-d}$ present in that integral around
the point $\mathbf{y}_2$. Writing down this multipole expansion
presents no problem, and in symmetric-trace-free (STF) form it
reads\,\footnote{The expansion is STF because $\Delta r^{2-d}=0$ in
$d$ dimensions (in the sense of functions). See Appendix
\ref{Formulae} for a compendium of $d$-dimensional formulae on STF
expansions. See also Eq.~(\ref{eqD6}) in Appendix \ref{Littleg}.}
\begin{equation}
r_1^{2-d}=\sum_{\ell=0}^{+\infty}\frac{(-)^\ell}{\ell!}\partial_L
\left(\frac{1}{r_{12}^{1+\varepsilon}}\right)r_2^\ell n_2^L\ .
\label{r1ofr2}
\end{equation}
The multipole expansion being then correctly taken into account, we
obtain
\begin{equation}
-\frac{\tilde{k}}{4\pi}\int_{B_2^{(d)}(s)}\frac{d^d\mathbf{x}}
{r_1^{d-2}}F^{(d)}(\mathbf{x})=-\frac{1}{1+\varepsilon}\sum_{p,q}
\frac{s^{p+\ell+3+(q+1)\varepsilon}}{p+\ell+3+(q+1)\varepsilon}
\sum_{\ell=0}^{+\infty}\frac{(-)^\ell}{\ell!}\partial_L
\left(\frac{1}{r_{12}^{1+\varepsilon}}\right)\bigl<n_2^L
\mathop{f}_2{}_{p,q}^{(\varepsilon)}\bigr>\,.
\label{Int2}
\end{equation}
As we can see, simple poles $\sim 1/\varepsilon$ will occur into our
two local integrals, as determined by (\ref{Int1}) and (\ref{Int2}),
only for the ``critical'' values $p=-2$ and $p=-\ell-3$ respectively.

Next we replace the explicit expressions (\ref{Int1}) and (\ref{Int2})
into the formula (\ref{DP1result}) we had for the ``difference''. As
expected we find that the divergencies when $s\rightarrow 0$, some
value $\varepsilon\neq 0$ being given, cancel out between
Eqs.~(\ref{Int1})-(\ref{Int2}) and the remaining terms in
(\ref{DP1result}), so that the result is finite for any
$\varepsilon\neq 0$. Furthermore, we find that if we neglect terms of
order $\mathcal{O}(\varepsilon)$, the only contributions which remain
are the ones coming from the poles (and their associated finite part),
\textit{i.e.}, for the latter critical values $p=-2$ in the case of
singularity 1 and $p=-\ell-3$ in the case of singularity 2. The other
contributions in (\ref{Int1}) and (\ref{Int2}) have a finite limit when
$\varepsilon\rightarrow 0$ which is therefore cancelled by the
corresponding terms in Hadamard's regularization. As a result we
obtain the following closed-form expression for the difference, which
will constitute the basis of all the practical calculations of the
present paper,
\begin{eqnarray}
\mathcal{D}P(1)=&-&\frac{1}{\varepsilon (1+\varepsilon)}\sum_{q_0\leq
q\leq q_1}\left(\frac{1}{q}+\varepsilon \bigl[\ln
r_1'-1\bigr]\right)\bigl<\mathop
{f}_1{}_{-2,q}^{(\varepsilon)}\bigr>\nonumber\\
&-&\frac{1}{\varepsilon
(1+\varepsilon)}\sum_{q_0\leq q\leq
q_1}\left(\frac{1}{q+1}+\varepsilon\ln s_2\right)
\sum_{\ell=0}^{+\infty}\frac{(-)^\ell}{\ell!}\partial_L
\left(\frac{1}{r_{12}^{1+\varepsilon}}\right)\bigl<
n_2^L\mathop{f}_2{}_{-\ell-3,q}^{(\varepsilon)}\bigr>\nonumber\\
&+&\mathcal{O}(\varepsilon)\,.
\label{DP1total}
\end{eqnarray}
Notice that (\ref{DP1total}) depends on the two ``constants'' $\ln
r_1'$ and $\ln s_2$. As we shall check these $\ln r_1'$ and $\ln s_2$
will exactly cancel out the same constants present in the
``pure-Hadamard'' calculation, so that the dimensionally regularized
acceleration will be finally free of the constants $r_1'$ and
$s_2$. Note also that the coefficients ${}_1f_{p,q}^{(\varepsilon)}$
and ${}_2f_{p,q}^{(\varepsilon)}$ in $d$ dimensions depend on the
length scale $\ell_0$ associated with dimensional regularization
[see Eq.~(\ref{l0})]. Taking this dependence into account one can
verify that $r_1'$ and $s_2$ in (\ref{DP1total}) appear only in the
combinations $\ln (r_1'/\ell_0)$ and $\ln (s_2/\ell_0)$.

Let us give also (without proof) the formula for the difference
between the \textit{gradients} of potentials, \textit{i.e.},
\begin{equation}
\mathcal{D}\partial_iP(1)\equiv\partial_iP^{(d)}(\mathbf{y}_1)-
(\partial_iP)_1\,.
\label{DdiP1}
\end{equation}
The formula is readily obtained by the same method as before, and we
have
\begin{eqnarray}
\mathcal{D}\partial_iP(1)=&-&\frac{1}{\varepsilon} \sum_{q_0\leq q\leq
q_1}\left(\frac{1}{q}+\varepsilon\ln
r_1'\right)\bigl<n_1^i\mathop{f}_1{}_{-1,q
}^{(\varepsilon)}\bigr>\nonumber\\ &-&\frac{1}{\varepsilon
(1+\varepsilon)}\sum_{q_0\leq q\leq
q_1}\left(\frac{1}{q+1}+\varepsilon\ln s_2\right)
\sum_{\ell=0}^{+\infty}\frac{(-)^\ell}{\ell!}\partial_{iL}
\left(\frac{1}{r_{12}^{1+\varepsilon}}\right)\bigl<
n_2^L\mathop{f}_2{}_{-\ell-3,q}^{(\varepsilon)}\bigr>\nonumber\\
&+&\mathcal{O}(\varepsilon)\,.
\label{DdiP1total}
\end{eqnarray}
Formulae (\ref{DP1total}) and (\ref{DdiP1total}) correspond to the
difference of Poisson integrals. But we have already discussed that we
shall need also the difference of inverse d'Alembertian integrals at
the 1PN order. To express as simply as possible the 1PN-accurate
generalizations of Eqs.~(\ref{DP1total}) and (\ref{DdiP1total}), let us
define two \textit{functionals} $\mathcal{H}$ and $\mathcal{H}_i$
which are such that their actions on any $d$-dimensional function
$F^{(d)}$ is given by the R.H.S.'s of Eqs.~(\ref{DP1total}) and
(\ref{DdiP1total}),
\textit{i.e.}, so that
\begin{subequations}
\label{calFFi}
\begin{eqnarray}
\mathcal{D}P(1)&=&\mathcal{H}\left[F^{(d)}\right]\,,
\label{calF}\\
\mathcal{D}\partial_iP(1)&=&\mathcal{H}_i\left[F^{(d)}\right]\,.
\label{calFi}
\end{eqnarray}
\end{subequations}
The difference of 1PN-retarded potentials and gradients of potentials
is denoted
\begin{subequations}
\label{DRdiR}
\begin{eqnarray}
\mathcal{D}R(1)&\equiv&R^{(d)}(\mathbf{y}_1)-(R)_1\,,
\label{DR1}\\
\mathcal{D}\partial_iR(1)&\equiv&\partial_iR^{(d)}(\mathbf{y}_1)
-(\partial_iR)_1\,,
\label{DdiR1}
\end{eqnarray}
\end{subequations}
where in 3 dimensions the potential $R(\mathbf{x}')$ is defined by
Eq.~(\ref{RPQ}) and the regularized values $(R)_1$ and
$(\partial_iR)_1$ follow from (\ref{P1'}), (\ref{diP1}),
(\ref{Q1}), and where in $d$ dimensions $R^{(d)}(\mathbf{y}_1)$ and
$\partial_iR^{(d)}(\mathbf{y}_1)$ are given by (\ref{Rd}), (\ref{PQd}),
(\ref{dPQd}). With this notation we now have our result, which will be
stated without proof, that the difference in the case of such
1PN-expanded potentials reads in terms of the above defined functionals
$\mathcal{H}$ and $\mathcal{H}_i$ as
\begin{subequations}
\label{DRdiRresult}
\begin{eqnarray}
\mathcal{D}R(1)&=&\mathcal{H}\left[F^{(d)}+\frac{r_1^2}{2c^2(4-d)}
\partial_t^2F^{(d)}\right]
-\frac{3}{4c^2}\bigl<\mathop{k}_1{}_{-4}\bigr>
+\mathcal{O}\left(\frac{1}{c^4}\right),
\label{DR1result}\\
\mathcal{D}\partial_iR(1)&=&\mathcal{H}_i\left[F^{(d)}
-\frac{r_1^2}{2c^2(d-2)}\partial_t^2F^{(d)}\right]
-\frac{1}{4c^2}\bigl<n_1^i\mathop{k}_1{}_{-3}\bigr>
+\mathcal{O}\left(\frac{1}{c^4}\right).
\label{DdiR1result}
\end{eqnarray}
\end{subequations}
These formulae involve some ``effective'' functions which are to be
inserted into the functional brackets of $\mathcal{H}$ and
$\mathcal{H}_i$. Beware of the fact that the effective functions are
not the same in the cases of a potential and the gradient of that
potential. Note the presence, besides the main terms
$\mathcal{H}[\cdots]$ and $\mathcal{H}_i[\cdots]$, of some extra
terms, purely of order 1PN, in Eqs.~(\ref{DRdiRresult}). These terms
are made of the average of some coefficients ${}_1k_p$ of the powers
$r_1^p$ in the expansion when $r_1\rightarrow 0$ of the
\textit{second-time-derivative} of $F$, namely $\partial_t^2F$. They
do not seem to admit a simple interpretation. They are important to
get the final correct result.

\subsection{Distributional derivatives in $d$ dimensions}
\label{DistrDeriv}

Let us end this Section by explaining in more details how we dealt
with distributional derivatives in $d$ dimensions. First, it is clear
that if we were dealing with $d$-dimensional integrals of the type
\begin{equation}
I \equiv \int d^d \mathbf{x} \, \varphi_{ij} (\mathbf{x})
\, \partial_{ij} u_1\,,
\label{I}
\end{equation}
where $\varphi_{ij} (\mathbf{x})$ is some (formally) everywhere smooth
function of $\mathbf{x} \in \mathbb{R}^d$, with fast enough decay at
infinity, and where $u_1 \equiv \Delta^{-1} \bigl(-4 \pi
\delta_1^{(d)}\bigr)$ is the elementary Newtonian potential in $d$
dimensions [see Eqs.~(\ref{u1}) above], we should, in a
straightforward $d$-continuation of Schwartz distributional
derivatives, consider that $\partial_{ij} u_1$ contains, besides an
``ordinary'' singular function $\partial_{ij}
(u_1)_{\vert_\text{ord}}$ (treated as a pseudo-function in the sense
of Schwartz), a distributional part proportional to $\delta^{(d)}
(\mathbf{x} - \mathbf{y}_1)$. In other words, we would write
\begin{subequations}\label{du1v1}\begin{eqnarray}
\partial_{ij} (u_1)&=&\partial_{ij} (u_1)_{\bigl|_\text{ord}}-\frac{4
\pi}{d}\delta_{ij}\delta^{(d)} (\mathbf{x}-\mathbf{y}_1)\,,\\
\partial_{ijkl} (v_1)&=&\partial_{ijkl} (v_1)_{\bigl|_\text{ord}}
-\frac{4\pi}{d(d+2)}\Bigl(\delta_{ij}\delta_{kl}+\delta_{ik}
\delta_{jl}+\delta_{il}\delta_{jk}\Bigr)\delta^{(d)} (\mathbf{x}
-\mathbf{y}_1)\,,
\end{eqnarray}\end{subequations}
where the indication ``ord'' refers to the ``ordinary''
(pseudo-function) part of the repeated derivative. We have also added
the corresponding result for the fourth derivatives of the ``less
singular'' kernel $v_1 \equiv \Delta^{-1} \, u_1$, Eqs.~(\ref{v1}).
Note that the decompositions above of $\partial_{ij} \, u_1$ or
$\partial_{ijkl} \, v_1$ into ``ordinary'' and ``distributional''
pieces arise because of our working in ($d$-dimensional)
$\mathbf{x}$-space, and of explicitly computing some derivatives, say
as $\partial_{ij} \, (r_1^n)_{\vert_\text{ord}} = [n \, \delta_{ij} +
n(n-2) \, n_1^i \, n_1^j] \, r_1^{n-2}$. If we were working in the
($d$-dimensional) Fourier-transform space $\mathbf{k}$ (which is where
dimensional continuation is most clearly defined \cite{Collins}), the
corresponding decomposition would be simply algebraic: \textit{e.g.}
$k^i k^j / \mathbf{k}^2 \equiv k^{\langle ij \rangle} / \mathbf{k}^2 +
d^{-1} \, \delta_{ij}$, where $k^{\langle ij \rangle}$ denotes the STF
part of $k^{ij} \equiv k^i k^j$.

The decompositions (\ref{du1v1}) are clearly needed when dealing with
simple integrals of the type (\ref{I}) (with a smooth $\varphi
(\mathbf{x})$) to ensure consistency with the requirement that one may
integrate by parts (which is one of the defining properties of dim.
reg. \cite{Collins}), and we shall therefore employ it, when
applicable. On the other hand, most of the singular integrals that we
have to deal with look like (\ref{I}) but contain a \textit{singular}
function $\varphi (\mathbf{x})$, of the type of Eq.~(\ref{Fd}). It is,
however, a very simplifying feature of dim. reg. that when considering
integrals like (\ref{I}) with some \textit{singular} $\varphi
(\mathbf{x})$ we can simply ignore any distributional contributions
$\propto \delta^{(d)} (\mathbf{x} - \mathbf{y}_1)$ or its
derivatives. Indeed, as long as the integer $q$ in the powers
$r_1^{p+q\varepsilon}$ present in (\ref{Fd}) is different from zero
(which is precisely the case of all delicate terms involving several
propagators shrinking towards a particle world-line), the ``singular''
expansion (\ref{Fd}) can be considered, in dim. reg., as defining a
sufficiently smooth function [by taking both $q\varepsilon$ and $N$
large enough in (\ref{Fd})] which \textit{vanishes}, as well as its
derivatives, at $\mathbf{x} = \mathbf{y}_1$. Therefore, all the
``dangerous'' terms of the form $\text{Sing}_F^{(d)}
(\mathbf{x})\delta^{(d)} (\mathbf{x} - \mathbf{y}_1)$ unambiguously
vanish in dim. reg.

Let us now consider the consequences of this fact for the time
derivatives occurring in expansions such as Eqs.~(\ref{DRdiRresult}).
The distributional time-derivatives, acting in our present example on
$u_1$ or $v_1$, \textit{i.e.}, on functions of $r_1^i\equiv
x^i-y_1^i(t)$, can be treated in a simple way from the rule
$\partial_t=-v_1^i\partial_i$ applicable to the purely distributional
part of the derivative. For instance we can write
\begin{subequations}\label{dtu1v1}\begin{eqnarray}
\partial_t^2(u_1)&=&\partial_t^2(u_1)_{\bigl|_\text{ord}}
-\frac{4\pi}{d}\mathbf{v}_1^2\delta^{(d)}(\mathbf{x}-\mathbf{y}_1)\,,\\
\partial_t^2\partial_{ij}(v_1)&=&\partial_t^2\partial_{ij}
(v_1)_{\bigl|_\text{ord}}-\frac{4\pi}{d(d+2)}\Bigl(\delta_{ij}
\mathbf{v}_1^2+2v_1^iv_1^j\Bigr)\delta^{(d)}
(\mathbf{x}-\mathbf{y}_1)\,.
\end{eqnarray}
\end{subequations}
We have checked using these formulae that all the $d$-dimensional terms
coming from second-order derivatives of potentials, taken in the
distributional sense (for instance the term $\hat W_{ij}\,
\partial_{ij}V$ in the source of the $\hat
X$-potential\,\footnote{Since this term is to be computed at the 1PN
order, not only does it contain second-order derivatives of $u_1$, but
also fourth-order derivatives acting on $v_1$.}) yield the
\textit{same} purely distributional contributions, in the limit
$\varepsilon \rightarrow 0$, as the ones that would be computed using
what we called above a ``pure Schwartz'', three-dimensional computation
of such contributions [to ``smooth'' integrals (\ref{I})]. On the other
hand, the extended version of distributional derivatives introduced in
\cite{Blanchet:2000nu} does yield some specific additional
contributions, two of which were already mentioned in
\cite{Blanchet:2000ub} and are reported in the \textit{items} (i) and
(ii) of Section \ref{additional} above, and a third one (also included
in \cite{Blanchet:2000ub}) which comes in connection with the second
time-derivatives in our formulae for the difference,
Eqs.~(\ref{DRdiRresult}).

Let us indicate here that the distributional second-time-derivatives in
$d$ dimensions have been obtained by using the following
(generalizations of) Gel'fand-Shilov formulae \cite{gelfand}, valid for
general functions $F^{(d)}(\mathbf{x})$ admitting some expansions of
the type (\ref{Fd}): namely, for the spatial derivative,
\begin{equation}
\partial_iF^{(d)}={\partial_iF^{(d)}}_{\bigl|_\text{ord}}+
\Omega_{d-1}\sum_{\ell=0}^{+\infty}\frac{(-)^\ell}{\ell
!}\bigl<n_1^{iL}\mathop{f}_1{}_{-\ell-2,-1}^{(\varepsilon)}
\bigr>\partial_L\delta_1^{(d)}+1\leftrightarrow 2\,,
\label{gelfand}
\end{equation}
where $\partial_L\delta_1^{(d)}$ is the $\ell$-th partial derivative
of the $d$-dimensional Dirac delta-function at the point 1 ($L\equiv
i_1i_2\cdots i_\ell$) and where the angular average is performed over
the $(d-1)$-dimensional sphere having total volume $\Omega_{d-1}$;
and, concerning the time derivative,
\begin{equation}
\partial_tF^{(d)}={\partial_tF^{(d)}}_{\bigl|_\text{ord}}
-\Omega_{d-1}\sum_{\ell=0}^{+\infty}\frac{(-)^\ell}{\ell
!}\bigl<n_1^L(n_1v_1)\mathop{f}_1{}_{-\ell-2,-1}^{(\varepsilon)}
\bigr>\partial_L\delta_1^{(d)}+1\leftrightarrow 2\,.
\label{gelfandt}
\end{equation}
{}From the latter formula one deduces the second time-derivative in a
way similar to Eqs.~(\ref{Dtt}). We have indicated in the \textit{item}
(vii) of Section \ref{additional} the correction it leads to when
comparing with the extended-Hadamard prescription for the second
time-derivative, and we have subtracted it from
$\mathbf{a}_1^\text{BF}$ to define the pure Hadamard-Schwartz result
(\ref{defpH}). Therefore, we consistently do not need to include such
an effect into the differences $\mathcal{D} P(1)$ discussed here.

Finally we are now in position to obtain the supplement of
acceleration $\mathcal{D}\mathbf{a}_1$ induced by dimensional
regularization, which is composed of the sum of all the differences of
potentials and their gradients computed by means of the generals
formulae of (\ref{DP1total}), (\ref{DdiP1total}) and
(\ref{DRdiRresult}). The term $\mathcal{D}\mathbf{a}_1$ when added to
the ``pure-Hadamard-Schwartz'' acceleration defined by (\ref{defpH}),
gives our result for the dimensionally regularized (``dr'')
acceleration
\begin{equation}
\mathbf{a}_1^\text{dr} = \mathbf{a}_1^\text{pHS} + {\mathcal D}
\mathbf{a}_1~~\text{and}~~1\leftrightarrow 2\,.
\end{equation}
More details on the practical computation of $\mathcal{D}\mathbf{a}_a$
(which parts of the potentials contribute; what is the diagrammatic
picture) will be given in Section \ref{DimRegStat}.

\section{Dimensional regularization of the equations of motion}
\label{Results}
\subsection{Structure of the dimensionally regularized equations of
motion}
\label{Structure}
The preceding Section has explained the method we used to compute the
dimensionally regularized equations of motion as the sum ($a = 1,2$;
considered modulo 2)
\begin{equation}
\label{eq5.1}
\mathbf{a}_a^\text{dr} [\varepsilon , \ell_0] =
\mathbf{a}_a^\text{pHS} [r'_a , s_{a+1}] +
{\mathcal D} \mathbf{a}_a
[r'_a , s_{a+1} ; \varepsilon , \ell_0]\,,
\end{equation}
where the label ``pHS'' refers to the ``pure Hadamard-Schwartz''
definition of the acceleration (\textit{i.e.}, the ``raw'' result of
\cite{Blanchet:2000ub}, after subtraction of the additional
contributions quoted in Section \ref{additional} above,
Eq.~(\ref{defpH})), and where ${\mathcal D} \mathbf{a}_a$ is the
difference induced when using dimensional continuation as
regularization method, instead of Hadamard's one. A first check on our
results will be that, as indicated in (\ref{eq5.1}), the four
regularization parameters (with dimension of length), $r'_1 , r'_2 ,
s_1 , s_2$, that enter the Hadamard method must cancel between
$\mathbf{a}_a^\text{pHS}$ and ${\mathcal D} \mathbf{a}_a$ to leave a
result for the dimensionally regularized accelerations
$\mathbf{a}_a^\text{dr}$ which depends only on the two regularization
parameters of dimensional continuation: $\varepsilon \equiv d-3$ and
the basic length scale $\ell_0$ entering Newton's constant in $d$
dimensions, $G = G_N \ell_0^{\varepsilon}$, where we recall that $G_N$
denotes the usual three-dimensional Newton constant.

The dimensionally regularized acceleration (\ref{eq5.1}) has the
structure
\begin{eqnarray}
\label{eq5.2}
\mathbf{a}_a^\text{dr} [\mathbf{y}_{12} , \mathbf{v}_1 , \mathbf{v}_2]
&= &\mathbf{a}_{\text{N}a} [\mathbf{y}_{12}] + \mathbf{a}_{1\text{PN}a}
[\mathbf{y}_{12} , \mathbf{v}_1 , \mathbf{v}_2] \nonumber \\
&+ &
\mathbf{a}_{2\text{PN}a} [\mathbf{y}_{12} , \mathbf{v}_1 ,
\mathbf{v}_2] + \mathbf{a}_{2.5\text{PN}a} [\mathbf{y}_{12} ,
\mathbf{v}_1 , \mathbf{v}_2] + \mathbf{a}_{3\text{PN}a}
[\mathbf{y}_{12} , \mathbf{v}_1 , \mathbf{v}_2]\,,
\end{eqnarray}
where we denote $\mathbf{y}_{12}\equiv \mathbf{y}_1 -
\mathbf{y}_2$. The 3PN term (which is the only one to have a pole at
$\varepsilon = 0$) has a tensor structure of the form (say for the
first particle, $a=1$)
\begin{equation}
\label{eq5.3}
A \, \mathbf{n}_{12} + B' \, \mathbf{v}_1 - B'' \, \mathbf{v}_2\,,
\end{equation}
where, as usual, $\mathbf{n}_{12}\equiv\mathbf{y}_{12}/r_{12}$ denotes
the unit vector directed from particle 2 to particle 1. The scalar
coefficients $A$, $B'$, $B''$ entering the equation of motion of
$\mathbf{y}_1$ can be decomposed in powers of the masses, say
\begin{subequations}
\label{eq5.4}
\begin{eqnarray}
A &=& \sum_{1 \leq n_1 + n_2 \leq 4} c_{n_1 n_2} (\mathbf{v}_1 ,
\mathbf{v}_2 , \mathbf{n}_{12} , \ln r_{12}) \, \frac{G^{n_1 + n_2} \,
m_1^{n_1} \, m_2^{n_2}}{c^6 \, r_{12}^{n_1 + n_2 + 1}} \,,
\label{eq5.4a}\\
B' &=& \sum_{1 \leq n_1 + n_2 \leq 3} c'_{n_1 n_2}
(\mathbf{v}_1 , \mathbf{v}_2 , \mathbf{n}_{12} , \ln r_{12}) \,
\frac{G^{n_1 + n_2} \, m_1^{n_1} \, m_2^{n_2}}{c^6 \, r_{12}^{n_1 +
n_2 + 1}}\,,\label{eq5.4b}\\
B'' &=& \sum_{1 \leq n_1 + n_2 \leq 3}
c''_{n_1 n_2} (\mathbf{v}_1 , \mathbf{v}_2 , \mathbf{n}_{12} , \ln
r_{12}) \, \frac{G^{n_1 + n_2} \, m_1^{n_1} \, m_2^{n_2}}{c^6 \,
r_{12}^{n_1 + n_2 + 1}}\,,\label{eq5.4c}
\end{eqnarray}
\end{subequations}
where $n_1$ and $n_2$ are natural integers, with the restrictions
indicated. Note that, in Eqs.~(\ref{eq5.4}), we have conventionally
factored out an integer power of the ``full'' ($d$-dimensional)
gravitational constant $G$, and a corresponding integer power of
$r_{12}$. This creates a mismatch between the usual 3-dimensional
dimension of, say, $c_{n_1 n_2}^{(d=3)}$ and the dimension of $c_{n_1
n_2}$. Using $G = G_N \ell_0^{\varepsilon}$ one sees that it is the
combination $\ell_0^{(n_1 + n_2) \varepsilon} c_{n_1 n_2}$ which has
the same dimension as $c_{n_1 n_2}^{(d=3)}$. Alternatively said, the
ensuing fact that $r_{12}^{(n_1 + n_2) \varepsilon} c_{n_1 n_2}$ has
the same dimension as $c_{n_1 n_2}^{(d=3)}$ implies, as indicated in
Eqs.~(\ref{eq5.4}), a dependence of $c_{n_1 n_2}$ on $\ln r_{12}$ when
$\varepsilon \rightarrow 0$. Notice also that in Eq.~(\ref{eq5.3}) we
have introduced separate notations for the coefficient of
$\mathbf{v}_1$ and that of $\mathbf{v}_2$. Actually, the Poincar\'e
invariance of the equations of motion imposes the restriction $B' =
B''$ so that the last two terms in Eq.~(\ref{eq5.3}) are proportional
to the relative velocity $\mathbf{v}_{12} \equiv \mathbf{v}_1 -
\mathbf{v}_2$. [Note, however, that $B'$ is not a function of
$\mathbf{v}_{12}$ only; it depends both on $\mathbf{v}_1$ and
$\mathbf{v}_2$.] Because the calculation of the separate contributions
$\mathbf{a}_a^\text{pHS}$ and $\mathcal{D} \mathbf{a}_a$ to the
equations of notion breaks the over-all Poincar\'e invariance of the
formalism, our computation of the separate pieces
$\mathbf{a}_a^\text{pHS}$ and $\mathcal{D} \mathbf{a}_a$ will involve
partial contributions to $B'$ and $B''$ that do not coincide. It is
only at the end of the calculation that the equality $B\equiv B' =
B''$ will be satisfied, so that finally
\begin{equation}
\label{eq5.3'}
\mathbf{a}_{3\text{PN}1} = A \, \mathbf{n}_{12} + B \,
\mathbf{v}_{12}\,.
\end{equation}

Most of the coefficients $c_{n_1 n_2}$, $c'_{n_1 n_2}$, $c''_{n_1
n_2}$ entering the 3PN acceleration are well behaved when $\varepsilon
\rightarrow 0$, in the sense that their evaluation never involves any
poles $\propto 1/\varepsilon$. By this we mean that whatever be the
(reasonable) way of decomposing the integral giving a coefficient in
separate contributions, the latter contributions do not involve poles
$\propto 1/\varepsilon$. The subset of coefficients whose evaluation
involves poles coincides with the set of ``delicate'' coefficients in
the Hadamard regularization, namely the nine coefficients contributing
to terms of the following form in the acceleration of the first
particle:
\begin{eqnarray}
\label{eq5.5}
&&\frac{G^4}{c^6 \, r_{12}^5}\Bigl[ c_{31} \, m_1^3 \, m_2 + c_{22} \,
m_1^2 \, m_2^2 + c_{13} \, m_1 \, m_2^3 \Bigr] \mathbf{n}_{12}
\nonumber\\
&&\qquad + \frac{G^3 \, m_1^2 \, m_2}{c^6 \, r_{12}^4}
\Bigl[ c_{21} (\mathbf{v}_1 , \mathbf{v}_2) \mathbf{n}_{12} \, +
c'_{21} (\mathbf{v}_1 , \mathbf{v}_2) \mathbf{v}_1 - c''_{21}
(\mathbf{v}_1 , \mathbf{v}_2)\, \mathbf{v}_2\Bigr] \nonumber\\
&&\qquad
+ \frac{G^3 \, m_2^3}{c^6 \, r_{12}^4} \Bigl[c_{03} (\mathbf{v}_1 ,
\mathbf{v}_2) \, \mathbf{n}_{12} + c'_{03} (\mathbf{v}_1 ,
\mathbf{v}_2)\, \mathbf{v}_1 - c''_{03} (\mathbf{v}_1 ,
\mathbf{v}_2)\, \mathbf{v}_2 \Bigr]\,.
\end{eqnarray}
The first three terms in Eq.~(\ref{eq5.5}) do not depend on velocities
and will be referred to as the \textit{static} delicate contributions,
by contrast to the \textit{kinetic} delicate contributions involving
the velocity-dependent coefficients $c_{21}$, $c'_{21}$, $c''_{21}$,
$c_{03}$, $c'_{03}$, and $c''_{03}$ (they depend on $\mathbf{v}_1$,
$\mathbf{v}_2$ and also on $\mathbf{n}_{12}$).

\subsection{Pure-Hadamard-Schwartz static contributions to the
equations of motion}
\label{PureHadStat}
Correspondingly to the decomposition (\ref{eq5.1}) of the equations of
motion, the dimensionally regularized static
contributions\,\footnote{As explained above, we consider only the
``delicate'' ones. In the present case, this means that we do not
consider the contribution $c_{04}^\text{dr} = 16 + {\mathcal O}
(\varepsilon)$, unambiguously obtained from the test-mass limit $m_1
\ll m_2$.} $c_{31}^\text{dr}$, $c_{22}^\text{dr}$, $c_{13}^\text{dr}$
to the acceleration of the first particle can be written as the sum
($m+n= 4$, $m\geq 1$, $n\geq 1$)
\begin{equation}
\label{eq5.7}
c_{mn}^\text{dr} [\varepsilon] = c_{mn}^\text{pHS} [r'_1 , s_2] +
{\mathcal D} c_{mn} [r'_1 , s_2 , \varepsilon]\,.
\end{equation}
In this subsection, we discuss the explicit evaluation of the pure
Hadamard-Schwartz static coefficients $c_{mn}^\text{pHS}$.

As explained in the previous Section, the pHS static
contributions $c_{mn}^\text{pHS} [r'_1 , s_2]$ are obtained from the
results reported in \cite{Blanchet:2000ub} by undoing two things.
First, the ``BF'' results reported there for $\mathbf{a}_1$
(Eq.~(7.16) of \cite{Blanchet:2000ub}) were expressed in terms of the
three parameters $r'_1$, $r'_2$ and $\lambda$, instead of the two pure
Hadamard parameters $r'_1$ and $s_2$ more relevant for the present
purpose. The introduction of the parameter $\lambda$ was motivated by
requiring that the full set of equations of motion (which \textit{a
priori} depended on four independent regularizing parameters $r'_1$,
$r'_2$, $s_1$, $s_2$) admit a conserved energy. This led to the link,
Eqs.~(7.9) in \cite{Blanchet:2000ub},\,\footnote{We use here the link
corresponding to the ``particular'' improved distributional derivative
$\text{D}_i[F]$ given by (\ref{distrpart}) above. Another derivative,
the ``correct'' one, was also considered in \cite{Blanchet:2000ub} and
shown to yield equivalent equations of motion. The pure Hadamard
result does not depend on this choice because we shall subtract below
the specific additional contributions coming from the distributional
derivative $\text{D}_i[F]$.}
\begin{equation}
\label{eq5.8}
\ln \left(\frac{r'_2}{s_2}\right) = \frac{159}{308} + \lambda \,
\frac{m_1 + m_2}{m_2}\,.
\end{equation}
When inserting (\ref{eq5.8}) in the expression of
$\mathbf{a}_1^\text{BF}[r'_1, r'_2 , \lambda]$ we find, as it should
be, that the result simplifies to an expression depending only on the
two pure Hadamard parameters $r'_1$ and $s_2$. This leads to the
following net results from \cite{Blanchet:2000ub},
\begin{subequations}\label{eq5.9}\begin{eqnarray}
c_{31}^\text{BF} [r'_1 , s_2] &= & - \frac{3187}{1260}
+ \frac{44}{3} \ln \left( \frac{r_{12}}{r'_1} \right) , \\
c_{22}^\text{BF} [r'_1 , s_2] &= &\frac{34763}{210} -
\frac{41}{16} \, \pi^2\,,
\\
c_{13}^\text{BF} [r'_1 , s_2] &= &\frac{1565}{9} -
\frac{41}{16} \,
\pi^2 - \frac{44}{3} \ln \left( \frac{r_{12}}{s_2} \right).
\end{eqnarray}\end{subequations}
Second, Ref.~\cite{Blanchet:2000ub} obtained their results for the
equations of motion by adding to the pure Hadamard-Schwartz
contributions 7 additional corrections, imposed by their
extended-Hadamard regularization and explained in Section
\ref{additional} above: see the \textit{items} (i)-(vii) there. Note
that these various corrections affect the ``delicate'' contributions to
$\mathbf{a}_1$, in general both the static and kinetic ones, but only
five of them contribute to the static part. These are the self term
(\ref{deltaself}), the Leibniz term (\ref{deltaLeibniz}), the
$V$-correction given by (\ref{deltaV}), the EOM non-distributivity
(\ref{deltaEOM}), and the distributional time-derivative one
(\ref{deltatime}). Following Eq.~(\ref{defpH}), and focussing on the
static contributions, we now \textit{subtract} these static terms from
the result (\ref{eq5.9}) in order to get the looked-for pure Hadamard
contributions:
\begin{subequations}\label{eq5.14}\begin{eqnarray}
c_{31}^\text{pHS} [r'_1 , s_2] &= &c_{31}^\text{BF} [r'_1 , s_2] -
\frac{779}{210}\,, \\
c_{22}^\text{pHS} [r'_1 , s_2] &=
&c_{22}^\text{BF} [r'_1 , s_2] + \frac{97}{210}\,, \\
c_{13}^\text{pHS} [r'_1 , s_2] &= &c_{13}^\text{BF} [r'_1 , s_2] - 5 +
\frac{88}{9} - \frac{151}{9} + \frac{2}{15}\,,
\end{eqnarray}\end{subequations}
\textit{i.e.}, explicitly,
\begin{subequations}\label{eq5.15}\begin{eqnarray}
c_{31}^\text{pHS} [r'_1 , s_2] &= &- \frac{1123}{180} + \frac{44}{3}
\ln \left( \frac{r_{12}}{r'_1} \right) , \\
c_{22}^\text{pHS} [r'_1 , s_2] &= &166 - \frac{41}{16} \, \pi^2\,,\\
c_{13}^\text{pHS} [r'_1 , s_2] &= &\frac{7291}{45} - \frac{41}{16} \,
\pi^2 - \frac{44}{3} \ln \left( \frac{r_{12}}{s_2} \right).
\end{eqnarray}\end{subequations}
Note in passing that though the coefficient $c_{22}$ does not contain
regularization logarithms, its evaluation involves many intermediate
logarithmic divergencies that cancel in the final result. Such
``cancelled logs'' lead to as much ambiguity in the final result than
uncancelled ones that explicitly depend on an arbitrary regularization
scale such as $r'_1$ or $s_2$ in $c_{31}$ or $c_{13}$.

\subsection{Dimensionally regularized static contributions}
\label{DimRegStat}
We now turn to the evaluation of the ``dim.-reg. minus pure-Hadamard''
differences ${\mathcal D} c_{mn}$ in Eq.~(\ref{eq5.7}), coming from the
differences ${\mathcal D} \mathbf{a}_1$ in Eq.~(\ref{eq5.1}). We start
{}from the $d$-dimensional expression for the acceleration
$\mathbf{a}_1$ [see Eq.~(\ref{SmallAccel}) for a short-hand form],
which is itself expressed in terms of the $d$-dimensional elementary
potentials $V$, $V_i$, $K$, $\hat W_{ij}$, $\hat R_i$, $\hat X$, $\hat
Z_{ij}$, $\hat Y_i$ and $\hat T$ defined in Section \ref{FieldEq}. Each
elementary potential can be naturally decomposed in a ``compact'' (or,
equivalently, ``contact'') piece (whose source is compact,
\textit{i.e.}, involves the basic delta-function sources $\sigma$,
$\sigma_i$, $\sigma_{ij}$) and a ``non-compact'' one (whose source is
non-linearly generated and extends all over space). The potentials $V$,
$V_i$ and $K$ are purely ``compact'', $V = V^C$, $V_i = V_i^C$, $K =
K^C$, while all the other potentials admit a decomposition of the form
$\hat W_{ij} = \hat W_{ij}^C + W_{ij}^{NC}$, etc. For instance, the
``compact'' part of $\hat W_{ij}$ is defined by
\begin{equation}
\label{eq5.16}
\Box \, \hat W_{ij}^C = -4 \pi G \left( \sigma_{ij} - \frac{1}{d-2} \,
\delta_{ij} \, \sigma_{kk} \right) ,
\end{equation}
while its ``non-compact'' part is defined by
\begin{equation}
\label{eq5.17}
\Box \, \hat W_{ij}^{NC} = - \frac{1}{2} \, \frac{d-1}{d-2} \,
\partial_i V \, \partial_j V\,.
\end{equation}
A more complicated example is the potential $\hat X = \hat X^C + \hat
X^{NC}$ with
\begin{equation}
\label{eq5.18}
\Box \, \hat X^C = -4\pi G \left[ \frac{1}{d-2} \, V \sigma_{ii} + 2 \,
\frac{d-3}{d-1} \, \sigma_i \, V_i + \left( \frac{d-3}{d-1} \right)^2
\sigma \left( \frac{1}{2} \, V^2 + K \right)\right]\,,
\end{equation}
and
\begin{eqnarray}
\label{eq5.19}
\Box \, \hat X^{NC} &=&\hat W_{ij} \, \partial_{ij} V - 2 \,
\partial_i \, V_j \, \partial_j \, V_i + 2 \, V_i \, \partial_t \,
\partial_i \, V
\nonumber \\
&+ &\frac{1}{2} \, \frac{d-1}{d-2} \, V \partial_t^2 \, V +
\frac{d(d-1)}{4(d-2)^2} \, (\partial_t \, V)^2\,.
\end{eqnarray}
The $NC$ contribution can be further decomposed into the piece of
$\hat X^{NC}$ whose source is quadratic in compact potentials, namely
\begin{equation}
\label{eq5.21}
\Box \, \hat X^{VV} = \hat W_{ij}^C \, \partial_{ij} \, V - 2 \,
\partial_i \, V_j \, \partial_j \, V_i + \ \text{other} \ VV \
\text{terms}\,,
\end{equation}
and its ``cubically non-compact'' piece given by
\begin{equation}
\label{eq5.22}
\Box \, \hat X^{CNC} = \hat W_{ij}^{NC} \ \partial_{ij} \, V =
\Box^{-1} \left( - \frac{1}{2} \, \frac{d-1}{d-2} \, \partial_i \, V
\partial_j \, V \right) \partial_{ij} \, V\,.
\end{equation}
To get a feeling of the actual evaluation of the difference ${\mathcal
D} \mathbf{a}_1$ let us consider a specific contribution to
$\mathbf{a}_1$, say the term
\begin{equation}
\label{eq5.20}
a_1^i [\hat X] \equiv \frac{4}{c^4} \, (\partial_i \, \hat
X)_{\mathbf{x} = \mathbf{y}_1}\,.
\end{equation}
It can be decomposed into: (i) its ``compact'' piece $\mathbf{a}_1
[\hat X^C]$, (ii) its ``quadratically non-compact'' one $\mathbf{a}_1
[\hat X^{VV}]$, (iii) its ``cubically non-compact'' part $\mathbf{a}_1
[\hat X^{CNC}]$.

\begin{figure}[!ht]
\includegraphics[scale=.8]{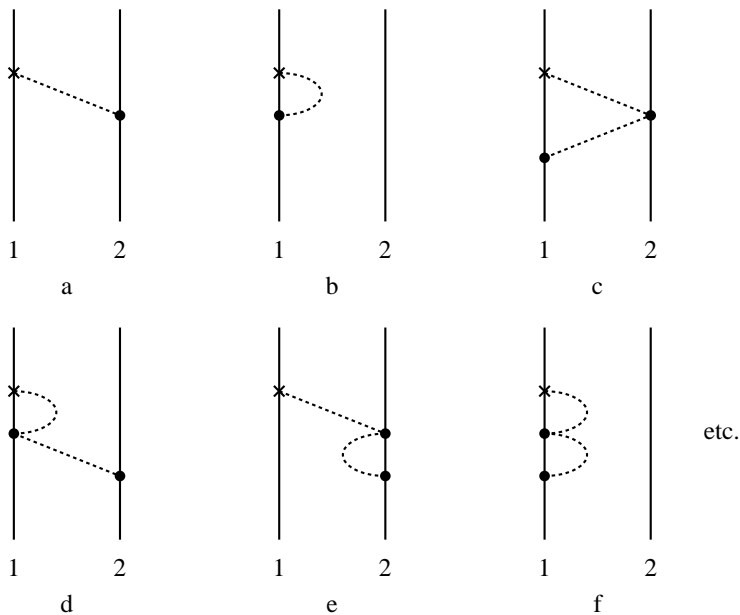}
\caption{Diagrams representing ``compact'' contributions to
acceleration $\mathbf{a}_1$. The dotted line represents $\Box^{-1}$,
the cross represents the field point $\mathbf{x}$ (here taken on the
first worldline), and the bullet represents either a source point or
(in the Figures below) an intermediate nonlinear vertex.}
\label{fig1}
\end{figure}

It is sometimes convenient to think of the various contributions to
$\mathbf{a}_1$ in terms of space-time diagrams. If we represent the
basic delta-function sources [proportional to $m_1 \, \delta
(\mathbf{x} - \mathbf{y}_1)$ and $m_2\, \delta (\mathbf{x} -
\mathbf{y}_2)$] as two world-lines and each propagator $\Box^{-1}$ as a
dotted line, a ``compact'' contribution to $\mathbf{a}_1$ will be
represented by one of the diagrams in Fig.~\ref{fig1}.
For instance, Fig.~\ref{fig1}a can represent a term $(\partial_i \,
V)_1$ in $a_1^i$ in which the (compact) source $\sigma$ of $V$ is
proportional to $m_2 \, \delta (\mathbf{x} - \mathbf{y}_2)$ and
involves no further powers of the masses, while Fig.~\ref{fig1}b
represents a \textit{self-action} term\,\footnote{While in usual
regularization schemes using dimensionful cut-offs (\textit{e.g.} small
length scales $s_1$, $s_2$) the self-action diagrams, such as
Fig.~\ref{fig1}b or Fig.~\ref{fig1}d, are the first divergencies that
one encounters and must then renormalize away,
dimensional regularization has the technically useful property of
setting all of these diagrams to zero. Indeed, when using
a time-symmetric propagator $\Box_\text{sym}^{-1} = \Delta^{-1} +
c^{-2} \, \partial_t^2 \, \Delta^{-2} + \cdots$ these diagrams are
seen to involve the coinciding-point limits of $\vert\mathbf{x} -
\mathbf{y}_1 \vert^{2-d+2n}$, which vanish when
$\mathbf{x}\rightarrow\mathbf{y}_1$ by dimensional continuation in
$d$.} in $(\partial_i \, V)_1$ with source proportional to $m_1 \,
\delta (\mathbf{x} - \mathbf{y}_1)$. By contrast, Fig.~\ref{fig1}c
might correspond to another term in $(\partial_i \, V)_1$ where the
compact source $\sigma$ is concentrated at $\mathbf{y}_2$, $\sigma_2 =
\tilde\mu_2 \, \delta (\mathbf{x} - \mathbf{y}_2)$, and where a part of
the ``effective mass'',
\begin{equation}
\tilde\mu_2 = \frac{2}{d-1}\,\frac{m_2 \, c}{\sqrt{-(g_{\rho \sigma})_2
\, v_2^{\rho} \, v_2^{\sigma}}} \, \frac{(d-2) + \mathbf{v}_2^2 /
c^2}{\sqrt{-(g)_2}}\,,
\end{equation}
contains, besides the overall factor $m_2$, another factor $m_1$. As
all the sources of $\hat X^C$ contain, besides some ``basic''
$\sigma_{\mu\nu}$, a potential ($V$, $V_i$, $V^2$ or $K$), the
diagrams contained in $a_1^i [\hat X^C]$ will be at least of the form
of Fig.~\ref{fig1}c, \ref{fig1}d, \ref{fig1}e, \ref{fig1}f, or will
involve a more complicated mass-dependence.

\begin{figure}[!ht]
\includegraphics[scale=.8]{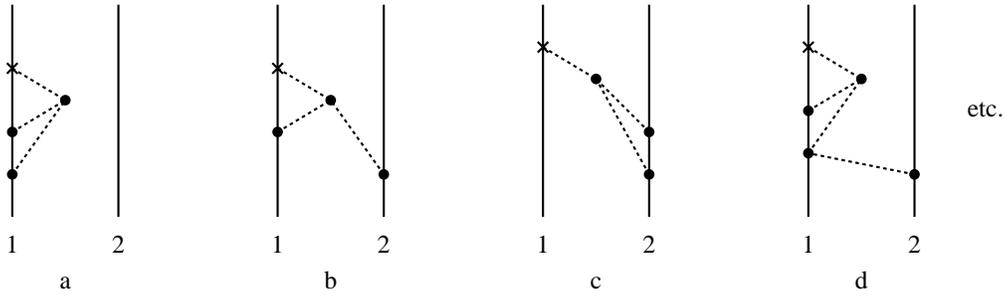}
\caption{Quadratically non-compact contributions to acceleration
$\mathbf{a}_1$.}
\label{fig2}
\end{figure}

\begin{figure}[!ht]
\includegraphics[scale=.8]{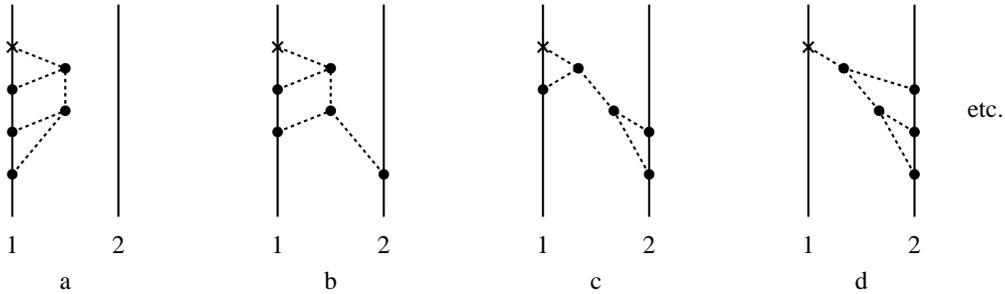}
\caption{Cubically non-compact contributions to acceleration
$\mathbf{a}_1$.}
\label{fig3}
\end{figure}

The quadratically non-compact terms $a_1^i [\hat X^{VV}]$ will then
contain diagrams of the type of Fig.~\ref{fig2}, while the cubically
non-compact term $a_1^i [\hat X^{CNC}]$ contains many subdiagrams of
the type sketched in Fig.~\ref{fig3}.

\begin{figure}[!ht]
\includegraphics[scale=.8]{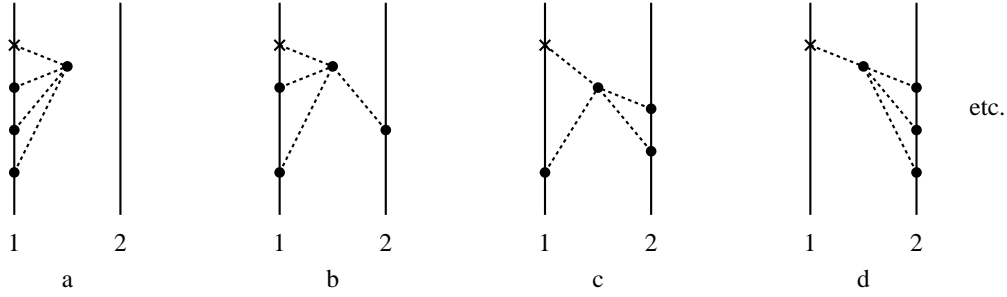}
\caption{Other non-linear contributions to acceleration
$\mathbf{a}_1$.}
\label{fig4}
\end{figure}

\noindent
The particular term (\ref{eq5.20}) that we considered contains only
diagrams of the general type of Figs.~\ref{fig1}, \ref{fig2} or
\ref{fig3}. Note, however, that there are also more non-linear
contributions to $a_1^i$, such as some terms in
\begin{equation}
a_1^i [\hat T] = \frac{16}{c^6} \, (\partial_i \, \hat T)_1\,,
\end{equation}
corresponding to diagrams of the type sketched in Fig.~\ref{fig4}.
Similarly to the diagrams Fig.~\ref{fig1}c or Fig.~\ref{fig2}d, all the
diagrams above can be modified by the presence of additional lines
propagating directly between the two world-lines and corresponding to
``potential'' modifications of compact-support sources.

\begin{figure}[!ht]
\includegraphics[scale=.8]{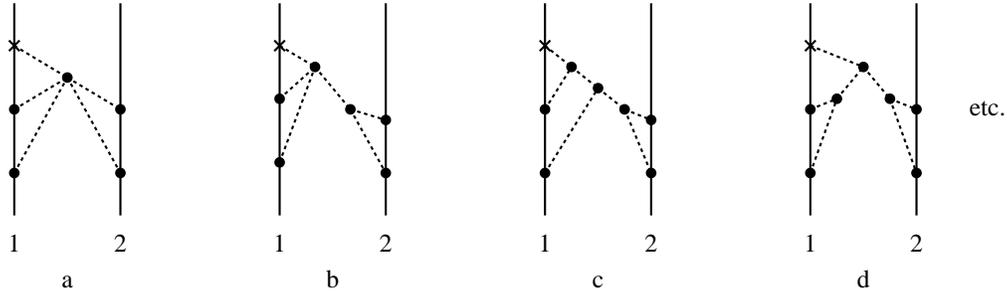}
\caption{Quartically non-compact contributions which do \textit{not}
occur in our calculation of acceleration $\mathbf{a}_1$ at the 3PN
order.}
\label{fig5}
\end{figure}

As underlined in Section~\ref{IteratedEinstein} above, the 3PN
equations of motion do \textit{not} involve ``quartically non-linear''
contributions corresponding to diagrams such as those of
Fig.~\ref{fig5}. Terms like $\Delta^{-1}\left(V^2 \partial_kV
\partial_k V\right)$ or $\Delta^{-1}\left(\partial_k V \partial_k \hat
X\right)$ are of this form, and they do occur in the 3PN acceleration
$\mathbf{a}_1$, but since they involve double contracted gradients, it
was possible to integrate them away thanks to rule (ii) of
Section~\ref{FieldEq}; see Eq.~(\ref{ExpFi}) in
Appendix~\ref{ExpFieldEq} below. On the other hand, terms of the form
$\partial_i V \partial_j V \hat W_{ij}$ or $\partial_i\hat W_{jk}
\partial_j\hat W_{ki}$ do not occur at the 3PN order
$\mathcal{O}(1/c^6)$, although they are of the third
post-\textit{Minkowskian} order $\mathcal{O}(G^4)$.

Drawing diagrams often helps to highlight the nature of the UV
singularities contained in the integrals they represent. As a rule of
thumb, the ``delicate'' diagrams, that might involve poles, or
cancelled poles, when $\varepsilon \rightarrow 0$ (corresponding to
logarithms, or cancelled logarithms, in $d=3$) are characterized by
the presence of a subdiagram containing three propagator lines that
can simultaneously shrink to zero size, as a subset of vertices
coalesce together on one of the two world-lines. Examples of such UV
dangerous diagrams are Fig.~\ref{fig2}d and Fig.~\ref{fig3}b [for
vertices coalescing towards $\bigl(t , \mathbf{y}_1 (t)\bigr)$] or
Fig.~\ref{fig3}d and Fig.~\ref{fig4}d [for vertices coalescing on the
second world-line]. The former diagrams can give poles proportional to
$m_1^2 \, m_2$ (with some velocity dependence, or some extra mass
dependence due to an extra line propagating between the two
world-lines), while the latter can give poles proportional to $m_2^3$
(possibly with some extra velocity or mass dependence). The reason why
three simultaneously shrinking propagators can yield poles as
$\varepsilon
\rightarrow 0$ is easy to see in the approximation where the
relativistic propagators
$\Box^{-1}$ are replaced by non-relativistic ones
$\Delta_{\mathbf{x},\mathbf{x}'}^{-1} =
-\frac{\tilde{k}}{4\pi}\vert\mathbf{x} -
\mathbf{x}'\vert^{-1-\varepsilon}$. Indeed, when three such
propagators shrink simultaneously, the overall integral contains a
subintegral of the form $\int d^{3+\varepsilon}\mathbf{x}\, (\vert
\mathbf{x} \vert^{-1-\varepsilon})^3 \sim \int_0^a \text{dr} \,
r^{-1-2\varepsilon} \sim a^{-2\varepsilon} / (-2\varepsilon)$.

On the other hand, beyond our obtaining a heuristic feeling of what
are the origins of the poles in $\mathbf{a}_1$, we did not use a
diagrammatic technique for evaluating the equations of motion. [Note,
however, that a generalization of the (2PN level) work
\cite{Damour:1996kt} would lead to a diagrammatic technique for
evaluating the Fokker Lagrangian of two point masses.] Our actual
computations used the techniques elaborated in the previous Sections.

We evaluated the contributions to the difference ${\mathcal D}
\mathbf{a}_1$ coming from all the terms in the expression for
$\mathbf{a}_1$ deduced from (\ref{eqGeod}) together with the complete
expanded forms (\ref{ExpPi})-(\ref{ExpFi}). However, as expected from
various arguments --- diagrammatic analysis, existence of (possibly
cancelled) logarithms in the corresponding $d=3$ evaluation --- most
of the terms lead to a vanishing difference ${\mathcal D}
\mathbf{a}_1$. The only terms that give non-vanishing contributions to
${\mathcal D} \mathbf{a}_1$ are the four terms given in
Eq.~(\ref{SmallAccel}),
\begin{eqnarray}
a_1^i [\hat X] &=& \frac{4}{c^4} \, (\partial_i \, \hat X)_1\,,
\quad a_1^i [\hat T] = \frac{16}{c^6} \, (\partial_i \, \hat T)_1 \,,
\nonumber\\
a_1^i [\hat R_i] &=& \frac{8}{c^4} \, \frac{d}{dt} \,
(\hat R_i)_1\,, \quad a_1^i [\hat Y_i] = \frac{16}{c^6} \,
\frac{d}{dt} \, (\hat Y_i)_1\,.
\label{eq5.23}\end{eqnarray}
Note that, for the contributions associated to $\hat X$ and $\hat R_i$,
one needs a 1PN-accurate treatment of both their respective sources and
the propagator $\Box^{-1}$. Apart from the compact support terms in the
sources for $\hat X$, $\hat T$, $\hat R_i$ and $\hat Y_i$ which lead to
zero difference, most of the non-compact terms do lead to some
non-vanishing contributions to the difference of acceleration
${\mathcal D} \mathbf{a}_1$.
We give in Tables~\ref{table1}-\ref{table4} the contributions to
${\mathcal D} c_{mn}$ associated to the various individual source
terms of the ``delicate'' potentials $\hat X$, $\hat T$, $\hat R_i$
and $\hat Y_i$, which were displayed in Section \ref{FieldEq},
Eqs.~(\ref{potentialEq}) [of course, we limit ourselves to non-compact
source terms]. In these tables, we use the simplifying notation
\begin{equation}
\ln_r \equiv \ln\left(\overline q\, r'_1 r_{12}\right),
\qquad
\ln_s \equiv \ln\left(\overline q\, s_2 r_{12}\right),
\qquad
\overline q \equiv 4\pi e^C\,,
\label{lnrlns}
\end{equation}
where $C = 0.577\ldots$ denotes the Euler constant.

{
\renewcommand{\arraystretch}{1.5}
\begin{table}[t]
\caption{Static contributions of ${\mathcal{D}}\partial_i\hat
X(1)$. All the results are presented modulo some neglected terms
$\mathcal{O}(\varepsilon)$. The ``principal'' part of a term
corresponds to the term $\Delta^{-1}$ in the 1PN symmetric propagator
$\Box_\text{1PN}^{-1} = \Delta^{-1} + c^{-2} \partial_t^2
\Delta^{-2}$, while the ``retarded'' part corresponds to the purely
1PN piece $c^{-2} \partial_t^2 \Delta^{-2}$. The ``extra term'' refers
to the last term in the R.H.S. of Eq.~(\ref{DdiR1result}). Note that,
in view of Eqs.~(\ref{eq5.23}), one must multiply the results by a
factor 4 in order to get the contributions to the coefficients
$\mathcal{D}c_{mn}$ in the equations of motion.}
\label{table1}
\begin{ruledtabular}
\begin{tabular}{c|c|c|c}
&$\frac{1}{4} \, {\mathcal D} c_{31}$&$\frac{1}{4} \, {\mathcal D}
c_{22}$&$\frac{1}{4} \, {\mathcal D} c_{13}$\\
\hline
$\hat W_{ij}\, \partial_{ij}V|_\text{principal}$&$\frac{4}{5}
+\frac{1}{6\varepsilon}
-\frac{1}{3}\ln_r$&$\frac{103}{200}
+\frac{1}{20\varepsilon}
-\frac{1}{10}\ln_r$&$-\frac{5}{18}
-\frac{1}{12\varepsilon}
+\frac{1}{6}\ln_s$\\
$\hat W_{ij}\, \partial_{ij}V|_\text{retarded}$&$-\frac{25}{18}
-\frac{1}{3\varepsilon} +\frac{2}{3}\ln_r$&$-\frac{2947}{1800}
-\frac{23}{60\varepsilon}
+\frac{23}{30}\ln_r$&$-\frac{53}{90}
-\frac{1}{12\varepsilon}
+\frac{1}{6}\ln_s$\\
$\frac{1}{2}
\left(\frac{d-1}{d-2}\right)
V \partial^2_t V|_\text{principal}$&$\frac{11}{18}
+\frac{1}{6\varepsilon}
-\frac{1}{3}\ln_r$&$\frac{11}{18}
+\frac{1}{6\varepsilon}
-\frac{1}{3}\ln_r$&$0$\\
$\frac{1}{2}
\left(\frac{d-1}{d-2}\right)
V \partial^2_t V|_\text{retarded}$&$\frac{11}{18}
+\frac{1}{6\varepsilon}
-\frac{1}{3}\ln_r$&$\frac{11}{18}
+\frac{1}{6\varepsilon}
-\frac{1}{3}\ln_r$&$0$\\
extra term&$-\frac{1}{6}$&$-\frac{13}{60}$&$0$\\
\hline
Total&$\frac{7}{15} +\frac{1}{6\varepsilon}
-\frac{1}{3}\ln_r$&$-\frac{7}{60}$&$-\frac{13}{15}
-\frac{1}{6\varepsilon} +\frac{1}{3}\ln_s$\\
\end{tabular}
\end{ruledtabular}
\end{table}}

{
\renewcommand{\arraystretch}{1.5}
\begin{table}[!ht]
\caption{Static contributions of ${\mathcal{D}}\partial_i\hat T(1)$.}
\label{table2}
\begin{ruledtabular}
\begin{tabular}{c|c|c|c}
&$\frac{1}{16} \, {\mathcal D} c_{31}$&$\frac{1}{16} \, {\mathcal D}
c_{22}$&$\frac{1}{16} \, {\mathcal D} c_{13}$\\
\hline
$\hat Z_{ij}\partial_{ij}V$&$-\frac{119}{900}
-\frac{1}{30\varepsilon}
+\frac{1}{15}\ln_r$&$\frac{1429}{900}
+\frac{11}{30\varepsilon}
-\frac{11}{15}\ln_r$&$\frac{28}{9}
+\frac{2}{3\varepsilon}
-\frac{4}{3}\ln_s$\\
$\frac{1}{8}
\left(\frac{d-1}{d-2}\right)^2
V^2\partial^2_tV$&$-\frac{19}{36}
-\frac{1}{6\varepsilon}
+\frac{1}{3}\ln_r$&$-\frac{119}{450}
-\frac{1}{15\varepsilon}
+\frac{2}{15}\ln_r$&$-\frac{19}{36}
-\frac{1}{6\varepsilon}
+\frac{1}{3}\ln_s$\\
$-\frac{1}{2}\left(\partial_t
V_i\right)^2$&$0$&$0$&$0$\\
$-\frac{(d-1)(d-3)}{4(d-2)^2}\,
V\partial^2_t K$&$0$&$0$&$0$\\
$-\frac{(d-1)(d-3)}{4(d-2)^2}\,
K\partial^2_t V$&$0$&$0$&$0$\\
$-\frac{1}{2}
\left(\frac{d-3}{d-2}\right)\hat
W_{ij}\partial_{ij}K$&$0$&$0$&$0$\\
\hline
Total&$-\frac{33}{50}
-\frac{1}{5\varepsilon}
+\frac{2}{5}\ln_r$&$\frac{397}{300}
+\frac{3}{10\varepsilon}
-\frac{3}{5}\ln_r$&$\frac{31}{12}
+\frac{1}{2\varepsilon}
-\ln_s$\\
\end{tabular}
\end{ruledtabular}
\end{table}}

{
\renewcommand{\arraystretch}{1.5}
\begin{table}[!ht]
\caption{Static contributions of ${\mathcal{D}}\frac{d\hat
R_i}{dt}(1)$. The ``principal'' part, ``retarded'' part and the
``extra term'' have the same meaning as in Table~\ref{table1}.}
\label{table3}
\begin{ruledtabular}
\begin{tabular}{c|c|c|c}
&$\frac{1}{8} \, {\mathcal D} c_{31}$&$\frac{1}{8} \, {\mathcal D}
c_{22}$&$\frac{1}{8} \, {\mathcal D} c_{13}$\\
\hline
$-\frac{d-1}{d-2}\, \partial_k V\partial_i
V_k|_\text{principal}$&$-\frac{31}{18}
-\frac{1}{3\varepsilon}
+\frac{2}{3}\ln_r$&$-\frac{31}{18}
-\frac{1}{3\varepsilon}
+\frac{2}{3}\ln_r$&$0$\\
$-\frac{d-1}{d-2}\, \partial_k V\partial_i
V_k|_\text{retarded}$&$\frac{43}{18}
+\frac{1}{3\varepsilon}
-\frac{2}{3}\ln_r$&$\frac{43}{18}
+\frac{1}{3\varepsilon}
-\frac{2}{3}\ln_r$&$0$\\
$-\frac{d(d-1)}{4(d-2)^2}\, \partial_t V \partial_i
V|_\text{principal}$&$\frac{5}{4}
+\frac{1}{4\varepsilon}
-\frac{1}{2}\ln_r$&$\frac{5}{4}
+\frac{1}{4\varepsilon}
-\frac{1}{2}\ln_r$&$0$\\
$-\frac{d(d-1)}{4(d-2)^2}\, \partial_t V \partial_i
V|_\text{retarded}$&$-\frac{7}{4}
-\frac{1}{4\varepsilon}
+\frac{1}{2}\ln_r$&$-\frac{7}{4}
-\frac{1}{4\varepsilon}
+\frac{1}{2}\ln_r$&$0$\\
extra term&$-\frac{1}{4}$&$-\frac{1}{4}$&$0$\\
\hline
Total&$-\frac{1}{12}$&$-\frac{1}{12}$&$0$\\
\end{tabular}
\end{ruledtabular}
\end{table}}

{
\renewcommand{\arraystretch}{1.5}
\begin{table}[!ht]
\caption{Static contributions of ${\mathcal{D}}\frac{d\hat
Y_i}{dt}(1)$.}
\label{table4}
\begin{ruledtabular}
\begin{tabular}{c|c|c|c}
&$\frac{1}{16} \, {\mathcal D} c_{31}$&$\frac{1}{16} \, {\mathcal D}
c_{22}$&$\frac{1}{16} \, {\mathcal D} c_{13}$\\
\hline
$\hat W_{kl}\, \partial_{kl} V_i$&$0$&$0$&$0$\\
$-\frac{1}{2}\left(\frac{d-1}{d-2}\right)
\partial_t\hat W_{ik}\, \partial_k V$&$\frac{65}{18}
+\frac{2}{3\varepsilon}
-\frac{4}{3}\ln_r$&$\frac{107}{45}
+\frac{5}{12\varepsilon}
-\frac{5}{6}\ln_r$&$\frac{71}{450}
+\frac{1}{60\varepsilon}
-\frac{1}{30}\ln_s$\\
$\partial_i\hat W_{kl}\, \partial_k V_l$&$\frac{257}{450}
+\frac{7}{60\varepsilon}
-\frac{7}{30}\ln_r$&$-\frac{149}{225}
-\frac{2}{15\varepsilon}
+\frac{4}{15}\ln_r$&$\frac{71}{450}
+\frac{1}{60\varepsilon}
-\frac{1}{30}\ln_s$\\
$-\partial_k\hat W_{il}\, \partial_l V_k$&$-\frac{257}{450}
-\frac{7}{60\varepsilon}
+\frac{7}{30}\ln_r$&$\frac{149}{225}
+\frac{2}{15\varepsilon}
-\frac{4}{15}\ln_r$&$-\frac{71}{450}
-\frac{1}{60\varepsilon}
+\frac{1}{30}\ln_s$\\
$-\frac{d-1}{d-2}\, \partial_k V \partial_i \hat R_k$&$\frac{2681}{900}
+\frac{19}{30\varepsilon}
-\frac{19}{15}\ln_r$&$\frac{3791}{900}
+\frac{53}{60\varepsilon}
-\frac{53}{30}\ln_r$&$-\frac{71}{450}
-\frac{1}{60\varepsilon}
+\frac{1}{30}\ln_s$\\
$-\frac{d(d-1)}{4 (d-2)^2}\, V_k\, \partial_i V \partial_k
V$&$-\frac{53}{100}
-\frac{1}{10\varepsilon}
+\frac{1}{5}\ln_r$&$-\frac{9}{2}
-\frac{1}{\varepsilon}
+2\ln_r$&$-\frac{33}{100}
-\frac{1}{10\varepsilon}
+\frac{1}{5}\ln_s$\\
$-\frac{d(d-1)^2}{8 (d-2)^3}\,
V\partial_t V\partial_i V$&$-\frac{9}{4}
-\frac{1}{2\varepsilon}
+\ln_r$&$\frac{43}{25}
+\frac{2}{5\varepsilon}
-\frac{4}{5}\ln_r$&$\frac{33}{100}
+\frac{1}{10\varepsilon}
-\frac{1}{5}\ln_s$\\
$-\frac{1}{2}\left(\frac{d-1}{d-2}\right)^2 V \partial_k V
\partial_k V_i$&$0$&$0$&$0$\\
$\frac{1}{2}\left(\frac{d-1}{d-2}\right) V
\partial^2_tV_i$&$-\frac{9}{2}
-\frac{1}{\varepsilon}
+2\ln_r$&$-\frac{9}{2}
-\frac{1}{\varepsilon}
+2\ln_r$&$0$\\
$2 V_k\, \partial_k\partial_t V_i$&$0$&$0$&$0$\\
$\frac{(d-1)(d-3)}{(d-2)^2}\, \partial_k K \partial_i
V_k$&$0$&$0$&$0$\\
$\frac{d(d-1)(d-3)}{4(d-2)^3}\,
\partial_t V\partial_i K$&$0$&$0$&$0$\\
$\frac{d(d-1)(d-3)}{4(d-2)^3}\,
\partial_i V\partial_t K$&$0$&$0$&$0$\\
\hline
Total&$-\frac{69}{100}
-\frac{3}{10\varepsilon}
+\frac{3}{5}\ln_r$&$-\frac{69}{100}
-\frac{3}{10\varepsilon}
+\frac{3}{5}\ln_r$&$0$\\
\end{tabular}
\end{ruledtabular}
\end{table}}

Summing up the separate non-vanishing contributions displayed in
Tables~\ref{table1}-\ref{table4}, we get the following total
differences
\begin{subequations}\label{eq5.28}\begin{eqnarray}
{\mathcal D} c_{31} &= &- \frac{22}{3\varepsilon} + \frac{44}{3} \ln
(\overline q \, r'_1 \, r_{12}) - \frac{102}{5} + {\mathcal O}
(\varepsilon)\,, \\
{\mathcal D} c_{22} &= &9 + {\mathcal
O} (\varepsilon) \\
{\mathcal D} c_{13} &= &
\frac{22}{3\varepsilon} - \frac{44}{3} \ln (\overline q \, s_2 \,
r_{12}) + \frac{568}{15} + {\mathcal O} (\varepsilon)\,.
\end{eqnarray}
\end{subequations}
Finally, adding Eqs.~(\ref{eq5.28}) to the pure Hadamard-Schwartz
result (\ref{eq5.15}), we get the dimensionally regularized static
contributions to $\mathbf{a}_1$:
\begin{subequations}
\label{eq5.29}
\begin{eqnarray}
c_{31}^\text{dr} &= &-\frac{22}{3\varepsilon} + \frac{44}{3} \ln
(\overline q \, r_{12}^2) - \frac{959}{36} + {\mathcal O}
(\varepsilon)\,, \\
c_{22}^\text{dr} &= &175 - \frac{41}{16} \,
\pi^2 + {\mathcal O} (\varepsilon)\,, \\
c_{13}^\text{dr} &= &
\frac{22}{3\varepsilon} - \frac{44}{3} \ln (\overline q \, r_{12}^2) +
\frac{1799}{9} - \frac{41}{16} \, \pi^2 + {\mathcal O} (\varepsilon)
\,.
\end{eqnarray}
\end{subequations}

As expected the two Hadamard regularization length scales $r'_1$ and
$s_2$ have cancelled between $c_{mn}^\text{pHS}$ and ${\mathcal D}
c_{mn}$ to leave a result which depends only on the dim. reg.
regularization parameter $\varepsilon = d-3$. One might be surprised by
the presence in $c_{mn}^\text{dr}$ of terms $\pm \frac{44}{3} \ln
(r_{12}^2)$ compared to corresponding terms $\pm \frac{44}{3} \ln
(r_{12})$ in $c_{mn}^\text{pHS}$, and by the absence of any
adimensionalizing length scale in these logarithms of $r_{12}^2$. These
two properties can be understood when one remembers from the discussion
above that the coefficients which have the same physical dimension as
$c_{mn}^{(d=3)}$ are the combinations $\ell_0^{(m+n)\varepsilon} \,
c_{mn}^\text{dr}$. In the present case, this means that
$\ell_0^{4\varepsilon} \, c_{mn}^\text{dr}$ are dimensionless. It is
easy to see, thanks to the pole terms $\mp 22/(3\varepsilon)$ and the
expansion $\ell_0^{4\varepsilon} \equiv \exp (4\varepsilon \ln \ell_0)
= 1 + 2 \varepsilon \ln \ell_0^2 + {\mathcal O} (\varepsilon^2)$, that
the combinations $\ell_0^{4\varepsilon} \, c_{31}^\text{dr}$ and
$\ell_0^{4\varepsilon} \, c_{13}^\text{dr}$ do indeed depend only on
the dimensionless quantities $\varepsilon$ and $\ln (r_{12}^2 /
\ell_0^2)$.

\section{Renormalization of the equations of motion}
\label{Renormalise}
\subsection{Poles in the dimensionally regularized bulk metric}
\label{Poles}
The first computation of the dimensional continuation of the 3PN
gravitational interaction of point masses was done in ADM coordinates
and resulted into a \textit{finite} (\textit{i.e.}, without
$1/\varepsilon$ poles) answer \cite{Damour:2001bu}. Our task in
analyzing the physical meaning of the harmonic-coordinates result
(\ref{eq5.29}) is to interpret the presence of $1/\varepsilon$ poles
in it. For this we have to remember that, as in Quantum Field Theory
(QFT), dimensional continuation is a \textit{regularization} method
which, like all regularization methods, transforms truly infinite
results, say containing $\int_0^{r_{12}} d^3\mathbf{x} / r_1^3$, into
finite, but ``large'' ones, which depend on some cut-off parameter,
\textit{e.g.} $\int_{s_1}^{r_{12}} d^3 \mathbf{x} / r_1^3 = 4 \pi \ln
(r_{12} / s_1)$ or $\int_0^{r_{12}} d^{3+\varepsilon} \mathbf{x} /
r_1^{3(1+\varepsilon)} \propto 1/\varepsilon$. Any regularization
must be followed by a \textit{renormalization} process which allows
one to absorb the cut-off dependent terms in some of the basic
\textit{bare parameters} of the theory.

In order to have a clearer understanding of the poles in the (static)
equations of motion (\ref{eq5.29}) [we shall prove below that our
discussion extends to the full, velocity-dependent equations of
motion], we need to analyze the presence of poles in the ``bulk''
metric, \textit{i.e.}, the metric $g_{\mu\nu} (\mathbf{x};
\mathbf{y}_1, \mathbf{y}_2)$ evaluated at a generic field point
$\mathbf{x}$, away from the two world-lines. Indeed, if we were
considering the gravitational field generated by regular
(\textit{i.e.}, non point-like) sources, a complete physical
description of their gravitational effects would necessitate the
simultaneous consideration of the bulk metric and of the equations of
motion of the (extended) sources. Similarly, in the present formal
study of two-point-like sources, we need to consider both the equations
of motion $\ddot{\mathbf{y}}_a = \mathbf{a}_a (\mathbf{y}_1 ,
\mathbf{y}_2, \mathbf{v}_1 , \mathbf{v}_2 )$ and the bulk metric
$g_{\mu\nu} \bigl(\mathbf{x} ; \mathbf{y}_1 (t) , \mathbf{y}_2 (t) ,
\mathbf{v}_1 (t) , \mathbf{v}_2 (t) \bigr)$.

\begin{figure}[!ht]
\includegraphics[scale=.8]{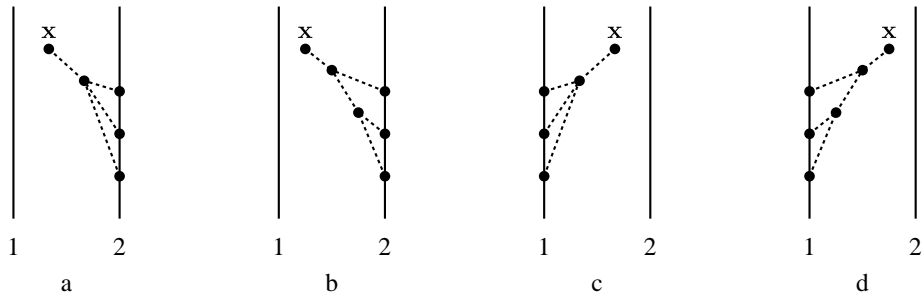}
\caption{Some divergent diagrams for the bulk metric. Here, contrary
to the previous Figures, the field point, labelled by an $\mathbf{x}$,
is detached from the first worldline.}
\label{fig6}
\end{figure}

It is here that the diagrammatic representation introduced above plays
a useful role in highlighting the structure of divergencies in the
equations of motion and in the bulk metric. Indeed, it is clear that
the divergent diagrams of the equations of motion of the first
particle, where the $1/\varepsilon$ pole is due to the presence of a
subdivergence induced by three propagators shrinking onto the second
world-line (such as in Fig.~\ref{fig3}d or Fig.~\ref{fig4}d), will
correspond to similar $1/\varepsilon$ poles in the bulk metric, for the
corresponding ``bulk diagrams'' where the special point marked by a
cross in the diagrams above (denoting the coincidence $\mathbf{x} =
\mathbf{y}_1$) is detached from the first world-line to end at an
arbitrary point in the bulk, as indicated in Fig.~\ref{fig6}.

\noindent
Evidently, in addition to such diagrams as Fig.~\ref{fig6}a and
\ref{fig6}b which will contain (at least) a factor $m_2^3$, there will
exist ``mirror diagrams'', containing a factor $m_1^3$, and obtained by
exchanging the labels 1 and 2. On the other hand, note that the bulk
poles $\propto m_1^3$ of the type of Fig.~\ref{fig6}c and \ref{fig6}d
do not (necessarily) correspond to poles in the equations of motion of
$\mathbf{y}_1$ because their coincidence limits $\mathbf{x} \rightarrow
\mathbf{y}_1$ induce diagrams of the type of Fig.~\ref{fig3}a or
\ref{fig4}a containing four shrinking propagators instead of three.
[Though such diagrams would exhibit worse divergences in a
dimensionful cut-off regularization schemes, they are generally non
dangerous in dimensional regularization because the integral $\int
d^{3+\varepsilon} \mathbf{x} / r_1^{4(1+\varepsilon)}$ has no pole as
$\varepsilon \rightarrow 0$.]

A careful analysis of the possible presence of poles in the various
potentials $V , V_i , \cdots , \hat T$ we use to parametrize the bulk
metric (aided by the structure of potentially dangerous terms sketched
in Fig.~\ref{fig6}) shows that, at the 3PN
approximation,\footnote{\textit{I.e.}, at order $c^{-8}$ in
$g_{00}$, $c^{-7}$ in $g_{0i}$, and $c^{-6}$ in $g_{ij}$.} such poles
can only be present in the 1PN-level expansion of $\hat X$ and in the
Newtonian-level approximation of $\hat T$. Drawing on the results of
\cite{Blanchet:2000ub} and \cite{FayeThesis} we can also see that all
velocity-dependent terms in the poles present in $\hat X_\text{1PN}$
and $\hat T$ (\textit{i.e.}, the terms proportional to $m_1^3 \,
v_1^2$ or $m_2^3 \, v_2^2$) exactly cancel in the combination $4 \,
\hat X / c^4 + 16 \, \hat T / c^6$ that matters for the bulk metric.
[This shows up, for instance, in Eq.~(7.1) of \cite{Blanchet:2000ub}
which implies that the divergencies linked to the second world-line,
characterized by the presence of $\ln s_2$, do not depend on
velocities. This shows up also in the absence of poles in the
velocity-dependent contributions to $\mathbf{a}_1$ proportional to
$m_2^3$, see Eqs.~(\ref{eq5.76c})-(\ref{eq5.76d}) below.] We are
therefore left with evaluating the poles present in the \textit{static
limit} $(\mathbf{v}_1 , \mathbf{v}_2 \rightarrow 0)$ of $\hat
X_\text{1PN}$ and $\hat T$. Clearly, from Fig.~\ref{fig6}, the poles
in $\hat X_\text{1PN}$ and $\hat T$ will come only from cubically
non-compact ($CNC$) sources. Finally, as we are only interested in the
pole part we can neglect the $d$-dependence of the coefficients in the
sources of $\hat X$ and $\hat T$ (which we indicate by using a symbol
$\simeq$). Thus, these poles can only come from
\begin{subequations}
\label{eq5.31ab}
\begin{eqnarray}
\hat X^{CNC}_\text{static} &=& \Box^{-1} \Bigl[\partial_{ij} \, V\,
\hat{W}_{ij}^{NC}\Bigr] \nonumber\\ &\simeq& \Box^{-1}
\Bigl[\partial_{ij} \, V \, \Box^{-1} \bigl(- \partial_i \, V
\partial_j \, V\bigr)\Bigr] \, ,\label{eq5.30}\\ \hat
T^{CNC}_\text{static} &\simeq& \Box^{-1} \biggl[ \frac{1}{2} \, V^2 \,
\partial_t^2 \, V + \partial_{ij} \, V\, \hat{Z}_{ij}^{NC} \biggr]
\nonumber \\ &\simeq & \Box^{-1} \biggl[\frac{1}{2} \, V^2 \,
\partial_t^2 \, V + \partial_{ij} \, V \, \Box^{-1} \bigl(-
2\partial_i \, V \partial_t \, V_j \bigr) \biggr]\,.
\label{eq5.31}
\end{eqnarray}
\end{subequations}
The static poles (involving factors $m_1^3$ or $m_2^3$) in
Eqs.~(\ref{eq5.31ab}) are then obtained by: (1) considering sources
involving three times $V_1$ or three times $V_2$ (where $V_a$ denotes
the piece $\propto m_a$ in $V$), (2) evaluating the time derivatives
in the static limit, using for instance
\begin{equation}
\label{eq5.32}
(\partial_t^2 \, V_a)_\text{static} = - a_a^j \, \partial_j \, V_a \,,
\end{equation}
and (3) expanding up to the required accuracy the (time-symmetric)
propagators according to $\Box^{-1} = \Delta^{-1} + c^{-2} \,
\partial_t^2 \, \Delta^{-2} + {\mathcal O} (c^{-4})$.

As an example among the simplest terms, let us consider the $m_1^3$
contribution coming from the first term on the R.H.S. of
Eq.~(\ref{eq5.31}),
\begin{equation}
\label{eq5.33}
\hat T^{m_1^3}_{\text{static} (1)} \simeq \frac{1}{2} \, \Delta^{-1}
[V_1^2 \, \partial_t^2 \, V_1]_\text{static} = - \frac{1}{2} \,
\Delta^{-1} [V_1^2 \, a_1^j \, \partial_j \, V_1] = - \frac{1}{6} \,
a_1^j \, \partial_j \, \Delta^{-1} [V_1^3]\,.
\end{equation}
Using $V_1 = 2\frac{d-2}{d-1} \, G \, \tilde{k} \, m_1 \, r_1^{2-d} +
{\mathcal O} (c^{-2}) \simeq G \, m_1 \, r_1^{-1-\varepsilon} +
{\mathcal O} (c^{-2})$ and $\Delta^{-1} \, r_1^{\lambda} =
r_1^{\lambda + 2} / [(\lambda + 2)(\lambda + d)]$, one finds that the
pole part of (\ref{eq5.33}) reads
\begin{equation}
\label{eq5.34}
\hat T^{m_1^3}_{\text{static} (1)} \simeq - \frac{1}{12 \varepsilon} \,
G^3 \, m_1^3 \, a_1^j \, \partial_j \, r_1^{-1-3\varepsilon}\,.
\end{equation}
Similarly an analysis of the second source term in Eq.~(\ref{eq5.31})
yields
\begin{equation}
\label{eq5.35}
\hat T^{m_1^3}_{\text{static} (2)} \simeq \frac{1}{3\varepsilon} \,
G^3 \, m_1^3 \, a_1^j \, \partial_j \, r_1^{-1-3\varepsilon}\,,
\end{equation}
so that the full (static) contribution of $\hat T$ is
\begin{equation}
\label{eq5.36}
\hat T^{m_1^3}_{\text{static}} \simeq \frac{1}{4\varepsilon} \, G^3 \,
m_1^3 \, a_1^j \, \partial_j \, r_1^{-1-3\varepsilon}\,.
\end{equation}
The analysis of the pole part in the static limit of $\hat X$,
Eq.~(\ref{eq5.30}), is more intricate because one must expand to 1PN
accuracy both $V_a \simeq G \, m_a \, r_a^{-1-\varepsilon} +
\frac{1}{2 c^2} \, G \, m_a \, \partial_t^2 \, r_a^{1-\varepsilon}$
and the propagator $\Box^{-1}$. This yields
\begin{equation}
\label{eq5.37}
\hat X^{m_1^3}_{\text{static}} \simeq - \frac{1}{12\varepsilon} \,
\frac{G^3}{c^2} \, m_1^3 \, a_1^j \, \partial_j \,
r_1^{-1-3\varepsilon}\,.
\end{equation}

Let us now consider the improved $V$-potential (\ref{calV}) that makes
up the essential part of $g_{00}$,
\begin{equation}
\mathcal{V} \equiv V -\frac{2}{c^2} \left(\frac{d-3}{d-2}\right) K
+\frac{4 \hat X}{c^4} +\frac{16 \hat T}{c^6} \simeq V +\frac{4 \hat
X}{c^4} +\frac{16 \hat T}{c^6}\,,
\label{eq5.38}\end{equation}
such that $g_{00} = - \exp (-2 {\mathcal V} / c^2) [1-8 \, V_i \, V_i
/ c^6 - 32 \, R_i \, V_i / c^8+\mathcal{O}(1/c^{10})]$, see
Eq. (\ref{g00}). Combining the results above, we find that the only
$1/\varepsilon$ poles in the bulk metric $g_{\mu\nu}
(\mathbf{x},\mathbf{y}_1,\mathbf{y}_2)$ show up in $g_{00}$ at the 3PN
level and are (when expressed in terms of the improved potential
$\mathcal{V}$, and after cancellation of $m_a^3 \, v_a^2 / \varepsilon$
terms between $\hat X$ and $\hat T$ of the following static form
\begin{equation}
\label{eq5.39}
\mathcal{V}(\mathbf{x}, \mathbf{y}_1 , \mathbf{y}_2) = V +
\frac{1}{c^2} \mathcal{V}_2 + \frac{1}{c^4} \mathcal{V}_4 +
\frac{1}{c^6} \biggl[\mathcal{V}'_6 + \frac{11}{3\varepsilon} \, \sum_a
G^3\, m_a^3 \, a_a^j \, \partial_j \, r_a^{-1-3\varepsilon}\biggr]\,,
\end{equation}
where $\mathcal{V}_2$, $\mathcal{V}_4$,$\mathcal{V}'_6$'s are finite
when $\varepsilon\rightarrow 0$. To understand better the structure of
result (\ref{eq5.39}) let us introduce the notation
\begin{equation}
\label{eq5.40}
\zeta_a^i \equiv + \frac{11}{3\varepsilon} \, \frac{G_N^2 \,
m_a^2}{c^6} \, a_a^i\,,
\end{equation}
where $G_N$ is the 3-dimensional Newton constant and $a_a^i$ the
$d$-dimensional acceleration of $y_a^i$. [This definition ensures that
$\zeta_a^i$ has the physical dimension of a length.] In terms of the
definition (\ref{eq5.40}), the result (\ref{eq5.39}) can be
equivalently written as
\begin{equation}
\label{eq5.41}
\mathcal{V} (\mathbf{x}, \mathbf{y}_1 , \mathbf{y}_2) = \sum_a \
\Bigl[V_a (\mathbf{x} - \mathbf{y}_a) + \zeta_a^j \, \partial_j \, V_a
(\mathbf{x} - \mathbf{y}_a)\Bigr] + \frac{1}{c^2} \mathcal{V}_2 +
\frac{1}{c^4} \mathcal{V}_4 + \frac{1}{c^6} \mathcal{V}_6\,,
\end{equation}
where the pole part is entirely contained in the terms proportional to
$\zeta_1^j$ and $\zeta_2^j$ [$\mathcal{V}_6$ here differs from
$\mathcal{V}'_6$ in Eq.~(\ref{eq5.39}) by some finite corrections when
$\varepsilon\rightarrow 0$].
The fact that poles appear only in $\mathcal{V}$, at order $c^{-6}$,
implies that there are no divergencies in the harmonic gauge conditions
(\ref{GaugeIdentities}) in the bulk. Indeed, (\ref{diVi}) needs
$\mathcal{V}$ at order $c^{-4}$ only, and (\ref{djWij}) at Newtonian
order only.

\subsection{Renormalization of poles by shifts of the world-lines}
\label{Shift}
Result (\ref{eq5.41}) indicates a simple way of renormalizing away
the poles present in the bulk metric. Indeed, the logic up to now has
been to describe in the simplest possible manner a gravitationally
interacting two-particle system, parametrized by the following
\textit{bare} parameters: $G^\text{bare}$, $m_1^\text{bare}$,
$m_2^\text{bare}$, $\mathbf{y}_1^\text{bare}$,
$\mathbf{y}_2^\text{bare}$, considered in everywhere-harmonic
coordinates, $\Gamma^{\lambda} \equiv g^{\alpha \beta} \,
\Gamma_{\alpha \beta}^{\lambda} = 0$. In particular, the internal
structure of each particle has been, up to now, entirely described by
a monopolar stress-energy distribution, \textit{i.e.}, $T_a^{\mu\nu}
\propto m_a^\text{bare} \, \delta (\mathbf{x} -
\mathbf{y}_a^\text{bare})$. In other words, we have set to zero any
higher multipolar structure. Eq.~(\ref{eq5.41}) is most simply
interpreted by saying that the non-linear interactions (see
Fig.~\ref{fig6}) dress each particle by a cloud of gravitational
energy which generates, at the 3PN order, a divergent \textit{dipole}
in the Newtonian-like potential. Therefore, to get a net, finite bulk
gravitational field we must endow each initial particle by an
infinite, bare dipole, corresponding to a counterterm $\Delta
T_a^{\mu\nu} \propto -m_a^\text{bare}\zeta_a^j \, \partial_j \, \delta
(\mathbf{x} - \mathbf{y}_a)$, which will cancel the non-linearly
generated one (\ref{eq5.41}). An equivalent, but technically simpler
way of endowing each particle by a bare structure able to cancel the
dipolar pole terms in (\ref{eq5.41}) is simply to say that the central
\textit{bare} world-lines used in our derivations up to now, henceforth
denoted as $\mathbf{y}_a^\text{bare}$, can be decomposed in a finite
\textit{renormalized} part $\mathbf{y}_a^\text{ren}$ and a formally
infinite shift $\bm{\xi}_a$ involving a pole $\propto 1/\varepsilon$,
\begin{equation}
\label{eq5.42}
\mathbf{y}_a^\text{bare} \equiv \mathbf{y}_a^\text{ren} +
\bm{\xi}_a\,, \quad \mathbf{v}_a^\text{bare} \equiv
\mathbf{v}_a^\text{ren} + \dot{\bm{\xi}}_a\,.
\end{equation}
The gravitational potential of two point particles [$\propto \delta
(\mathbf{x} - \mathbf{y}_a^\text{bare})$] is then
\begin{eqnarray}
\label{eq5.430}
\mathcal{V} (\mathbf{x}, \mathbf{y}_1^\text{bare} ,
\mathbf{y}_2^\text{bare}) &=& \sum_a \ \Bigl[V_a (\mathbf{x} -
\mathbf{y}_a^\text{ren} - \bm{\xi}_a) + \zeta_a^j \, \partial_j \, V_a
(\mathbf{x} - \mathbf{y}_a^\text{ren} - \bm{\xi}_a)\Bigr] \\
&+&
\frac{1}{c^2} \, \mathcal{V}_2 (\mathbf{x}, \mathbf{y}_a^\text{ren} +
\bm{\xi}_a) + \frac{1}{c^4} \, \mathcal{V}_4 (\mathbf{x},
\mathbf{y}_a^\text{ren} + \bm{\xi}_a) + \frac{1}{c^6} \, \mathcal{V}_6
(\mathbf{x}, \mathbf{y}_a^\text{ren} + \bm{\xi}_a)\,. \nonumber
\end{eqnarray}
Assuming that the vector $\bm{\xi}_a$ is of 3PN order [\textit{i.e.},
$\bm{\xi}_a=\mathcal{O}(1/c^6)$], we can rewrite Eq.~(\ref{eq5.430}) as
\begin{eqnarray}
\label{eq5.43}
\mathcal{V} (\mathbf{x}, \mathbf{y}_1^\text{bare} ,
\mathbf{y}_2^\text{bare}) &=& \sum_a \ \Bigl[V_a (\mathbf{x} -
\mathbf{y}_a^\text{ren}) + (\zeta_a^j-\xi_a^j) \, \partial_j \, V_a
(\mathbf{x} - \mathbf{y}_a^\text{ren})\Bigr] \\
&+& \frac{1}{c^2} \,
\mathcal{V}_2 (\mathbf{x}, \mathbf{y}_a^\text{ren}) + \frac{1}{c^4} \,
\mathcal{V}_4 (\mathbf{x}, \mathbf{y}_a^\text{ren}) + \frac{1}{c^6} \,
\mathcal{V}_6 (\mathbf{x}, \mathbf{y}_a^\text{ren}) +
\mathcal{O}\left(\frac{1}{c^8}\right), \nonumber
\end{eqnarray}
which makes it clear that the potential will be finite (at 3PN
accuracy) when $\varepsilon \rightarrow 0$ if we choose
\begin{equation}
\label{eq5.44}
\bm{\xi}_a = \bm{\zeta}_a + {\mathcal
O}\left(\frac{\varepsilon^0}{c^6}\right),
\end{equation}
where by ${\mathcal O}(\varepsilon^0/c^6)$ we mean a term finite when
$\varepsilon \rightarrow 0$ and of the 3PN order. We shall henceforth
refer to $\bm{\xi}_a$ in Eq.~(\ref{eq5.44}) as a \textit{shift} of the
$a^\text{th}$ world-line. The reasoning above shows that the
introduction of such shifts, at the 3PN order and having the pole
structure (\ref{eq5.40}), is \textit{necessary} to renormalize away the
poles present in the \textit{bulk metric}. It remains to show that
these shifts are also \textit{sufficient} to renormalize away the poles
present in the \textit{equations of motion}.

The effect of 3PN-level shifts $\bm{\xi}_a$ on the equations of motion
is easy to obtain. Indeed, the equations of motion we computed above
concern the original, \textit{bare} world-lines
$\mathbf{y}_a^\text{bare}$. For the first particle, they had the
structure (in dimensional regularization)
\begin{eqnarray}
\label{eq5.45}
\ddot{\mathbf{y}}_1^\text{bare} &= &\mathbf{a}_1^\text{dr}
(\mathbf{y}_{12}^\text{bare} , \mathbf{v}_1^\text{bare} ,
\mathbf{v}_2^\text{bare}) \nonumber \\
&=
&\mathbf{a}_\text{N1}^\text{dr} (\mathbf{y}_{12}^\text{bare}) +
\mathbf{a}_\text{1PN1}^\text{dr} (\mathbf{y}_{12}^\text{bare},
\mathbf{v}_1^\text{bare} , \mathbf{v}_2^\text{bare}) \nonumber \\
&+
&\mathbf{a}_\text{2PN1}^\text{dr} (\mathbf{y}_{12}^\text{bare},
\mathbf{v}_1^\text{bare} , \mathbf{v}_2^\text{bare}) +
\mathbf{a}_\text{2.5PN1}^\text{dr} (\mathbf{y}_{12}^\text{bare},
\mathbf{v}_1^\text{bare} , \mathbf{v}_2^\text{bare}) \nonumber \\
&+
&\mathbf{a}_\text{3PN1}^\text{dr} (\mathbf{y}_{12}^\text{bare},
\mathbf{v}_1^\text{bare} , \mathbf{v}_2^\text{bare})\,,
\end{eqnarray}
where $\mathbf{y}_{12}^\text{bare}\equiv \mathbf{y}_1^\text{bare} -
\mathbf{y}_2^\text{bare}$. Here $\mathbf{a}_\text{N1}^\text{dr}$
denotes the dimensionally continued Newtonian-level acceleration,
\begin{equation}
\label{eq5.46}
a_\text{N1}^{i (\text{dr})} = \partial_i \, V_2 (\mathbf{y}_{12}) = f
\, G \, m_2 \, \tilde k\,\partial_i \, r_{12}^{2-d}\,,
\end{equation}
where by a slight abuse of notation we pose $\partial_i = \partial /
\partial \, y_{12}^i$, where $G \equiv G_N \, \ell_0^{\varepsilon}$
denotes the $d$-dimensional gravitational constant, and where the
$d$-dependent correcting factors
\begin{equation}
\label{eq5.47}
f \equiv 2 \, \frac{d-2}{d-1} = \frac{1+\varepsilon}{1 +
\varepsilon/2}\,, \qquad \tilde k \equiv \frac{\Gamma \left(
\frac{d-2}{2} \right)}{\pi^{\frac{d-2}{2}}} = \frac{\Gamma \left(
\frac{1 + \varepsilon}{2} \right)}{\pi^{\frac{1 + \varepsilon}{2}}} \,,
\end{equation}
tend to 1 as $\varepsilon \rightarrow 0$, but will play a significant
role below.

When inserting the redefinitions (\ref{eq5.42}) into (\ref{eq5.45})
one easily finds that the \textit{renormalized} equations of motion,
\textit{i.e.}, the equations for $\mathbf{y}_a^\text{ren}$, read
[using only $\bm{\xi}_a = {\mathcal O} (1/c^6)$ at this stage]
\begin{equation}
\label{eq5.48}
\ddot{\mathbf{y}}_1^\text{ren} = \mathbf{a}_1^\text{ren}
(\mathbf{y}_{12}^\text{ren}, \mathbf{v}_1^\text{ren} ,
\mathbf{v}_2^\text{ren})\,,
\end{equation}
where
\begin{equation}
\label{eq5.49}
\mathbf{a}_1^\text{ren} (\mathbf{y}_{12} , \mathbf{v}_1 ,
\mathbf{v}_2) = \mathbf{a}_1^\text{dr} (\mathbf{y}_{12} , \mathbf{v}_1,
\mathbf{v}_2) + \delta_{\bm{\xi}} \, \mathbf{a}_1 (\mathbf{y}_{12},
\mathbf{v}_1, \mathbf{v}_2)
+ {\mathcal O} \left(\frac{1}{c^8}\right),
\end{equation}
with
\begin{equation}
\label{eq5.50}
\delta_{\bm{\xi}} \, \mathbf{a}_1 (\mathbf{y}_{12}, \mathbf{v}_1,
\mathbf{v}_2) = (\xi_1^j - \xi_2^j)
\, \partial_j \, \mathbf{a}_\text{N1}^\text{dr} (\mathbf{y}_{12}) -
\ddot{\bm \xi}_1\,.
\end{equation}

Let us note that the effect on the equations of motion of a (3PN) shift
of the world-lines, Eq. (\ref{eq5.50}), is technically identical to the
effect on the equations of motion of the restriction to the world-lines
of a 3PN-level coordinate transformation, say $x_\text{new}^i = x^i -
\epsilon^i (\mathbf{x},t)$ and $t_\text{new} = t - c^{-1}\epsilon^0
(\mathbf{x},t)$. Indeed, a coordinate transformation has two effects:
(i) it changes the bulk metric by $\delta_{\epsilon} \, g_{\mu\nu} (x)
= \mathcal{L}_{\epsilon} \, g_{\mu\nu} (x)$, where
$\mathcal{L}_{\epsilon}$ denotes the Lie derivative along
$\epsilon^\mu$, and (ii) it induces a shift of the world-lines $y_{a \,
\text{new}}^i = y_a^i - \epsilon_a^i+ c^{-1}\epsilon_a^0 v_a^i$ (plus
non-linear terms in $\epsilon^\mu$), where we denote the coordinate
change at the location of the $a$-th particle by $\epsilon_a^i(t)
\equiv [\epsilon^i (\mathbf{x},t)]_{\mathbf{x} = \mathbf{y}_a}$.
Because of the diffeomorphism invariance of the total action, the
effect (i) does not change the action,\footnote{Actually, one should
consider, as \textit{e.g.} in \cite{Damour:1996kt}, a
non-generally-covariant gauge-fixed action. But the ``double zero''
nature of the gauge-fixing term, say $\propto \sqrt{-g} \, g_{\mu\nu}
\, \Gamma^{\mu} \, \Gamma^{\nu}$, ensures that it does not contribute
to first order in $\epsilon^\mu$.} so that the net effect of a
coordinate transformation on the equations of motion reduces to the
effect (\ref{eq5.50}) of the following shift induced on the
world-lines:
\begin{equation}
\xi_a^i = \epsilon_a^i - c^{-1}\epsilon_a^0 \,v_a^i~~\hbox{+
non-linear terms}\,.
\label{xia}\end{equation}
The coordinate transformations considered in \cite{Blanchet:2000ub},
see Eq.~(6.11) there, were of the type $\epsilon_\mu
(x)=c^{-6}\,\partial_\mu\bigl(\sum_a\kappa_a/r_a\bigr)$, where the
$\kappa_a$'s are some coefficients, so that we see that the latter
induced shift reduces at the 3PN order to the (regularization of the)
purely spatial coordinate transformation evaluated on the world-line:
\begin{equation}
\xi_a^i = \epsilon_a^i + \mathcal{O}\left(\frac{1}{c^8}\right).
\label{xia'}\end{equation}
We have checked that the formula given by Eq.~(6.15) in
\cite{Blanchet:2000ub} for the coordinate transformation of the
acceleration of the particle 1 gives exactly the same result as the
one computed from the effect of the shift (\ref{eq5.50}). [The
agreement extends to $d$ dimensions if we consider the straightforward
extension of the latter coordinate transformation $\epsilon^\mu$ to
$d$ dimensions, namely $\epsilon_\mu
(x)=c^{-6}\,\partial_\mu\bigl(\sum_a\kappa_a\,
\tilde{k}\,r_a^{2-d}\bigr)$.]

Note that the coordinate transformations $\epsilon^\mu(x)$ were
considered in \cite{Blanchet:2000ub} only in terms of their effects,
Eqs. (\ref{xia})-(\ref{xia'}), on the equations of motion. This was
sufficient to prove, for instance, that the two constants $r_1'$ and
$r_2'$ in the 3PN equations of motion are not physical, because they
can be gauged away in 3 dimensions and therefore will never appear in
any physical result. However, we remark that the extension to $d$
dimensions of the coordinate transformation $\epsilon^\mu(x)$ of the
\textit{bulk} metric, say $\epsilon_\mu
(x)=c^{-6}\,\partial_\mu\bigl(\sum_a\kappa_a\,
\tilde{k}\,r_a^{2-d}\bigr)$ (with coefficients
$\kappa_a\propto\varepsilon^{-1}$, as needed to remove the poles in the
equations of motion), does not lead to a bulk metric free of poles.
Indeed, assuming $\kappa_a\propto\varepsilon^{-1}$, we see that the
pole in the spatial coordinate transformation $\epsilon_i(x)$ would
then induce a pole in the spatial components of the metric,
$\delta_{\epsilon} \,
g_{ij}=\partial_i\epsilon_j+\partial_j\epsilon_i+\cdots$, but this is
inadmissible because we have proved above Eqs.~(\ref{eq5.31ab}) that,
at 3PN order, only the time-time component of the bulk metric contained
a pole. A bulk coordinate transformation of the type above can then
remove the poles in the time-time component of the bulk metric only at
the price of creating a pole in the initially pole-free spatial metric.
We shall leave to future work a complete clarification of the
possibility of using, within our dim. reg. context, a coordinate
transformation to induce the shifts (\ref{eq5.44}). For the time being,
what is important is that our introduction above of shifts of the
world-lines (\textit{a priori} unconnected to any coordinate
transformation) is a consistent way of renormalizing away the poles in
the metric, and that its effect on the equations of motion,
Eq.~(\ref{eq5.50}), is identical to the transformations of the
acceleration obtained in Ref.~\cite{Blanchet:2000ub}.

It remains now to show that the same world-line shifts (\ref{eq5.44})
that renormalize away the poles in the bulk metric, Eq.~(\ref{eq5.43}),
do renormalize away also the poles present in the original bare
equations of motion [see (\ref{eq5.29}) for the static contributions
and (\ref{eq5.76}) below for the kinetic ones]. For this purpose let us
consider a shift of the more general form
\begin{equation}
\label{eq5.51}
\xi_a^i = \frac{e(d)}{\varepsilon} \, \frac{G_N^2 \, m_a^2}{c^6} \,
a_a^i\,,
\end{equation}
where $e(d)$ represents a certain numerical coefficient depending on
$d$, and where $a_a^i$ denotes the $d$-dimensional acceleration of
$y_a^i$ given by its Newtonian approximation (\ref{eq5.46}) [but, for
notational simplicity, we henceforth drop the label $N$ on such
accelerations entering 3PN effects]. Inserting (\ref{eq5.51}) into
(\ref{eq5.50}) yields (for the index $a=1$)
\begin{equation}
\label{eq5.52}
\delta_{\xi} \, a_1^i = \frac{e(d)}{\varepsilon} \, \frac{G_N^2}{c^6}
\, \Bigl[(m_1^2 - m_2^2) \, a_2^j \, \partial_j \, a_1^i - \, m_1^2 \,
v_{12}^{jk} \, \partial_{jk} \, a_1^i \Bigr]\,,
\end{equation}
where $v_{12}^j \equiv v_1^j - v_2^j$ and $v_{12}^{jk} \equiv v_{12}^j
v_{12}^k$ [and also, as before, $\partial_j \equiv \partial /\partial
y_{12}^j$]. Before further evaluating (\ref{eq5.52}) by inserting the
explicit expression (\ref{eq5.46}) for the acceleration, we shall
consider some simple but important consequences of the structure
(\ref{eq5.52}).

\subsection{Link to the general class of harmonic equations of
motion}
\label{LambdaDetermined}
As recalled in the Introduction, previous work on the 3PN equations of
motion in \textit{harmonic coordinates} has shown that these equations
necessarily belonged to a three-parameter class of equations of
motion, say
\begin{equation}
\label{eq5.53}
\ddot{\mathbf{y}}_a^{(d=3)} = \mathbf{a}_a^\text{BF} (\mathbf{y}_{12},
\mathbf{v}_1 , \mathbf{v}_2 ; \lambda , r'_1 , r'_2)\,.
\end{equation}
The dimensionless parameter $\lambda$ could not be determined by the
previous work in harmonic coordinates. However, comparison with the
work in ADM coordinates, has shown
\cite{Damour:2000ni,deAndrade:2000gf} that, \textit{if there were
consistency} between the two calculations one should have the
following link between $\lambda$ and the corresponding ADM ``static
ambiguity'' parameter $\omega_s$,
\begin{equation}
\label{eq5.54}
\lambda = - \frac{3}{11} \, \omega_s - \frac{1987}{3080}\,.
\end{equation}
If dimensional regularization is a fully consistent regularization
scheme for classical perturbative gravity, we then expect that the
dim. reg. determination of $\omega_s$ in ADM coordinates
\cite{Damour:2001bu}, namely $\omega_s^\text{dr ADM} = 0$, should lead
to a dim. reg. direct determination of $\lambda$ (in harmonic
coordinates) of $\lambda^\text{dr harmonic} = - 1987/3080$. We will
turn to this verification in a moment.

The two other parameters, denoted above $r'_1$, $r'_2$, entering the
general ``parametric'' harmonic equations of motion (\ref{eq5.53}) have
the dimension of length and have the character of gauge parameters.
Indeed, they can be chosen at will (except that one cannot set them to
zero) by the effect of shifts of the world-line, induced for instance
[but not necessarily, \textit{cf.} a discussion in Section \ref{Shift}
above] by some gauge transformations. In the way they were originally
introduced \cite{Blanchet:2000ub}, the two parameters $r_1'$ and $r_2'$
can be interpreted as some infinitesimal radial distances used as
cut-offs when the field point tends towards the two singularities
$\mathbf{y}_1$ and $\mathbf{y}_2$. Therefore in principle $\ln r_1'$
and $\ln r_2'$ should initially be thought of as being (formally)
infinite. However, it is trivial to show that by a (formally infinite)
gauge transformation, involving the logarithmic ratios $\ln
(r_1''/r_1')$ and $\ln (r_2''/r_2')$, where $r_1''$ and $r_2''$ denote
any two \textit{finite} length scales, one can replace everywhere
$r_1'$, $r_2'$ by the finite scales $r_1''$, $r_2''$. By this process
it is therefore possible to identify the two sets of scales and thereby
to think of the scales $r_1'$, $r_2'$ as being in fact finite, as
implicitly done in Ref. \cite{Blanchet:2000ub}. In the language of
renormalization theory, the original (infinitesimal) scales $r_1'$ and
$r_2'$ would be referred to as \textit{Hadamard-regularization} scales
entering the computation of divergent Poisson integrals [see Section
\ref{HadamardPoisson} above], while the (finite) scales $r_1''$ and
$r_2''$ would be referred to as the arbitrary \textit{renormalization}
scales entering the final, renormalized harmonic-coordinates equations
of motion. In the present paper, in order to remain close to the
notation used in \cite{Blanchet:2000ub}, we shall keep the notation
$r_1'$ and $r_2'$, but interpret them as arbitrary finite constants,
which means that we shall \textit{identify} them with the finite
renormalization length scales $r_1''$ and $r_2''$. In other words, the
scales $r_1'$, $r_2'$ used in the present Section should in principle
be distinguished from the scales $r_1'$, $r_2'$ used in Section
\ref{HadamardPoisson} above. [Remember, in this respect, that the
regularization scales $r_1'$, $r_2'$ have disappeared when computing
the dim. reg. equations of motion.]

With our notation, and still focussing on the static contributions to
the equations of motion, the ``parametric'' equations of motion
(\ref{eq5.53}) imply the following structure for the static
coefficients $c_{mn}$:
\begin{subequations}
\label{cStaticLambda}
\begin{eqnarray}
\label{eq5.55}
c_{31}^\text{BF} (\lambda , r'_1 , r'_2) &=& \frac{44}{3} \ln \left(
\frac{r_{12}}{r'_1} \right) - \frac{3187}{1260}\,,\\
\label{eq5.56}
c_{22}^\text{BF} (\lambda , r'_1 , r'_2) &=& \frac{34763}{210} -
\frac{44}{3} \, \lambda - \frac{41}{16} \, \pi^2\,,\\
\label{eq5.57}
c_{13}^\text{BF} (\lambda , r'_1 , r'_2) &=& - \frac{44}{3} \ln
\left( \frac{r_{12}}{r'_2} \right) + \frac{10478}{63} - \frac{44}{3}
\, \lambda - \frac{41}{16} \, \pi^2\,.
\end{eqnarray}\end{subequations}
It will be convenient to replace the parameter $\lambda$ by the
parameter $\omega_s$, using (\ref{eq5.54}) as a defining one-to-one map
between $\lambda$ and $\omega_s$. With this change of notation the
static coefficients become
\begin{subequations}
\label{cStaticOmega}
\begin{eqnarray}
\label{eq5.58}
c_{31}^\text{BF} (\omega_s , r'_1 , r'_2) &=& \frac{44}{3} \ln \left(
\frac{r_{12}}{r'_1} \right) - \frac{3187}{1260}\,,\\
\label{eq5.59}
c_{22}^\text{BF} (\omega_s , r'_1 , r'_2) &=& 175 + 4 \, \omega_s -
\frac{41}{16} \, \pi^2\,,\\
\label{eq5.60}
c_{13}^\text{BF} (\omega_s , r'_1 , r'_2) &=& - \frac{44}{3} \ln
\left( \frac{r_{12}}{r'_2} \right) + \frac{110741}{630} + 4 \,
\omega_s - \frac{41}{16} \, \pi^2\,.
\end{eqnarray}
\end{subequations}
Note that there are two combinations of the three coefficients
$c_{mn}^\text{BF}$ which do not depend on $\ln r_{12}$, namely
$c_{22}^\text{BF}$, and the combination $c_{31}^\text{BF} +
c_{13}^\text{BF}$, or even better the combination
\begin{equation}
\label{eq5.61}
c_{31}^\text{BF} + c_{13}^\text{BF} - c_{22}^\text{BF} =
\frac{44}{3} \ln \left( \frac{r'_2}{r'_1} \right) - \frac{7}{4}\,,
\end{equation}
which depends neither on $\ln r_{12}$ nor on $\omega_s$ (or
$\lambda$), and which contains, like for $c_{22}^\text{BF}$, simpler
looking rational numbers.

We now come back to the effect of the general shift (\ref{eq5.51}) on
the dim. reg. equations of motion. Let us first focus on the static
terms. We recall that the (dim. reg.) \textit{renormalized} equations
of motion had necessarily the form (\ref{eq5.49}). By projecting the
latter equation along the static terms $c_{mn}$, with $m+n = 4$
[recalling Eq.~(\ref{eq5.5})], it will induce a result for the
\textit{renormalized} static coefficients of the form
\begin{equation}
\label{eq5.62}
c_{mn}^\text{ren} = c_{mn}^\text{dr} (\varepsilon) + \delta_{\bm{\xi}
(\varepsilon)} c_{mn}\,,
\end{equation}
where the $\delta_{\bm{\xi} (\varepsilon)} c_{mn}$'s are the static
coefficients corresponding to $\delta_{\bm{\xi}} \, a_1^i$,
Eq.~(\ref{eq5.50}). When choosing $\xi_a^i(\varepsilon)$ of the form
(\ref{eq5.51}), we see from Eq.~(\ref{eq5.52}) that $\delta_{\bm{\xi}}
\, c_{mn}$ is simply obtained by projecting the first term on the
R.H.S. of (\ref{eq5.52}). Remembering that $\mathbf{a}_2 \propto m_1$
and $\mathbf{a}_1 \propto m_2$, we see that the latter term contains
the factor $(m_1^2 - m_2^2) \, m_1 \, m_2 = m_1^3 \, m_2 - m_1 \,
m_2^3$. Therefore, without doing any further calculation, we see that
the shifts $\delta_{\xi} \, c_{mn}$ have the special properties:
$\delta_{\bm{\xi}} \, c_{22} = 0$ and $\delta_{\bm{\xi}} \, c_{31} +
\delta_{\bm{\xi}} \, c_{13} = 0$. In other words, a shift of the
world-lines of the type (\ref{eq5.51}) leaves invariant both $c_{22}$
and the combination $c_{31} + c_{13}$ (as well therefore as the
combination $c_{31} + c_{13} - c_{22}$ considered above). As a
consequence, we can compute without effort from our previous
regularized (but unrenormalized) dim. reg. results (\ref{eq5.29}) the
following two combinations of the $\xi^i
(\varepsilon)$-\textit{renormalized} static coefficients:
\begin{subequations}\begin{eqnarray}
\label{eq5.63}
&&c_{22}^\text{ren} = c_{22}^\text{dr} = 175 - \frac{41}{16} \, \pi^2
\,,\\
\label{eq5.64}
&&c_{31}^\text{ren} + c_{13}^\text{ren} - c_{22}^\text{ren} =
c_{31}^\text{dr} + c_{13}^\text{dr} - c_{22}^\text{dr} = - \frac{7}{4}
\,.
\end{eqnarray}\end{subequations}

By comparing (\ref{eq5.63}) with Eq.~(\ref{eq5.59}) we discover that
our present calculation using dimensional regularization in harmonic
coordinates necessarily implies that
\begin{equation}
\label{eq5.65}
\omega_s = 0 ~\Longleftrightarrow ~\lambda = - \frac{1987}{3080}\,.
\end{equation}
This nicely confirms the previous determination of $\omega_s$ by a dim.
reg. calculation in ADM-type coordinates \cite{Damour:2001bu}. We think
that our present harmonic-coordinates dim. reg. result calculation is
important in proving the consistency of dimensional regularization, and
thereby in confirming the physical significance of the result
(\ref{eq5.65}). A recent calculation \cite{itoh1,itoh2} has also
confirmed independently the result (\ref{eq5.65}) by means of a
completely different method based on surface integrals, and aimed at
describing compact (strongly-gravitating) objects.

By comparing (\ref{eq5.64}) with (\ref{eq5.61}),
we further see that
\begin{equation}
\label{eq5.66}
r'_1 = r'_2~~~\hbox{[in the case of the dim. reg. shift
(\ref{eq5.51})]}\,.
\end{equation}
Contrary to (\ref{eq5.65}) which represents the determination of a
physical parameter (having an invariant meaning), the result
(\ref{eq5.66}) has no invariant physical significance.
Eq.~(\ref{eq5.66}) is simply a consequence of our particular choice for
the shift vector (\ref{eq5.51}), in which we assumed that $e(d)$ is a
purely numerical coefficient, independent on any properties indexed by
the particles' labels 1 and 2. In summary the particular shift
(\ref{eq5.51}) yields some equations of motion which are physically
equivalent to a subclass of the general equations of motion considered
in \cite{Blanchet:2000ub}, characterized by the constraint
(\ref{eq5.66}).

Next we relate the common length scale (\ref{eq5.66}) to the basic
length scale $\ell_0$ entering dimensional regularization. For doing
this we need to fully specify the value of the shift, \textit{i.e.}, to
choose a specific coefficient $e(d) = e(3) + \varepsilon \, e'(3) +
{\mathcal O} (\varepsilon^2)$ in Eq.~(\ref{eq5.51}). We already know
{}from Eq.~(\ref{eq5.40}) that the coefficient $e(d)$ in
Eq.~(\ref{eq5.51}) must tend to $11/3$ when $d \rightarrow 3$, if the
$\xi$-shift is to remove the poles in the bulk metric. As in quantum
field theory (QFT) we could then define the \textit{Minimal
Subtraction} (MS) shift as
\begin{equation}
\label{eq5.67}
\xi_{a \text{MS}}^i \equiv \frac{11}{3 \varepsilon} \, \frac{G_N^2 \,
m_a^2}{c^6} \, (a_a^i)^{d=3}\,.
\end{equation}
However, as is well-known in QFT, such a MS subtraction has the
unpleasant feature of leaving some logarithms of $\pi$ and the Euler
constant in the renormalized results. These numbers come from the
expansion of the Gamma function and the associated dimension-dependent
powers of $\pi$ entering the $d$-dimensional Green function. In our
context, these numbers showed up in Eq.~(\ref{eq5.29}) in the guise of
the combination
\begin{equation}
\ln (\overline q) \equiv \ln (4 \pi e^C) = C + \ln (4 \pi)\,.
\end{equation}
Like in QFT, this leads us to
consider the following \textit{modified minimal subtraction}
($\overline{\text{MS}}$) shift,
\begin{equation}
\label{eq5.68}
\xi_{a \overline{\text{MS}}}^i \equiv \frac{11}{3 \varepsilon} \,
\frac{G_N^2 \, \tilde k^2 \, m_a^2}{c^6} \, a_a^i\,,
\end{equation}
which differs from the MS shift (\ref{eq5.67}) by the explicit factor
of $\tilde{k}^2$ it contains, and by the use of the $d$-dimensional
(Newtonian) acceleration given by Eq.~(\ref{eq5.46}). The inclusion of
two explicit powers of $\tilde k$ in the coefficient $e(d)$ entering
Eq.~(\ref{eq5.51}), \textit{i.e.} the definition
$e_{\overline{\text{MS}}} (d) = \frac{11}{3}\, \tilde{k}^2$, means,
when remembering that $a_a^i$, Eq.~(\ref{eq5.46}), contains one power
of $\tilde k$, that the static terms in Eq.~(\ref{eq5.52}) will have
four powers of $\tilde k$ and the kinetic terms three. The overall
factor $\tilde k^4$ in the static terms is natural because these terms
are of order $G^4$ and the $\mathbf{x}$-space gravitational propagator
in $d$ dimensions always includes the combination $G \, \tilde k \,
\vert \mathbf{x}-\mathbf{x}' \vert^{2-d}$. Finally, using the fact that
the expansion of $\tilde k (d)$ near $d=3$ reads
\begin{equation}
\label{eq5.69}
\tilde k (d) \equiv
\frac{\Gamma\left(\frac{1+\varepsilon}{2}\right)}{\pi^{\frac{1
+\varepsilon}{2}}} = 1 - \frac{1}{2} \, \varepsilon \ln \overline q +
{\mathcal O} (\varepsilon^2)\,,
\end{equation}
it is easy to see that the $\overline{\text{MS}}$ shift defined by
(\ref{eq5.68}) will cancel the $\ln \overline q$ in the bare
dim. reg. results of Eq.~(\ref{eq5.29}). Finally, we find that the
evaluation of Eq.~(\ref{eq5.52}) for the specific
$\overline{\text{MS}}$ shift, given by $e_{\overline{\text{MS}}} (d) =
\frac{11}{3}\, \tilde{k}^2$, yields for the
$\overline{\text{MS}}$-renormalized static coefficients
\begin{subequations}\label{eq5.70abc}\begin{eqnarray}
\label{eq5.70}
c_{31}^{\overline{\text{MS}}} &=& \frac{44}{3} \ln \left(
\frac{r_{12}}{\ell_0} \right) - \frac{35}{36}\,,\\
\label{eq5.71}
c_{22}^{\overline{\text{MS}}} &=& 175 - \frac{41}{16} \, \pi^2\,,\\
\label{eq5.72}
c_{13}^{\overline{\text{MS}}} &=& - \frac{44}{3} \ln \left(
\frac{r_{12}}{\ell_0} \right) + \frac{1568}{9} - \frac{41}{16} \,
\pi^2\,,
\end{eqnarray}\end{subequations}
where the reader can note that the $\ln (r_{12}^2)$ entering the bare
dim. reg. result (\ref{eq5.29}) have been transformed into $\ln
(r_{12} / \ell_0)$ through the $\varepsilon$-expansion of the factor
$r_{12}^{-5-2\varepsilon}$ present in $\delta_{\bm{\xi}} \, a_1^i$.

We
already discussed above the comparison of two simple combinations of
the dim. reg. results (\ref{eq5.70abc}) with the
Hadamard-regularization results (\ref{cStaticLambda}). It is easy to
see that the remaining independent combination, say (\ref{eq5.70}), is
fully consistent with its counterpart (\ref{eq5.55}), and allows one
to relate the basic renormalization length scale $\ell_0$ entering
dim. reg. to the common length scale (\ref{eq5.66}) entering the
general equations of motion of \cite{Blanchet:2000ub}:
\begin{equation}
\label{eq5.73}
\ln \left( \frac{r'_1}{\ell_0} \right) = \ln \left(
\frac{r'_2}{\ell_0} \right) = - \frac{327}{3080}~~~~\hbox{(for the
$\overline{\text{MS}}$ renormalization)}\,.
\end{equation}
Evidently, the precise values one gets for $r'_1$ and $r'_2$ depend on
the precise choice of the compensating shift.

Let us now remark that in fact one can recover exactly, provided of
course that the crucial result (\ref{eq5.65}) holds, the general
``dissymmetric'' class of equations of motion of
\cite{Blanchet:2000ub}, \textit{i.e.}, the general parametric result
(\ref{cStaticLambda}) or (\ref{cStaticOmega}) with $r'_1 \neq r'_2$.
For this purpose it suffices to consider a slightly more general shift
than the one assumed in the $\overline{\text{MS}}$ regularization.
Namely, consider a shift of the same form as (\ref{eq5.51}), but in
which one allows the $d$-dependent coefficient $e(d)$ to depend on the
label of the particle in question, that is
\begin{equation}
\xi_a^i = \frac{e_a(d)}{\varepsilon} \, \frac{G_N^2 \, m_a^2}{c^6} \,
a_a^i\,,
\label{shifteta}\end{equation}
where now $e_1(d)$ and $e_2(d)$ are allowed to be different from each
other. The most general way of parametrizing such dissymmetric $e_a(d)$
[however constrained by $e_a(3) = 11/3$] is
\begin{equation}
e_a(d) = \frac{11}{3}\tilde{k}^2\left[1-2\varepsilon \rho_a
+ \mathcal{O}(\varepsilon^2)\right]\, ,
\label{ead}
\end{equation}
with two independent numerical coefficients $\rho_a$. It is then
easily checked that the shift (\ref{shifteta}) defined by the
particular choice
\begin{equation}
\rho_a = \ln\left(\frac{r'_a}{ \ell_0}\right)+\frac{327}{3080} \,,
\label{rhoa}
\end{equation}
transforms the dim. reg. equations of motion into the general
($r'_a$-dependent) family of solutions obtained in
\cite{Blanchet:2000ub}. If we suppose that the constraints
(\ref{eq5.73}) hold, then $\rho_a = 0$ and we recover the shift
assumed in the $\overline{\text{MS}}$ regularization. On the other
hand, note than one can reach even more general classes of
renormalized harmonic equations of motion in dim. reg. (as one could
have also done in Hadamard regularization). Indeed, we could use the
freedom indicated in Eq.~(\ref{eq5.44}) of adding \textit{arbitrary}
finite parts to the shifts. Anyway, the result that the shift
(\ref{shifteta})-(\ref{rhoa}) gives equivalence between the
dim. reg. and the extended-Hadamard 3PN accelerations [we check in
Section \ref{Kinetic} below that the kinetic terms work also],
constitutes the main result of the present paper (Theorem~2 in
the Introduction).

The two approaches we have discussed here are of course equivalent:
choosing some dim. reg. basic length scale $\ell_0$ and some specific,
simplifying dim. reg. shift (such as the $\overline{\text{MS}}$ one),
and then determining the values of the scales $r'_a$ for which the
dim. reg. results match with the Hadamard reg. ones; or arbitrarily
choosing some Hadamard scales $r'_a$ and then determining the
corresponding general dissymmetric dim. reg. shift
(\ref{shifteta})-(\ref{rhoa}), in terms of the chosen $r'_a$'s. What
is important is that we have checked that the \textit{three}
renormalized dim. reg. static coefficients (\ref{eq5.29}) are fully
compatible with the \textit{three} extended-Hadamard reg. static
coefficients (\ref{cStaticLambda}) or (\ref{cStaticOmega}), and that
their comparison yields \textit{one and only one} physical result,
namely: $\lambda = -\frac{1987}{3080}$.

\subsection{Kinetic terms and check of the consistency of
dimensional regularization}
\label{Kinetic}
Up to now we have verified the following aspects of the consistency of
a dim. reg. treatment of the 3PN dynamics of two point particles: (1)
consistency between the shift (\ref{eq5.40}) needed to renormalize the
bulk metric and the shift (\ref{eq5.68}) [or
(\ref{shifteta})-(\ref{ead})] needed to renormalize the equations of
motion; (2) consistency between the three finite, renormalized
dim. reg. static coefficients (\ref{eq5.70abc}) and the general
three-dimensional ones (\ref{cStaticLambda}), \cite{Blanchet:2000ub};
and (3) consistency between the present dim. reg. value of $\lambda$
and the previously derived dim. reg. value of $\omega_s$ in the
ADM-Hamiltonian \cite{Damour:2001bu}. It remains, however, to check
that the velocity-dependent terms in the renormalized
dim. reg. equations of motion do agree with their analogs in the
harmonic-coordinates equations of motion of
\cite{Blanchet:2000ub}. This will in particular prove that the
dim. reg. equations of motion are Lorentz invariant.

In the notation of Eq.~(\ref{eq5.5}) above, we need to consider the
values of the velocity-dependent coefficients $c_{21} (\mathbf{v}_1 ,
\mathbf{v}_2 , \mathbf{n}_{12})$, $c'_{21} (\mathbf{v}_1 ,
\mathbf{v}_2 , \mathbf{n}_{12})$, $c''_{21} (\mathbf{v}_1 ,
\mathbf{v}_2 , \mathbf{n}_{12})$, $c_{03} (\mathbf{v}_1 , \mathbf{v}_2,
\mathbf{n}_{12})$, $c'_{03} (\mathbf{v}_1 , \mathbf{v}_2 ,
\mathbf{n}_{12})$, and $c''_{03} (\mathbf{v}_1 , \mathbf{v}_2 ,
\mathbf{n}_{12})$. In Ref.~\cite{Blanchet:2000ub}, they were shown
to take the following parametric forms, which actually depend only on
the regularization scale $r'_1$ but not on $r'_2$ nor $\lambda$:
\begin{subequations}
\label{eq5.78}
\begin{eqnarray}
c_{21}^{\text{BF}}(r'_1)&=&-22
\left[v_{12}^2 - 5 (n_{12}v_{12})^2\right]
\ln\left(\frac{r_{12}}{r'_1}\right)
+ \frac{48197}{840}\, v_1^2
-\frac{36227}{420} (v_1 v_2)
+ \frac{36227}{840}\, v_2^2
\nonumber\\
&&- \frac{45887}{168} (n_{12}v_1)^2
+ \frac{24025}{42} (n_{12}v_1)(n_{12}v_2)
-\frac{10469}{42} (n_{12}v_2)^2
\,,\qquad\label{eq5.78a}\\
c'^{\text{BF}}_{21}(r'_1)
= c''^{\text{BF}}_{21}(r'_1)&=&
-44 (n_{12}v_{12})
\ln\left(\frac{r_{12}}{r'_1}\right)
+ \frac{31397}{420}(n_{12}v_1)
- \frac{36227}{420} (n_{12}v_2)
\,,\label{eq5.788b}\\
c_{03}^{\text{BF}}&=&18 (v_1 v_2)
- 9\, v_2^2 - (n_{12}v_1)^2 + 2(n_{12}v_1)(n_{12}v_2)
+\frac{43}{2}(n_{12}v_2)^2\,,\label{eq5.78c}\\
c'^{\text{BF}}_{03}
= c''^{\text{BF}}_{03}&=&4(n_{12}v_1) + 5 (n_{12}v_2)\,.
\label{eq5.78d}
\end{eqnarray}
\end{subequations}

As explained in Eq.~(\ref{eq5.1}), the dim. reg. expressions of
these coefficients can be computed as the sum of pure
Hadamard-Schwartz (pHS) contributions, $c_{mn}^\text{pHS}$ and
$c'^{\text{pHS}}_{mn}$, and the ``$\text{dr} - \text{pHS}$''
differences, ${\mathcal D} c_{mn}$, ${\mathcal D} c'_{mn}$.
The calculation of the pHS contributions has been explained in Section
\ref{additional} above, and we get the following results from
Eqs.~(\ref{defpH}) and (\ref{eq5.78}):
{\allowdisplaybreaks
\begin{subequations}
\label{eq5.74}
\begin{eqnarray}
c_{21}^\text{pHS}&=&- 22
\left[v_{12}^2 - 5(n_{12}v_{12})^2\right]
\ln\left(\frac{r_{12}}{r'_1} \right)
+ \frac{10639}{168}\, v_1^2
- \frac{5879}{60} (v_1 v_2)
+ \frac{5843}{120}\, v_2^2
\nonumber\\
&&- \frac{50885}{168} (n_{12}v_1)^2
+ \frac{1892}{3} (n_{12}v_1)(n_{12}v_2)
- \frac{3325}{12} (n_{12}v_2)^2
\,,\label{eq5.74a}\\
c'^{\text{pHS}}_{21}&=&- 44 (n_{12}v_{12})
\ln\left(\frac{r_{12}}{r'_1}\right)
+\frac{7279}{84} (n_{12}v_1)
- \frac{5879}{60} (n_{12}v_2)
\,,\label{eq5.74b}\\
c''^{\text{pHS}}_{21}&=&- 44 (n_{12}v_{12})
\ln\left(\frac{r_{12}}{r'_1}\right)
+\frac{5189}{60} (n_{12}v_1)
- \frac{5843}{60} (n_{12}v_2)
\,,\label{eq5.74c}\\
c_{03}^\text{pHS}&=&18 (v_1 v_2) -
\frac{64}{7}\, v_2^2 - (n_{12}v_1)^2 + 2 (n_{12}v_1)(n_{12}v_2)
+\frac{311}{14}(n_{12}v_2)^2\,,\label{eq5.74d}\\
c'^{\text{pHS}}_{03}&=&4(n_{12}v_1) + 5 (n_{12}v_2)\,,\label{eq5.74e}\\
c''^{\text{pHS}}_{03}&=&4(n_{12}v_1) + \frac{37}{7} (n_{12}v_2)
\,.\label{eq5.74f}
\end{eqnarray}
\end{subequations}}$\!\!$
Secondly, the method explained in Sections \ref{Diff} and
\ref{DimRegStat} for computing the differences ${\mathcal D}
\mathbf{a}_a$ is found (after doing calculations similar to those
reported in the Tables of ``static'' contributions above) to lead to
{\allowdisplaybreaks
\begin{subequations}
\label{eq5.75}
\begin{eqnarray}
{\mathcal D} c_{21}&=&\left[
\frac{11}{\varepsilon}-\frac{33}{2}
\ln\left(\overline q\, r'^{4/3}_1 r_{12}^{2/3}\right)\right]
\left[v_{12}^2 -5 (n_{12}v_{12})^2
\right]
+\frac{499}{42}\, v_1^2
- \frac{359}{15}(v_1 v_2)
\nonumber\\
&&+ \frac{184}{15}\, v_2^2
- \frac{2957}{42} (n_{12}v_1)^2
+ \frac{425}{3} (n_{12}v_1) (n_{12}v_2)
- \frac{217}{3} (n_{12}v_2)^2
+ {\mathcal O} (\varepsilon)\,,\qquad
\label{eq5.75a}\\
{\mathcal D} c'_{21}&=&\left[
\frac{22}{\varepsilon}
-33\ln\left(\overline q\, r'^{4/3}_1 r_{12}^{2/3}\right)
\right](n_{12}v_{12})
+ \frac{499}{21}(n_{12}v_1)
- \frac{359}{15}(n_{12}v_2)
+ {\mathcal O} (\varepsilon)\,,
\label{eq5.75b}\\
{\mathcal D} c''_{21}&=&\left[
\frac{22}{\varepsilon}
-33\ln\left(\overline q\, r'^{4/3}_1 r_{12}^{2/3}\right)
\right](n_{12}v_{12})
+ \frac{359}{15}(n_{12}v_1)
- \frac{368}{15}(n_{12}v_2)
+ {\mathcal O} (\varepsilon)\,,
\label{eq5.75c}\\
{\mathcal D} c_{03}&=&\frac{1}{7}\, v_2^2
- \frac{5}{7} (n_{12}v_2)^2\,,\label{eq5.75d}\\
{\mathcal D} c'_{03}&=&0\,,\label{eq5.75e}\\
{\mathcal D} c''_{03}&=&-\frac{2}{7}(n_{12}v_2)
\,.\label{eq5.75f}
\end{eqnarray}
\end{subequations}}$\!\!$
Together with Eqs.~(\ref{eq5.28}) above, these equations give the full
difference between the dimensionally regularized and the pure Hadamard
accelerations, and they constitute the main new input of the present
work. The bare dim. reg. results, $c_{mn}^\text{dr} = c_{mn}^\text{pHS}
+ {\mathcal D} c_{mn}$, read therefore {\allowdisplaybreaks
\begin{subequations}
\label{eq5.76}
\begin{eqnarray}
c_{21}^\text{dr}&=&\left[\frac{11}{\varepsilon}
-\frac{33}{2}
\ln\left(\overline q\, r_{12}^2\right)\right]
\left[v_{12}^2 - 5 (n_{12}v_{12})^2\right]
+ \frac{1805}{24}\, v_1^2
- \frac{1463}{12} (v_1 v_2)
\nonumber\\
&&+ \frac{1463}{24}\, v_2^2
-\frac{8959}{24} (n_{12}v_1)^2
+ \frac{2317}{3} (n_{12}v_1) (n_{12}v_2)
- \frac{4193}{12} (n_{12}v_2)^2
\,,\qquad\label{eq5.76a}\\
c'^{\text{dr}}_{21}
= c''^{\text{dr}}_{21}&=&\left[
\frac{22}{\varepsilon}
- 33 \ln\left(\overline q\, r_{12}^2\right)
\right](n_{12}v_{12})
+ \frac{1325}{12}(n_{12}v_1)
- \frac{1463}{12} (n_{12}v_2)
\,,\label{eq5.76b}\\
c_{03}^\text{dr}&=&18 (v_1 v_2)
- 9\, v_2^2 - (n_{12}v_1)^2 + 2(n_{12}v_1)(n_{12}v_2)
+\frac{43}{2}(n_{12}v_2)^2\,,\label{eq5.76c}\\
c'^{\text{dr}}_{03}
= c''^{\text{dr}}_{03}&=&4(n_{12}v_1) + 5 (n_{12}v_2)\,.\label{eq5.76d}
\end{eqnarray}
\end{subequations}}$\!\!$
Note that in the final result the equality between $c'_{mn}$ and
$c''_{mn}$, \textit{i.e.}, between $B'$ and $B''$ in Eq.~(\ref{eq5.3}),
is recovered. The bare dim. reg. kinetic coefficients (\ref{eq5.76})
contain poles $\propto 1/\varepsilon$ but do not depend anymore of the
arbitrary Hadamard regularization scale $r'_1$ which appeared in
(\ref{eq5.74}). As in the case discussed above of the static
coefficients the previous kinetic coefficients do not involve any
adimensionalizing length scales in the logarithms of $r_{12}$ they
contain. This is consistent with the fact that it is the combinations
$\ell_0^{3 \varepsilon} \, c_{mn}^\text{dr}$ and $\ell_0^{3
\varepsilon} \, c'^{\text{dr}}_{mn}$ which have the same physical
dimension as their $d=3$ counterparts.

Finally, given a specific choice of shift, say the
$\overline{\text{MS}}$ one, Eq.~(\ref{eq5.68}), the
\textit{renormalized} kinetic coefficients are obtained by adding to
(\ref{eq5.76}) the velocity-dependent part of the effect of the shift,
\textit{i.e.}, the second term on the R.H.S. of Eq.~(\ref{eq5.52})
[with, say, $e^{\overline{\text{MS}}} (d) = (11/3) \, \tilde
k^2$]. Our final $(\overline{\text{MS}})$ results for the renormalized
kinetic coefficients are found to be
{\allowdisplaybreaks
\begin{subequations}
\label{eq5.77}
\begin{eqnarray}
c_{21}^{\overline{\text{MS}}}&=&-22
\left[v_{12}^2 - 5 (n_{12}v_{12})^2\right]
\ln\left(\frac{r_{12}}{\ell_0}\right)
+ \frac{1321}{24}\, v_1^2
-\frac{979}{12} (v_1 v_2)
\nonumber\\
&&+ \frac{979}{24}\, v_2^2
- \frac{6275}{24} (n_{12}v_1)^2
+ \frac{1646}{3} (n_{12}v_1)(n_{12}v_2)
-\frac{2851}{12} (n_{12}v_2)^2
\,,\qquad\label{eq5.77a}\\
c'^{\overline{\text{MS}}}_{21}
= c''^{\overline{\text{MS}}}_{21}&=&
-44 (n_{12}v_{12})
\ln\left(\frac{r_{12}}{\ell_0}\right)
+ \frac{841}{12}(n_{12}v_1)
- \frac{979}{12} (n_{12}v_2)
\,,\label{eq5.77b}\\
c_{03}^{\overline{\text{MS}}}&=&18 (v_1 v_2)
- 9\, v_2^2 - (n_{12}v_1)^2 + 2(n_{12}v_1)(n_{12}v_2)
+\frac{43}{2}(n_{12}v_2)^2\,,\label{eq5.77c}\\
c'^{\overline{\text{MS}}}_{03}
= c''^{\overline{\text{MS}}}_{03}&=&4(n_{12}v_1) + 5 (n_{12}v_2)
\,.\label{eq5.77d}
\end{eqnarray}
\end{subequations}}

\noindent
When comparing these results with the ones of
Ref.~\cite{Blanchet:2000ub}, Eqs.~(\ref{eq5.78}) above, one remarkably
finds that our previously derived link (\ref{eq5.73}) is necessary and
sufficient for ensuring the full compatibility between the renormalized
dim. reg. results and the corresponding Hadamard reg. ones. Note that
the rational coefficients entering the dim. reg. results are often
simpler than the coefficients entering the equations of motion of
\cite{Blanchet:2000ub}.

The results Eq.~(\ref{eq5.77}) complete our check of the full
consistency of the dim. reg. evaluation of the 3PN equations of motion,
and the proof of Theorems~1 and 2 stated in the Introduction.

\section{Conclusions}
\label{Conclusions}
We have used dimensional regularization (\textit{i.e.} analytic
continuation in the spatial dimension $d$) to determine the spacetime
metric and the equations of motion (EOM) in \textit{harmonic
coordinates}, of two, gravitationally interacting, point masses, at the
third post-Newtonian (3PN) order of General Relativity. Our starting
point consisted in writing the 3PN-accurate metric $g_{\mu \nu}(x)$ in
terms of a certain number of ``elementary potentials''
$V,\,V_i,\,\hat{W}_{ij},\,\cdots$, satisfying a hierarchy of
inhomogeneous d'Alembert equations of the form $\Box\text{(potential) =
source}$. The sources of the latter equations contain both ``compact''
terms, \textit{i.e.}, in the present case \textit{contact} terms of
the form
$\mathcal{F}[V(\mathbf{x}),V_i(\mathbf{x}),\hat{W}_{ij}(\mathbf{x}),
\cdots] \delta^{(d)}(\mathbf{x}-\mathbf{y}_1)$, and nonlinearly
generated ``non compact'' terms of the typical form, say,
$\partial(\text{potential})\partial(\text{potential})$. This
representation of the 3PN metric, as well as the associated iterative
way of solving for the potentials [using the time-symmetric Green's
function $ {\Box}^{-1} = \Delta^{-1} + c^{-2}\Delta^{-2}\partial_t^2 +
{\cal O} (c^{-4})$] is a direct generalization of the one used in
Ref.~\cite{Blanchet:2000ub}. However, it has been crucial for our work
to determine (in Section \ref{FieldEq}) the dependence upon the
dimension $d$ of the coefficients appearing in this representation, as
well as the $d$-dependence of the kernels expressing the operators
$\Delta^{-1}$ and $\Delta^{-2}$ in $\mathbf{x}$-space.

By studying the structure of the iterative solution for the metric, and
that of the corresponding EOM (which are conveniently pictured by means
of diagrams, see Figs.~\ref{fig1}-\ref{fig8}), we determined, in the
form of a Laurent expansion in $\varepsilon \equiv d-3$, the pole part
of the metric $g_{\mu\nu}(x)$, and the pole and finite parts of the
EOM, namely $\mathbf{a}_a = A_a(\mathbf{y}_b,\mathbf{v}_b)$ where
$a,b=1,2$ and $\mathbf{v}_b\equiv d\mathbf{y}_b/dt$. [See, however,
Appendix \ref{Littleg} where the basic quadratically non-linear kernel
$\textsl{g}(\mathbf{x} , \mathbf{y}_1 , \mathbf{y}_2)$ is computed in
any $d$ dimensions, not necessarily close to 3.] Our calculations
relied heavily on previous work in $d=3$ \cite{Blanchet:2000ub}, and
were technically implemented in two steps (at least for the
determination of the EOM, which are more delicate; the determination of
the pole part of the metric uses only the second step).

(i) A first step consisted of subtracting from the final, published
results for the EOM \cite{Blanchet:2000ub}, seven contributions that
were specific consequences of the use of an extension of the Hadamard
regularization method \cite{Blanchet:2000nu,Blanchet:2000cw} (which
included an extension of the Schwartz notion of distributional
derivative). The result of this first step is referred to as the
``pure Hadamard-Schwartz'' evaluation of the EOM.

(ii) The second step was the evaluation of the \textit{difference}
between the dimensional regularization of each contribution to the EOM
(written in terms of the iterative solutions for the various
potentials $V,\,V_i,\,\hat{W}_{ij},\,\cdots$), and the corresponding
``pure Hadamard-Schwartz'' contribution obtained in the first step.
This difference is obtained, similarly to the method used in
\cite{Damour:2001bu}, by splitting the $d$-dimensional integral into
several pieces, and by carefully analyzing the terms due to the
neighborhoods of the two singular points $\mathbf{y}_1,\,\mathbf{y}_2$
(including possible $d$-dimensional distributional contributions).

Concerning the ``bulk'' metric $g_{\mu\nu}(x)$, at a field point away
{}from the singular particle world-lines $\mathbf{y}_a(t)$, we derived
only the pole part, that is the coefficient of $1/\varepsilon$ in the
Laurent expansion of $g_{\mu\nu}(x; \varepsilon)$. We found that at the
3PN order only the time-time component of the metric $g_{00}(x)$
contained a pole [see Eq.~(\ref{eq5.39})]. For the EOM we derived both
the pole part and the finite part, \textit{i.e.} $\mathbf{a}_a \sim
\varepsilon^{-1} + \varepsilon^0 + \mathcal{O}(\varepsilon)$. The parts
of the EOM for which the regularization was delicate are given by the
nine coefficients $c_{31},\,\cdots,\,c''_{03}$ defined in
Eq.~(\ref{eq5.5}). Our complete results for the
dimensionally-regularized values of these nine ``delicate''
coefficients are given in Eqs. (\ref{eq5.29}) and (\ref{eq5.76}).

We proved that the pole parts of both the metric and the EOM can be
``renormalized away'' by suitable \textit{shifts of the world-lines} of
the form $\mathbf{y}_a^\text{bare} = \mathbf{y}_a^\text{ren} +
\bm{\xi}_a(\varepsilon)$, where $\mathbf{y}_a^\text{bare}$ is the
original world-line on which are initially concentrated the
$\delta$-function sources representing the point masses, where the
shifts $\bm{\xi}_a(\varepsilon) \sim \varepsilon^{-1} + \varepsilon^0 +
\mathcal{O}(\varepsilon)$ are of the 3PN order, and where the EOM of
the renormalized world-line $\mathbf{y}_a^\text{ren}$ is
\textit{finite} as $\varepsilon\rightarrow 0$. The general form of the
needed shifts is given by Eq.~(\ref{eq5.44}) with (\ref{eq5.40}). The
renormalized EOM corresponding to the ``modified Minimal Subtraction''
scheme (\ref{eq5.68}) are given by Eqs.~(\ref{eq5.70abc}) and
(\ref{eq5.77}).

The finite renormalized 3PN-accurate EOM obtained by using the general
(two-parameter) renormalization shift (\ref{shifteta})-(\ref{rhoa})
were shown to be \textit{equivalent} to the final (three-parameter) EOM
of \cite{Blanchet:2000ub} if and only if the Hadamard-undetermined
dimensionless parameter $\lambda$ which entered the latter equations
takes the unique value $\lambda= - \frac{1987}{3080}$. This value is in
agreement with the result of a previous dimensional-regularization
determination of the Arnowitt-Deser-Misner Hamiltonian (in ADM-like
coordinates) \cite{Damour:2001bu}, which led to the unique
determination of the ADM analogue of $\lambda$, namely $\omega_s = 0$.
The value for $\lambda$ or $\omega_s$ is also in agreement with the
recent work \cite{itoh1,itoh2} which derived the 3PN equations of
motion in harmonic gauge using a surface-integral approach. Our result
provides an important check of the consistency of dimensional
regularization because our calculations are very different from the
ones of \cite{Damour:2001bu}, notably we use a different coordinate
system and a different method for iterating Einstein's field equations.
However, the applicability of our general approach to higher
post-Newtonian orders remains unexplored.

Finally, the present work opens the way to a
dimensional-regularization determination of the several unknown
dimensionless parameters that were shown to enter the
Hadamard-regularization of the 3PN binary's energy flux (in harmonic
coordinates) \cite{Blanchet:2001aw,Blanchet:2001ax}. The completion of
the 3PN energy flux is urgent in view of its importance in determining
the gravitational waveforms emitted by inspiralling black hole
binaries, which are primary targets for the international network of
interferometric gravitational wave detectors LIGO/VIRGO/GEO.

\acknowledgments
Most of the algebraic calculations reported in this paper were done
with the help of the software \textsl{Mathematica}. T.D. would like
to thank the Kavli Institute for Theoretical Physics for hospitality
(under the partial support of the National Science Foundation Grant
No. PHY99-07949) while this work was completed.

\appendix
\section{The $d$-dimensional metric and geodesic equation}
\label{ExpFieldEq}
We give in this Appendix several expanded expressions which are too
lengthy to be included in the body of the article. The expanded form
of the metric~(\ref{metric}) is easier to compare with the literature,
and notably with Eqs.~(3.24) of Ref.~\cite{Blanchet:2000ub}:
{\allowdisplaybreaks
\begin{subequations}
\label{metricExp}
\begin{eqnarray}
g_{00}&=&-1 +\frac{2}{c^2}\, V
-\frac{2}{c^4}\left[V^2+2\left(\frac{d-3}{d-2}\right) K\right]
+\frac{8}{c^6}\left[\hat X +V_i V_i
+\frac{1}{6}\, V^3 +\left(\frac{d-3}{d-2}\right) V K\right]\nonumber\\
&&+\frac{32}{c^8}\left[
\hat T -\frac{1}{2}\, V \hat X
+\hat R_i V_i -\frac{1}{2}\, V V_i V_i
-\frac{1}{48}\, V^4 +\frac{1}{4}\left(\frac{d-3}{d-2}\right) K V^2
-\frac{1}{4}\left(\frac{d-3}{d-2}\right)^2K^2\right]\nonumber\\
&&+{\mathcal{O}}\left(\frac{1}{c^{10}}\right),
\label{g00Exp}\\
g_{0i}&=&-\frac{4}{c^3}\, V_i
-\frac{8}{c^5}\left[\hat R_i
-\frac{1}{2}\left(\frac{d-3}{d-2}\right)V V_i\right]
-\frac{16}{c^7}\Biggl[
\hat Y_i+\frac{1}{2}\, \hat W_{ij}V_j\nonumber\\
&&+\frac{1}{4}\left(1+\frac{1}{(d-2)^2}\right)V^2 V_i
-\frac{1}{2}\left(\frac{d-3}{d-2}\right) V \hat R_i
+\frac{1}{2}\left(\frac{d-3}{d-2}\right)^2 K V_i\Biggr]
+{\mathcal{O}}\left(\frac{1}{c^{9}}\right),
\label{g0iExp}\\
g_{ij}&=&\delta_{ij}\biggl\{
1 +\frac{2}{(d-2)c^2}\, V
+\frac{2}{(d-2)^2 c^4}\left[V^2 -2 (d-3) K\right]\nonumber\\
&&+\frac{8}{c^6}\left[
\frac{\hat X}{d-2}
+\frac{V_k V_k}{d-2}
+\frac{V^3}{6(d-2)^3}
-\frac{(d-3)}{(d-2)^3}\, V K\right]
\biggr\}\nonumber\\
&&+\frac{4}{c^4}\, \hat W_{ij}
+\frac{16}{c^6}\left[
\hat Z_{ij} +\frac{V\hat W_{ij}}{2(d-2)} -V_i V_j\right]
+{\mathcal{O}}\left(\frac{1}{c^8}\right).
\label{gijExp}
\end{eqnarray}
\end{subequations}}$\!\!$
The inverse metric is such that $g_{\mu\nu}g^{\nu\rho} =
\delta_\mu^\rho$ in $d+1$ space-time dimensions. In terms of the
modified Newtonian potential $\mathcal{V}$ defined in Eq.~(\ref{calV})
above, it reads:
\begin{subequations}
\label{invMetric}
\begin{eqnarray}
g^{00}&=&-e^{2\mathcal{V}/c^2}\left(1 -\frac{8 V_i V_i}{c^6}
-\frac{32 \hat R_i V_i}{c^8}\right)
+{\mathcal{O}}\left(\frac{1}{c^{10}}\right),
\label{gup00}\\
g^{0i}&=&-e^{\frac{(d-3)\mathcal{V}}{(d-2)c^2}}
\left\{\frac{4 V_i}{c^3}\left[
1+\frac{1}{2}\left(\frac{d-1}{d-2}\ \frac{V}{c^2}\right)^2\right]
+\frac{8\hat R_i}{c^5}
+\frac{16}{c^7}\left[\hat Y_i
-\frac{1}{2}\hat W_{ij}V_j\right]\right\}
+{\mathcal{O}}\left(\frac{1}{c^9}\right),\nonumber\\
\label{gup0i}\\
g^{ij}&=&e^{-\frac{2\mathcal{V}}{(d-2)c^2}}\left\{\delta_{ij}
-\frac{4}{c^4} \hat W_{ij}
-\frac{16}{c^6} \left[\hat Z_{ij}
+\frac{1}{2(d-2)}\, \delta_{ij} V_k V_k
\right]\right\}
+{\mathcal{O}}\left(\frac{1}{c^8}\right).
\label{gupij}
\end{eqnarray}
\end{subequations}
Note the change of signs in the exponentials [with respect to the
covariant metric~(\ref{metric})], in front of $\hat W_{ij}V_j$ in
Eq.~(\ref{gup0i}), as well as for the ${\mathcal{O}}(1/c^4)$ and
${\mathcal{O}}(1/c^6)$ terms in Eq.~(\ref{gupij}). Note also that the
$V_iV_j$ contribution to $g_{ij}$ has disappeared in the inverse
spatial metric $g^{ij}$. The full post-Newtonian expansion of this
inverse metric reads: {\allowdisplaybreaks
\begin{subequations}
\label{invMetricExp}
\begin{eqnarray}
g^{00}&=&-1 -\frac{2}{c^2}\, V
-\frac{2}{c^4}\left[V^2-2\left(\frac{d-3}{d-2}\right) K\right]
-\frac{8}{c^6}\left[\hat X -V_i V_i
+\frac{V^3}{6} -\left(\frac{d-3}{d-2}\right) V K\right]\nonumber\\
&&-\frac{32}{c^8}\left[
\hat T +\frac{1}{2}\, V \hat X
-\hat R_i V_i -\frac{1}{2}\, V V_i V_i
+\frac{V^4}{48} -\frac{1}{4}\left(\frac{d-3}{d-2}\right) K V^2
+\frac{1}{4}\left(\frac{d-3}{d-2}\right)^2K^2\right]\nonumber\\
&&+{\mathcal{O}}\left(\frac{1}{c^{10}}\right),
\label{gup00Exp}\\
g^{0i}&=&-\frac{4}{c^3}\, V_i
-\frac{8}{c^5}\left[\hat R_i
+\frac{1}{2}\left(\frac{d-3}{d-2}\right)V V_i\right]
-\frac{16}{c^7}\Biggl[
\hat Y_i-\frac{1}{2}\, \hat W_{ij}V_j\nonumber\\
&&+\frac{1}{4}\left(1+\frac{1}{(d-2)^2}\right)V^2 V_i
+\frac{1}{2}\left(\frac{d-3}{d-2}\right) V \hat R_i
-\frac{1}{2}\left(\frac{d-3}{d-2}\right)^2 K V_i\Biggr]
+{\mathcal{O}}\left(\frac{1}{c^{9}}\right),
\label{gup0iExp}\\
g^{ij}&=&\delta_{ij}\biggl\{
1 -\frac{2}{(d-2)}\, \frac{V}{c^2}
+\frac{2}{(d-2)^2 c^4}\left[V^2 +2 (d-3) K\right]\nonumber\\
&&-\frac{8}{c^6}\left[
\frac{\hat X}{d-2}
+\frac{V_k V_k}{d-2}
+\frac{V^3}{6(d-2)^3}
+\frac{d-3}{(d-2)^3}\, V K\right]
\biggr\}\nonumber\\
&&-\frac{4}{c^4}\, \hat W_{ij}
-\frac{16}{c^6}\left[
\hat Z_{ij} -\frac{1}{2(d-2)}\, V\hat W_{ij}\right]
+{\mathcal{O}}\left(\frac{1}{c^8}\right).
\label{gupijExp}
\end{eqnarray}
\end{subequations}}$\!\!$
The determinant $g \equiv \det g_{\mu\nu}$ of the metric is a useful
quantity, notably to compute the ``gothic'' metric
$\mathfrak{g}^{\mu\nu} \equiv \sqrt{-g}\, g^{\mu\nu}$, which is the
natural variable when using the harmonic-coordinate system. The
simplest way to compute it is to use the exponential
form~(\ref{metric}) of the metric, and to perform a cofactor expansion
across both the first line and the first column:
\begin{equation}
\det g_{\mu\nu} = g_{00} \det g_{ij} -\sum_{k=1}^d\sum_{l=1}^d
(-)^{k+l} g_{0k}\, g_{0l} \det(g_{i\neq k~j\neq l})\,.
\label{defDet}
\end{equation}
Since $g_{ij} = \exp\left[\frac{2\mathcal{V}}{(d-2)c^2}\right]
\times\left[\delta_{ij} +{\mathcal{O}}(1/c^4)\right]$, the determinant
of the $(d-1)\times(d-1)$ matrix $g_{i\neq k~j\neq l}$ reads
\begin{eqnarray}
\det(g_{i\neq k~j\neq l})&=&e^\frac{2(d-1)\mathcal{V}}{(d-2)c^2}
\det(\delta_{i\neq k~j\neq l}) +{\mathcal{O}}(1/c^4) \nonumber\\
&=&e^\frac{2(d-1)\mathcal{V}}{(d-2)c^2} \delta_{kl}
+{\mathcal{O}}(1/c^4)\,.
\label{detSubMetric}
\end{eqnarray}
Therefore, the determinant of the full metric is given by
\begin{equation}
g\equiv \det g_{\mu\nu} = g_{00} \det g_{ij}
-e^\frac{2(d-1)\mathcal{V}}{(d-2)c^2} (g_{0i})^2
+{\mathcal{O}}\left(\frac{1}{c^{10}}\right),
\label{fullDet}
\end{equation}
where we have used the fact that $g_{0i} = {\mathcal{O}}(1/c^3)$. Note
that this formula suffices to compute $g$ up to order
${\mathcal{O}}(1/c^8)$ included if one knows the spatial metric
$g_{ij}$ up to this same order. At the 3PN order, we get easily
\begin{eqnarray}
g&=&-e^\frac{4\mathcal{V}}{(d-2)c^2}\biggl\{\left(1 -\frac{8}{c^6}\,
V_i V_i\right)
\det\left[
\delta_{ij} +\frac{4}{c^4}\, \hat W_{ij}
+\frac{16}{c^6}\left(\hat Z_{ij} -V_i V_j
+\frac{1}{2(d-2)}\, \delta_{ij} V_k V_k\right)
\right]\nonumber\\
&&\hphantom{-e^\frac{4\mathcal{V}}{(d-2)c^2}\biggl\{}+
\frac{16}{c^6}\, V_i V_i
\biggr\}
+{\mathcal{O}}\left(\frac{1}{c^8}\right)\nonumber\\
&=&-e^\frac{4\mathcal{V}}{(d-2)c^2}
\left[1 +\frac{4}{c^4}\, \hat W
+\frac{16}{c^6}\left(\hat Z
+\frac{1}{d-2}\, V_i V_i\right)
\right]+{\mathcal{O}}\left(\frac{1}{c^8}\right),
\label{g}
\end{eqnarray}
where we used the expansion $\det(\openone +M) = 1 +\mathop{\text{Tr}}
M +{\mathcal{O}}(M^2)$, valid for any matrix $M$ whose entries are
small with respect to 1. We can now compute the square root of this
determinant, and give its full post-Newtonian expansion:
\begin{subequations}
\label{sqrtg}
\begin{eqnarray}
\sqrt{-g} &=&
e^\frac{2\mathcal{V}}{(d-2)c^2}
\left[1 +\frac{2}{c^4}\, \hat W
+\frac{8}{c^6}\left(\hat Z
+\frac{1}{d-2}\, V_i V_i\right)
\right]+{\mathcal{O}}\left(\frac{1}{c^8}\right)
\label{sqrtgCompact}\\
&=&1 +\frac{2}{(d-2)}\, \frac{V}{c^2}
+\frac{2}{c^4}\left[
\hat W +\frac{V^2}{(d-2)^2} -\frac{2(d-3)}{(d-2)^2}\, K
\right]
+\frac{8}{c^6}\biggl[
\hat Z
+\frac{V_i V_i}{d-2}\nonumber\\
&&+\frac{\hat X}{d-2}
+\frac{V\hat W}{2(d-2)}
+\frac{V^3}{6(d-2)^3}
-\frac{d-3}{(d-2)^3}\, V K
\biggr]
+{\mathcal{O}}\left(\frac{1}{c^8}\right).
\label{sqrtgExpanded}
\end{eqnarray}
\end{subequations}
The gothic metric $\mathfrak{g}^{\mu\nu} \equiv \sqrt{-g}\, g^{\mu\nu}$
can now be written easily by combining Eqs.~(\ref{invMetric})
or~(\ref{invMetricExp}) with~(\ref{sqrtgCompact})
or~(\ref{sqrtgExpanded}). We shall not display here the explicit
results, since they were not directly useful for the present article.
Let us however quote the expression of the Ricci tensor in terms of the
gothic metric, in $d+1$ space-time dimensions and in any gauge:
\begin{eqnarray}
2 R_{\mu\nu}&=&-\mathfrak{g}_{\mu\alpha}\,
{\mathfrak{g}^{\alpha\beta}}_{,\beta\nu}
-\mathfrak{g}_{\nu\alpha}\,
{\mathfrak{g}^{\alpha\beta}}_{,\beta\mu}
\nonumber\\
&&+\left(\mathfrak{g}_{\mu\alpha}\,
\mathfrak{g}_{\nu\beta}
-\frac{1}{d-1}\, \mathfrak{g}_{\mu\nu}\,
\mathfrak{g}_{\alpha\beta}
\right)
\left(\mathfrak{g}^{\gamma\delta}\,
{\mathfrak{g}^{\alpha\beta}}_{,\gamma\delta}
-\mathfrak{g}_{\gamma\delta}\,
\mathfrak{g}^{\varepsilon\zeta}\,
{\mathfrak{g}^{\alpha\gamma}}_{,\varepsilon}\,
{\mathfrak{g}^{\beta\delta}}_{,\zeta}
+{\mathfrak{g}^{\alpha\beta}}_{,\gamma}\,
{\mathfrak{g}^{\gamma\delta}}_{,\delta}
\right)\nonumber\\
&&-\frac{1}{2}\,
\mathfrak{g}_{\alpha\beta}\,
\mathfrak{g}_{\gamma\delta}
\left(
{\mathfrak{g}^{\alpha\gamma}}_{,\mu}\,
{\mathfrak{g}^{\beta\delta}}_{,\nu}
-\frac{1}{d-1}\,
{\mathfrak{g}^{\alpha\beta}}_{,\mu}\,
{\mathfrak{g}^{\gamma\delta}}_{,\nu}
\right)
-\mathfrak{g}_{\mu\alpha}\,
\mathfrak{g}_{\nu\beta}\,
{\mathfrak{g}^{\alpha\gamma}}_{,\delta}\,
{\mathfrak{g}^{\beta\delta}}_{,\gamma}
\nonumber\\
&&+\mathfrak{g}_{\alpha\beta}
\left(
\mathfrak{g}_{\mu\gamma}\,
{\mathfrak{g}^{\beta\delta}}_{,\nu}
+\mathfrak{g}_{\nu\gamma}\,
{\mathfrak{g}^{\beta\delta}}_{,\mu}
\right)
{\mathfrak{g}^{\alpha\gamma}}_{,\delta}\ .
\label{RicciGoth}
\end{eqnarray}
As usual, a comma denotes partial derivation, and
$\mathfrak{g}_{\mu\nu}\equiv g_{\mu\nu}/\sqrt{-g}$ is the inverse of
$\mathfrak{g}^{\mu\nu}$. In terms of the gothic metric, the harmonic
gauge condition~(\ref{harmGauge}) takes a particularly simple form:
\begin{equation}
{\mathfrak{g}^{\mu\alpha}}_{,\alpha}
= -\sqrt{-g}\, g^{\alpha\beta}\Gamma^\lambda_{\alpha\beta} = 0\,.
\label{harmGoth}
\end{equation}
This is the reason why this gothic metric can be useful to write the
field equations. Note that several terms of Eq.~(\ref{RicciGoth})
vanish in this gauge, namely the first two (involving second
derivatives) and those proportional to
${\mathfrak{g}^{\gamma\delta}}_{,\delta}$ in the second line.
Nevertheless, this expression for $R_{\mu\nu}$ is slightly more
complicated than the one we used in Section~\ref{FieldEq} above,
Eq.~(\ref{RicciHarm}), which does not depend explicitly on the spatial
dimension $d$. It should be noted that many equations given in the
book \cite{Schmutzer} are erroneous when $d\neq 3$ (\textit{i.e.},
when $d+1 = n \neq 4$, in this book's notation), including
Eq.~(I,~14,~30) in \cite{Schmutzer} which gives the Ricci tensor in
terms of the gothic metric.

Let us end this Appendix by displaying the full expansion of the
geodesic equation~(\ref{eqGeod}), or more precisely of the vectors
$P^i$ and $F^i$, quickly illustrated in Eqs.~(\ref{SmallExpPiFi}).
The following expressions are $d$-dimensional generalizations of
Eqs.~(3.35) of Ref.~\cite{Blanchet:2000ub}, and we keep the same
writing and order of the terms to ease the comparison. The ``linear
momentum'' $P^i$ reads
{\allowdisplaybreaks
\begin{eqnarray}
P^i&=&v^i
+\frac{1}{c^2}\left(
\frac{1}{2}\, v^2 v^i +\frac{d}{d-2}\, V v^i -4 V_i
\right)
+\frac{1}{c^4}\biggl[
\frac{3}{8}\, v^4 v^i
+\frac{3d-2}{2(d-2)}\, V v^2 v^i
-4 V_j v^i v^j\nonumber\\
&&-2 V_i v^2
+\frac{d^2}{2(d-2)^2}\, V^2 v^i
-\frac{4}{d-2}\, V V_i
+4 \hat W_{ij} v^j
-8 \hat R_i
-\frac{2 d (d-3)}{(d-2)^2}\, K v^i
\biggr]\nonumber\\
&&+\frac{1}{c^6}\biggl[
\frac{5}{16}\, v^6 v^i
+\frac{3(5 d -4)}{8(d-2)}\, V v^4 v^i
-\frac{3}{2}\, V_i v^4
-6 V_j v^i v^j v^2
+\frac{(3 d -2)^2}{4(d-2)^2}\, V^2 v^2 v^i\nonumber\\
&&+2\hat W_{ij} v^j v^2
+2\hat W_{jk} v^i v^j v^k
-\frac{2(2d-1)}{d-2}\, V V_i v^2
-\frac{4(2d-1)}{d-2}\, V V_j v^i v^j
-4 \hat R_i v^2\nonumber\\
&&-8 \hat R_j v^i v^j
+\frac{d^3}{6(d-2)^3}\, V^3 v^i
+\frac{4 d}{d-2}\, V_j V_j v^i
+\frac{4 d}{d-2}\, \hat W_{ij} V v^j
+\frac{4 d}{d-2}\, \hat X v^i\nonumber\\
&&+16 \hat Z_{ij} v^j
-2\,\frac{d(d-2)+2}{(d-2)^2}\, V^2 V_i
-8 \hat W_{ij} V^j
-\frac{8}{d-2} V \hat R_i
-16 \hat Y_i\nonumber\\
&&-\frac{(3d-2)(d-3)}{(d-2)^2}\, K v^2 v^i
-\frac{2 d^2(d-3)}{(d-2)^3}\, K V v^i
+\frac{8(d-3)}{(d-2)^2}\, K V_i
\biggr]
+{\mathcal{O}}\left(\frac{1}{c^8}\right).
\label{ExpPi}
\end{eqnarray}}$\!\!$
This $d$-dimensional expression actually allows us to understand
better some of the numerical coefficients found for $d=3$ in
Ref.~\cite{Blanchet:2000ub}. For instance, we find that a factor 33
comes from the expression $3(5 d -4)$, and that a factor 49 comes from
$(3 d -2)^2$. The full post-Newtonian expansion of the ``force'' $F^i$
is given by an even longer formula: {\allowdisplaybreaks
\begin{eqnarray}
F^i&=&\partial_i V
+\frac{1}{c^2}\left[
-V\partial_i V
+\frac{d}{2(d-2)}\, \partial_i V v^2
-4 \partial_i V_j v^j
-2\left(\frac{d-3}{d-2}\right)\partial_i K
\right]
\nonumber\\
&&+\frac{1}{c^4}\biggl[
\frac{3 d -2}{8(d-2)}\, \partial_i V v^4
-2 \partial_i V_j v^j v^2
+\frac{d^2}{2(d-2)^2}\, V\partial_i V v^2
+2 \partial_i\hat W_{jk} v^j v^k
\nonumber\\
&&-\frac{4}{d-2}\left(V_j\, \partial_i V v^j
+V\partial_i V_j v^j\right)
-8 \partial_i\hat R_j v^j
+\frac{1}{2}\, V^2 \partial_i V
+8 V_j\, \partial_i V_j
\nonumber\\
&&+4\partial_i\hat X
+2\left(\frac{d-3}{d-2}\right)\left(
K \partial_i V +V \partial_i K\right)
-\frac{d(d-3)}{(d-2)^2}\, \partial_i K v^2
\biggr]
\nonumber\\
&&+\frac{1}{c^6}\biggl[
\frac{1}{16}\left(\frac{5d-4}{d-2}\right)v^6\partial_i V
-\frac{3}{2}\, \partial_i V_j v^j v^4
+\frac{1}{8}\left(\frac{3d-2}{d-2}\right)^2 V\partial_i V v^4
+\partial_i\hat W_{jk} v^2 v^j v^k
\nonumber\\
&&-2\left(\frac{2d-1}{d-2}\right) V_j\, \partial_i V v^2 v^j
-2\left(\frac{2d-1}{d-2}\right) V \partial_i V_j v^2 v^j
-4 \partial_i \hat R_j v^2 v^j
\nonumber\\
&&+\frac{1}{4}\left(\frac{d}{d-2}\right)^3 V^2 \partial_i V v^2
+\frac{4d}{d-2}\, V_j \partial_i V_j v^2
+\frac{2d}{d-2}\, \hat W_{jk}\, \partial_i V v^j v^k
\nonumber\\
&&+\frac{2d}{d-2}\, V \partial_i \hat W_{jk} v^j v^k
+\frac{2d}{d-2}\, \partial_i\hat X v^2
+8\partial_i \hat Z_{jk} v^j v^k
-4\,\frac{d(d-2)+2}{(d-2)^2}\, V_j V \partial_i V v^j
\nonumber\\
&&-2\,\frac{d(d-2)+2}{(d-2)^2}\, V^2 \partial_i V_j v^j
-8 V_k \partial_i\hat W_{jk} v^j
-8 \hat W_{jk} \partial_i V_k v^j
-\frac{8}{d-2}\, \hat R_j \partial_i V v^j
\nonumber\\
&&-\frac{8}{d-2}\, V\partial_i\hat R_j v^j
-16 \partial_i \hat Y_j v^j
-\frac{1}{6}\, V^3 \partial_i V
-4 V_j V_j \partial_i V
+16\hat R_j \partial_i V_j
+16 V_j \partial_i\hat R_j
\nonumber\\
&&-8 V V_j \partial_i V_j
-4\hat X\partial_i V
-4 V \partial_i\hat X
+16\partial_i\hat T
-\frac{d^2(d-3)}{(d-2)^3}\, K \partial_i V v^2
-2\left(\frac{d-3}{d-2}\right) K V \partial_i V
\nonumber\\
&&+\frac{8(d-3)}{(d-2)^2}\, K \partial_i V_j v^j
-\frac{(3d-2)(d-3)}{4(d-2)^2}\, \partial_i K v^4
-\frac{d^2(d-3)}{(d-2)^3}\, \partial_i K V v^2
\nonumber\\
&&-\frac{d-3}{d-2}\, V^2\partial_i K
+\frac{8(d-3)}{(d-2)^2}\, \partial_i K V_j v^j
-4\left(\frac{d-3}{d-2}\right)^2 K \partial_i K
\biggr]
+{\mathcal{O}}\left(\frac{1}{c^8}\right).
\label{ExpFi}
\end{eqnarray}}$\!\!$

\section{Useful formulae in \lowercase{$d$} dimensions}
\label{Formulae}
This appendix is intended to provide a compendium of (mostly
well-known) formulae for working in a space with $d$ dimensions. As
usual, though we shall motivate some formulae below by writing some
intermediate expressions which make complete sense only when $d$ is a
strictly positive integer, our final formulae are to be interpreted, by
complex analytic continuation, for a general complex dimension,
$d\in\mathbb{C}$. Actually one of the main sources of the power of
dimensional regularization is its ability to prove many results by
invoking complex analytic continuation in $d$.

We discuss first the volume of the sphere having $d-1$ dimensions
(\textit{i.e.}, embedded into Euclidean $d$-dimensional space). We
separate out the infinitesimal volume element in $d$ dimensions into
radial and angular parts,
\begin{equation}
d^d\mathbf{x}=r^{d-1}dr\,d\Omega_{d-1}\,,
\label{ddx}
\end{equation}
where $r=\vert\mathbf{x}\vert$ denotes the radial variable
(\textit{i.e.}, the Euclidean norm of $\mathbf{x}\in\mathbb{R}^d)$ and
$d\Omega_{d-1}$ is the infinitesimal solid angle sustained by the unit
sphere with $d-1$ dimensional surface. To compute the volume of the
sphere, $\Omega_{d-1}=\int d\Omega_{d-1}$, one notices that the
following $d$-dimensional integral can be computed both in Cartesian
coordinates, where it reduces simply to a Gaussian integral, and also,
using (\ref{ddx}), in spherical coordinates:
\begin{equation}
\int d^d\mathbf{x}\,e^{-r^2}=\left(\int dx\,
e^{-x^2}\right)^d=\pi^{\frac{d}{2}}=\Omega_{d-1}\int_0^{+\infty}dr\,
r^{d-1}e^{-r^2}=
\frac{1}{2}\Omega_{d-1}\Gamma\left(\frac{d}{2}\right),
\label{intd}
\end{equation}
where $\Gamma$ in the last equation denotes the Eulerian
function. This leads to the well known result
\begin{equation}
\Omega_{d-1}=\frac{2\pi^{\frac{d}{2}}}{
\Gamma\left(\frac{d}{2}\right)}\,.
\label{Omegad1}
\end{equation}
For instance one recovers the standard results $\Omega_2=4\pi$ and
$\Omega_1=2\pi$, but also $\Omega_0=2$, which
can be interpreted by remarking that the sphere with 0 dimension is
actually made of two points. If we parametrize the sphere
$\Omega_{d-1}$ in $d-1$ dimensions by means of $d-1$ spherical
coordinates $\theta_{d-1}$, $\theta_{d-2}$, $\cdots$, which are such
that the sphere $\Omega_{d-2}$ in $d-2$ dimensions is then
parametrized by $\theta_{d-2}$, $\theta_{d-3}$, $\cdots$, and so on
for the lower-dimensional spheres, then we find that the differential
volume elements on each of the successive spheres obey the recursive
relation
\begin{equation}
d\Omega_{d-1}=\left(\sin\theta_{d-1}\right)^{d-2}
d\theta_{d-1}d\Omega_{d-2}\,.
\label{recursiondOmega}
\end{equation}
Note that this implies
\begin{equation}
\frac{\Omega_{d-1}}{\Omega_{d-2}}=\int_0^\pi
d\theta_{d-1}\left(\sin\theta_{d-1}\right)^{d-2}=
\int_{-1}^{+1}dx\left(1-x^2\right)^{\frac{d-3}{2}}\,,
\label{ratioOmega}
\end{equation}
which can also be checked directly by using the explicit expression
(\ref{Omegad1}).

Next we consider the Dirac delta-function $\delta^{(d)}(\mathbf{x})$
in $d$ dimensions, which is formally defined, as in ordinary
distribution theory \cite{Schwartz}, by the following linear form
acting on the set $\mathcal{D}$ of smooth functions $\in
C^\infty(\mathbb{R}^d)$ with compact support:
$\forall\varphi\in\mathcal{D}$,
\begin{equation}
<\delta^{(d)},\varphi>\equiv\int
d^d\mathbf{x}\,\delta^{(d)}(\mathbf{x})\varphi(
\mathbf{x})=\varphi(\mathbf{0})\,,
\label{DiracBetter}
\end{equation}
where the brackets refer to the action of a distribution on
$\varphi\in\mathcal{D}$. Let us now check that the function
defined by
\begin{subequations}\begin{eqnarray}
u&=&\tilde{k}\,r^{2-d}\,,\\
\tilde{k}&=&\frac{\Gamma
\left(\frac{d-2}{2}\right)}{\pi^{\frac{d-2}{2}}}\,,
\label{u}
\end{eqnarray}\end{subequations}
[where $r$ is the radial coordinate in $d$ dimensions, such that
$r^2=\sum_{i=1}^d (x^i)^2$] is the ``Green's function'' of the
Poisson operator, namely that it obeys the distributional equation
\begin{equation}
\Delta u=-4\pi\delta^{(d)}(\mathbf{x})\,.
\label{Deltau}
\end{equation}
For any $\alpha\in\mathbb{C}$ we have $\Delta
r^\alpha=\alpha(\alpha+d-2)\,r^{\alpha-2}$, thus we see that $\Delta
u=0$ in the sense of functions. Let us formally compute its value in
the sense of distributions in $\mathbf{x}$-space. [The usual
verification of (\ref{Deltau}) is done in Fourier space.] We apply the
distribution $\Delta u$ on some test function $\varphi\in\mathcal{D}$,
use the definition of the distributional derivative to shift the
Laplace operator from $u$ to $\varphi$, compute the value of the
$d$-dimensional integral by removing a ball of small radius $s$
surrounding the origin [say $B(s)$], apply the fact that $\Delta u=0$
in the exterior of $B(s)$, use the Gauss theorem to transform the
result into a surface integral, and finally compute that integral by
inserting the Taylor expansion of $\varphi$ around the origin. The
proof of Eq.~(\ref{Deltau}) is thus summarized in the following steps:
\begin{eqnarray}
<\Delta u,\varphi>&=&<u,\Delta\varphi>\nonumber\\
&=&\lim_{s\rightarrow
0}\int_{\mathbb{R}^d\setminus
B(s)}d^d\mathbf{x}\,u\Delta\varphi\nonumber\\
&=&\lim_{s\rightarrow
0}\int_{\mathbb{R}^d\setminus
B(s)}d^d\mathbf{x}\,\partial_i\left[u\partial_i\varphi-\partial_iu
\,\varphi\right] \nonumber\\
&=&\lim_{s\rightarrow 0}\int
s^{d-1}d\Omega_{d-1}(-n_i)
\left[u\partial_i\varphi-\partial_iu\,\varphi\right] \nonumber\\
&=&\lim_{s\rightarrow 0}\int
s^{d-1}d\Omega_{d-1}(-n_i)\left[-\tilde{k}\,(2-d)s^{1-d}n_i\varphi(
\mathbf{0})\right] \nonumber\\
&=&\Omega_{d-1}\tilde{k}\,(2-d)\varphi(\mathbf{0}) \nonumber\\
&=&-4\pi\varphi(\mathbf{0})\,.
\label{Deltauphi}
\end{eqnarray}
In the last step we used the relation between $\tilde{k}$ and the
volume of the sphere, which is
\begin{equation}
\tilde{k}\,\Omega_{d-1}=\frac{4\pi}{d-2}\,.
\label{ktildeBis}
\end{equation}
{}From $u=\tilde{k}\,r^{2-d}$ one can next find the solution $v$
satisfying the equation $\Delta v=u$ (in a distributional sense),
namely
\begin{equation}
v=\frac{\tilde{k}\,r^{4-d}}{2(4-d)}\,.
\label{v}
\end{equation}
{}From (\ref{v}) we can then define a whole ``hierarchy'' of
higher-order functions $w$, $\cdots$ satisfying the Poisson equations
$\Delta w=v$, $\cdots$ in a distributional sense.

However, the latter hierarchy of functions $u$, $v$, $\cdots$ is better
displayed using some different, more systematic notation. This leads to
the famous Riesz kernels, here denoted $\delta_\alpha^{(d)}$, in
$d$-dimensional Euclidean space \cite{Riesz}. [These Euclidean kernels
differ from the Minkowski kernels $Z_A^{(d)}$, also introduced by
Riesz, and alluded to in the Introduction.] These kernels depend on a
complex parameter $\alpha\in\mathbb{C}$. They are defined by
\begin{subequations}
\begin{eqnarray}
\delta^{(d)}_\alpha\left(\mathbf{x}\right)&=&K_\alpha\,
r^{\alpha-d}\,,\\
K_\alpha&=&\frac{\Gamma\left(\frac{d-\alpha}{2}\right)}
{2^\alpha\pi^{\frac{d}{2}}\Gamma\left(\frac{\alpha}{2}\right)}\,.
\label{deltaalpha}
\end{eqnarray}\end{subequations}
For any $\alpha\in\mathbb{C}$, and also for any $d\in\mathbb{C}$, the
Riesz kernels satisfy the recursive relations
\begin{equation}
\Delta\delta^{(d)}_{\alpha+2}=-\delta^{(d)}_\alpha\,.
\label{recursiondeltaalpha}
\end{equation}
Furthermore, they obey also an interesting convolution relation, which
reads simply, with the chosen normalization of the coefficients
$K_\alpha$, as
\begin{equation}
\delta_\alpha^{(d)}\ast\delta_\beta^{(d)}
=\delta_{\alpha+\beta}^{(d)}\,.
\label{convolutiondeltaalpha}
\end{equation}
When $\alpha=0$ we recover the Dirac distribution in $d$ dimensions,
$\delta_0^{(d)}=K_0\,r^{-d}=\delta^{(d)}$ (the coefficient vanishes in
this case, $K_0=0$), and we have $u=4\pi\,\delta_2^{(d)}$,
$v=-4\pi\,\delta_4^{(d)}$, $\cdots$.

The convolution relation (\ref{convolutiondeltaalpha}) is nothing but
an elegant formulation of the Riesz formula in $d$ dimensions. To
check it let us consider the Fourier transform of $r^\alpha$ in $d$
dimensions,
\begin{equation}
\widetilde{f}_\alpha(\mathbf{k})\equiv\int
d^d\mathbf{x}\,|\mathbf{x}|^\alpha e^{-i\mathbf{k}.\mathbf{x}}\,.
\end{equation}
Using (\ref{ddx}) we can rewrite it as
\begin{equation}
\widetilde{f}_\alpha(\mathbf{k})=\int_0^{+\infty} dr
r^{\alpha+d-1}\int d\Omega_{d-1}e^{-i\mathbf{k}.\mathbf{x}}\,,
\label{radialang}\end{equation}
in which the angular integration can be performed as an application of
Eq.~(\ref{recursiondOmega}). This yields an expression depending on
the usual Bessel function,\footnote{We adopt for the Bessel function
the defining expression
$$J_\nu
(z)=\frac{\left(\frac{z}{2}\right)^\nu}{\Gamma\left(\nu+\frac{1}{2}
\right)\Gamma\left(\frac{1}{2}\right)}\int_{-1}^1
dx\left(1-x^2\right)^{\nu-\frac{1}{2}}e^{-i z x}\,.$$ To obtain
Eq.~(\ref{Fourier}) we employ the integration formula
$$\int_0^{+\infty}dz\, z^\mu J_\nu (z)=2^\mu\,\frac{
\Gamma\left(\frac{1+\mu+\nu}{2}
\right)}{\Gamma\left(\frac{1-\mu+\nu}{2}\right)}\,.$$}
\begin{equation}
\int d\Omega_{d-1}e^{-i\mathbf{k}.\mathbf{x}}=\Omega_{d-2}\int_0^\pi
d\theta_{d-1}\left(\sin\theta_{d-1}\right)^{d-2}e^{-i k\,r
\cos\theta_{d-1}}=\left(2\pi\right)^{\frac{d}{2}}
\left(k\,r\right)^{1-\frac{d}{2}}J_{\frac{d}{2}-1}(k\,r)\,,
\label{OmegaBessel}
\end{equation}
where $k\equiv\vert\mathbf{k}\vert$. The radial integration in
Eq.~(\ref{radialang}) is then readily done from using the previous
expression, and we obtain
\begin{equation}
\widetilde{f}_\alpha(\mathbf{k})=2^{\alpha+d}\,\pi^{
\frac{d}{2}}\,\frac{\Gamma\left(\frac{\alpha+d}{2}
\right)}{\Gamma\left(-\frac{\alpha}{2}\right)}\,k^{-\alpha-d}\,,
\label{Fourier}
\end{equation}
where the factor in front of the power $k^{-\alpha-d}$, say
$A_\alpha$, is checked from the Parseval theorem for the inverse
Fourier transform, which necessitates that $A_\alpha
A_{-\alpha-d}=(2\pi)^d$. Finally we can check the Riesz formula by
going to the Fourier domain, using the previous relations. The result,
\begin{equation}
\int d^d\mathbf{x}\,r_1^\alpha r_2^\beta=\pi^{\frac{d}{2}}
\frac{\Gamma\left(\frac{\alpha+d}{2}\right)
\Gamma\left(\frac{\beta+d}{2}\right)
\Gamma\left(-\frac{\alpha+\beta+d}{2}\right)}
{\Gamma\left(-\frac{\alpha}{2}\right)
\Gamma\left(-\frac{\beta}{2}\right)
\Gamma\left(\frac{\alpha+\beta+2d}{
2}\right)}\,r_{12}^{\alpha+\beta+d}\,,
\label{Riesz}
\end{equation}
is equivalent to Eq.~(\ref{convolutiondeltaalpha}).

A set of formulae concerning symmetric-trace-free (STF) multipole
expansions in $d$ dimensions is presented next, without proofs. We use
the multi-index notation $L=i_1\cdots i_\ell$; more generally the
notation is the same as in Appendix A of \cite{Blanchet:1986sp}. In
particular $\hat{n}_L$ denotes the STF projection of
$n_L=n_{i_1}\cdots n_{i_\ell}$, $[\frac{\ell}{2}]$ means the integer
part of $\frac{\ell}{2}$, $T_{\{i_1\cdots i_\ell\}}$ denotes the
(unnormalized, minimal) sum
of $T_{i_{\sigma (1)}\cdots i_{\sigma (\ell)}}$ where the $\sigma$'s
are permutations of the indices such that $T_{\{i_1\cdots i_\ell\}}$
is fully symmetric in $L$ (for convenience we do not normalize the
latter sum, for instance $\delta_{\{ij}n_{k\}}=\delta_{ij}n_{k}
+\delta_{ik}n_{j}+\delta_{jk}n_{i}$).
\begin{subequations}\begin{eqnarray}
n_L&=&\sum_{k=0}^{[\frac{\ell}{2}]}a_\ell^k\delta_{\{i_1i_2}\cdots
\delta_{i_{2k-1}i_{2k}}\hat{n}_{L-2K\}}\,,\\
\hat{n}_L&=&\sum_{k=0}^{[\frac{\ell}{2}]}
b_\ell^k\delta_{\{i_1i_2}\cdots
\delta_{i_{2k-1}i_{2k}}n_{L-2K\}}\,,
\label{nL}
\end{eqnarray}\end{subequations}
where the coefficients are
\begin{subequations}\begin{eqnarray}
a_\ell^k&=&\frac{1}{2^k}\frac{\Gamma\left(\frac{d}{2}+\ell-2k\right)}
{\Gamma\left(\frac{d}{2}+\ell-k\right)}\,,\\
b_\ell^k&=&\frac{(-)^k}{2^k}
\frac{\Gamma\left(\frac{d}{2}+\ell-k-1\right)}
{\Gamma\left(\frac{d}{2}+\ell-1\right)}\,.
\label{ablk}
\end{eqnarray}\end{subequations}
In particular (the brackets $\langle\,\rangle$ surrounding the indices
mean the STF projection)
\begin{subequations}\begin{eqnarray}
n_i\hat{n}_L&=&\hat{n}_{iL}+\frac{\ell}{d+2\ell-2} \delta_{i\langle
i_\ell}\hat{n}_{L-1\rangle}\,,\\
n_i\hat{n}_{iL}&=&\frac{d+\ell-2}{d+2\ell-2}\hat{n}_L\,.
\label{ninL}
\end{eqnarray}\end{subequations}
Spherical averages:
\begin{subequations}\begin{eqnarray}
\int \frac{d\Omega_{d-1}}{\Omega_{d-1}}n_{2P}&=&\frac{1}{2^p}
\frac{\Gamma\left(\frac{d}{2}\right)}
{\Gamma\left(\frac{d}{2}+p\right)}\delta_{\{i_1i_2}\cdots
\delta_{i_{2p-1}i_{2p}\}}\,,\\
\hat{f}_P\hat{g}_Q\int
\frac{d\Omega_{d-1}}{\Omega_{d-1}}n_{PQ}&=&\delta_{p,q}\frac{p!}{2^p}
\frac{\Gamma\left(\frac{d}{2}\right)}
{\Gamma\left(\frac{d}{2}+p\right)}\hat{f}_P\hat{g}_P\,.
\label{IntnPQ}
\end{eqnarray}\end{subequations}
STF decomposition of a scalar function:
\begin{subequations}\begin{eqnarray}
f\left(\mathbf{n}\right)&=&\sum_{\ell=0}^{+\infty}\hat{f}_L
\hat{n}_L\,,\\
\hat{f}_L&=&\frac{2^{\ell-1}\Gamma\left(\frac{d}{2}+\ell\right)}
{\ell!\pi^{\frac{d}{2}}}\int
d\Omega_{d-1}\hat{n}_Lf\left(\mathbf{n}\right).
\label{fL}
\end{eqnarray}\end{subequations}
Decomposition of a function $F(\mathbf{n}.\mathbf{n}')$ in terms of
Gegenbauer polynomials:\footnote{By definition, the Gegenbauer
polynomial $C_\ell^\gamma (x)$ is the coefficient of $\alpha^\ell$ in
the expansion
$$\left(1-2x\alpha+\alpha^2\right)^{-\gamma}=
\sum_{\ell=0}^{+\infty}C_\ell^\gamma (x)\alpha^\ell\,.$$ The
particular polynomial $P_\ell^{(d)}(x)\equiv
C_\ell^{\frac{d}{2}-1}(x)$ represents an appropriate generalization of
the Legendre polynomial in $d$ dimensions [indeed
$P_\ell^{(3)}(x)=P_\ell (x)$].}
\begin{subequations}\begin{eqnarray}
F(\mathbf{n}.\mathbf{n}')&=&\frac{\Gamma\left(d-2\right)}
{\Gamma\left(\frac{1}{2}\right)\Gamma\left(\frac{d-1}{2}\right)}
\sum_{\ell=0}^{+\infty}\frac{2^\ell\Gamma\left(\frac{d}{2}+\ell\right)}
{\Gamma\left(d+\ell-2\right)}
\hat{n}_L\hat{n}_L'\int_{-1}^{+1}dx\left(1-x^2\right)^{\frac{d-3}{2}}C_
\ell^{\frac{d}{2}-1}(x)F(x)\,,\nonumber\\
\\
\hat{n}_L\hat{n}_L'&=&\frac{\ell !}{2^\ell}
\frac{\Gamma\left(\frac{d}{2}-1\right)}
{\Gamma\left(\frac{d}{2}+\ell-1\right)}C_\ell^{\frac{d}{2}-1}
(\mathbf{n}.\mathbf{n}')\,.
\label{Fnnp}
\end{eqnarray}\end{subequations}
Integration formulae:
\begin{subequations}\begin{eqnarray}
\Delta^{-1} r^\alpha&=&\frac{r^{\alpha+2}}{(\alpha+2)(\alpha+d)}\,,\\
\Delta^{-n} r^\alpha&=&\frac{\Gamma\left(\frac{\alpha}{2}+1\right)
\Gamma\left(\frac{\alpha+d}{2}\right)}{\Gamma
\left(\frac{\alpha}{2}+n+1\right)
\Gamma\left(\frac{\alpha+d}{2}+n\right)}
\frac{r^{\alpha+2n}}{2^{2n}}\,,\\
\Delta^{-1}\left(\hat{n}_Lr^\alpha\right)&=&
\frac{\hat{n}_Lr^{\alpha+2}}{(\alpha-\ell+2)(\alpha+\ell+d)}\,.
\label{intform}
\end{eqnarray}\end{subequations}

\section{Explicit form of \lowercase{$\textsl{g}\equiv
\Delta^{-1}\Bigl(r_1^{2-d}\, r_2^{2-d}\Bigr)$} in
\lowercase{$d$} dimensions}
\label{Littleg}

\begin{figure}[!ht]
\includegraphics[scale=0.8]{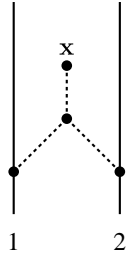}
\caption{Cubic-vertex diagram related to function $\textsl{g}\equiv
\Delta^{-1}\Bigl(r_1^{2-d}\, r_2^{2-d}\Bigr)$.}
\label{fig7}
\end{figure}

A very important technical fact which allowed one to compute
analytically the $d=3$ equations of motion is the possibility to
obtain explicitly the quadratically non linear potentials,
\textit{i.e.}, to evaluate in closed form the integrals appearing in
the PN expansion of the cubic-vertex diagram of Fig.~\ref{fig7}.

\noindent
At the lowest approximation in the $c^{-1}$ expansion, the diagram of
Fig.~\ref{fig7} leads, in $d=3$, to the integral
\begin{equation}
\label{eqD1}
\textsl{g}^{(d=3)} (\mathbf{x}) = \Delta^{-1} \left( \frac{1}{r_1 \,
r_2}\right),
\end{equation}
which was (probably) first evaluated by Fock in 1939 (``Sur le
mouvement des masses finies d'apr\`es la th\'eorie de la gravitation
einsteinienne'' \cite{Fock1939}), with the simple result
\begin{equation}
\label{eqD2}
\textsl{g}^{(d=3)} (\mathbf{x}) = \ln (r_1 + r_2 + r_{12})\,.
\end{equation}
Remembering that $r_1 \equiv \vert\mathbf{x} - \mathbf{y}_1 \vert$,
$r_2 \equiv \vert\mathbf{x} - \mathbf{y}_2 \vert$ and $r_{12} \equiv
\vert\mathbf{y}_1 - \mathbf{y}_2 \vert$ the combination $r_1 + r_2 +
r_{12}$ entering the logarithm in Eq.~(\ref{eqD2}) is simply seen to
be the perimeter of the triangle joining the three spatial points
$\mathbf{x}$, $\mathbf{y}_1$ and $ \mathbf{y}_2$ entering the
(Newtonian approximation of the) diagram of Fig.~\ref{fig7}. At the
${\mathcal O} (c^{-2})$ level of the PN expansion of Fig.~\ref{fig7},
there enter several new integrals which can be reduced to
\begin{equation}
\label{eqD3}
f^{(d=3)} = 2 \, \Delta^{-1} \, \textsl{g}^{(d=3)}\,,
\end{equation}
together with
\begin{equation}
\label{eqD4}
f_{12}^{(d=3)} = \Delta^{-1} \left( \frac{r_1}{r_2}
\right)~~\text{and}~~1 \leftrightarrow 2\,.
\end{equation}
The explicit evaluation of the integrals (\ref{eqD3}), (\ref{eqD4}) is
also possible, as was shown in
Refs.~\cite{Blanchet:1995fg},\cite{Blanchet:2000ub} (drawing on earlier
works \cite{schafer1987,Damour:1991ji}). In this Appendix we shall
explicitly evaluate the $d$-dimensional generalization of (\ref{eqD1}).
It will be clear, however, that our method can be rather
straightforwardly generalized to the ${\mathcal O} (c^{-2})$ diagrams
contained in Fig.~\ref{fig7}, \textit{i.e.}, to the $d$-dimensional
generalizations of (\ref{eqD3})-(\ref{eqD4}).

For our present purpose it will be more convenient not to include the
two factors of $\tilde k$ that accompany the two propagators issued
{}from 1 and 2 in Fig.~\ref{fig7}. We shall therefore define
\begin{equation}
\label{eqD5}
\textsl{g}(\mathbf{x}) \equiv \Delta^{-1} (r_1^{2-d} \, r_2^{2-d})\,.
\end{equation}
The method we present here consists of four basic steps: (i) expand the
integrand in series and construct a corresponding series for a
\textit{particular} solution $\textsl{g}_\text{part} = \Delta^{-1}\,
(r_1^{2-d} \, r_2^{2-d})_\text{part}$, (ii) resum the series to get an
explicit line-integral form of $\textsl{g}_\text{part}$, (iii) compute
$\Delta \, \textsl{g}_\text{part}$ in a distributional sense to
discover that it satisfies $\Delta \, \textsl{g}_\text{part} =
r_1^{2-d} \, r_2^{2-d} + S$ where $S$ is a distributional source
(localized along a line), and finally (iv) subtract $\Delta^{-1} S$
(which is given by another line-integral) from $\textsl{g}_\text{part}$
to get $\textsl{g}$ as a sum of line-integrals (which are expressible
in terms of one special function of one argument). What is crucial in
the argument is the uniqueness of the global solution (decaying at
infinity) of any (distributional) Poisson equation $\Delta
\, \varphi = \sigma$ when the (distributional) source decays fast
enough (or, at least, does not grow too fast)
at infinity. In our case, the sources $\sigma$ involved will
have fast-enough decay at infinity if we analytically continue $d$
toward large enough real parts (say $\Re\, [d] > 3$).

\begin{figure}[!ht]
\includegraphics[scale=0.8]{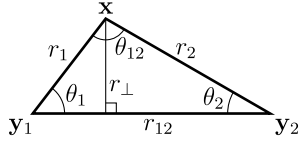}
\caption{Notation for various angles and distances, $\mathbf{y}_1$
and $\mathbf{y}_2$ denoting the positions of the two particles, and
$\mathbf{x}$ the field point.}
\label{fig8}
\end{figure}

There are several ways of implementing our method. For instance, we
could start by expanding $r_2^{2-d}$ in the source of (\ref{eqD5}) in
powers of $r_1$, such an expansion being valid only in a neighborhood
of $\mathbf{y}_1$. Namely, we have the $d$-dimensional generalization
of the familiar $d=3$ Legendre-polynomial expansion of $\vert
\mathbf{x} - \mathbf{y}_2 \vert^{-1}$ near $\mathbf{x} = \mathbf{y}_1$
(more precisely in the ball $r_1 < r_{12}$)
\begin{equation}
\label{eqD6}
r_2^{2-d} = r_{12}^{2-d} \sum_{\ell \geq 0} \left( \frac{r_1}{r_{12}}
\right)^{\ell} P_{\ell}^{(d)} (c_1)\,.
\end{equation}
Here, we denoted for visual clarity $P_{\ell}^{(d)} (x) \equiv
C_{\ell}^{\frac{d}{2}-1} (x)$, where $C_{\ell}^{\gamma} (x)$ is a
Gegenbauer polynomial such that $C_{\ell}^{1/2} (x) = P_{\ell}^{(3)}
(x)$ is the usual $d=3$ Legendre polynomial [see also Appendix
\ref{Formulae} above]. The quantity $c_1$ in (\ref{eqD6}) denotes the
cosine of the angle $\theta_1$ between $\mathbf{x} - \mathbf{y}_1$ and
$\mathbf{y}_2 - \mathbf{y}_1$. The notation we shall use is summarized
in Fig.~\ref{fig8}.

\noindent
When inserting the \textit{local} expansion (\ref{eqD6}) into the
source of (\ref{eqD5}) we are led to solving (locally) an equation of
the form $\Delta \, \textsl{g}_\text{loc} = \sum a_{\ell} \, r_1^{\ell
+ 2 - d}\, P_{\ell}^{(d)} (c_1)$. However, using the general formula
\begin{equation}
\label{eqD7}
\Delta (r^{\lambda} \, \hat n_L) = (\lambda - \ell) (\lambda + d-2 +
\ell) \, r^{\lambda - 2} \, \hat n_L\,,
\end{equation}
we know a particular solution of $\Delta \, \varphi = r^{\lambda} \,
\hat n_L$, namely
\begin{equation}
\label{eqD8}
\Delta^{-1} (r^{\lambda} \, \hat n_L) = \frac{r^{\lambda + 2} \, \hat
n_L}{(\lambda + 2 - \ell) (\lambda + d + \ell)}\,.
\end{equation}
The formulae (\ref{eqD7})-(\ref{eqD8}) apply to any source with fixed
multipolarity $(\ell)$ and a power law dependence on a radius. In
particular, they apply when $r \rightarrow r_1$, $\lambda \rightarrow
\ell + 2 -d$ and $\hat n_L \equiv n_1^{\langle i_1 \ldots i_{\ell}
\rangle} \rightarrow P_{\ell}^{(d)} (c_1)$ (because a generalized
Legendre polynomial is just proportional to the contraction of an
STF-projected multi unit vector $\hat n^L$ onto a fixed ``$z$''
direction; see Appendix \ref{Formulae}). This leads to a corresponding
expansion of a \textit{local} solution $\textsl{g}_{\text{loc} \, 1}$
(near $\mathbf{y}_1$) of $\Delta \, \textsl{g}_{\text{loc} \, 1} =
r_1^{2-d} \, r_2^{2-d}$ of the form
\begin{equation}
\label{eqD9}
\textsl{g}_{\text{loc} \, 1} = \frac{r_{12}^{2-d} \, r_1^{3-d}}{2
(4-d)} \sum_{\ell \geq 0} \frac{1}{\ell + 1} \, \frac{r_1^{\ell +
1}}{r_{12}^{\ell}} \, P_{\ell}^{(d)} (c_1)\,.
\end{equation}
In order to proceed further, we now need to resum the expansion
(\ref{eqD9}). This is done by a trick introduced, in a similar context
of resummation of multipolar expansions containing extra
$\ell$-dependent denominators, by Ref.~\cite{Damour:1993qi}. One
introduces some radial-integration operators $R_\alpha \, [\phi] \,
(\mathbf{r}) = \int_0^1 d\lambda \, \lambda^{\alpha} \, \phi (\lambda
\, \mathbf{r})$. For instance, in the context of (\ref{eqD9}), one
replaces $r_1^{\ell} / (\ell + 1)$ by $R_0 [r_1^{\ell}] = \int_0^1
d\lambda \, (\lambda \, r_1)^{\ell}$ or equivalently $r_1^{\ell + 1} /
(\ell + 1)$ by $\int_0^{r_1} d\ell_1 \, \ell_1^{\ell}$. This transforms
back the multipolar series appearing in (\ref{eqD9}) into the original
``Legendre'' series entering Eq.~(\ref{eqD6}). This allows one to write
$\textsl{g}_{\text{loc} \, 1}$ as a simple line-integral:
\begin{eqnarray}
\label{eqD10}
\textsl{g}_{\text{loc} \, 1} &= &\frac{r_1^{3-d}}{2(4-d)} \int_0^{r_1}
d\ell_1\, \vert\mathbf{y}_{\ell_1} - \mathbf{y}_2 \vert^{2-d} \nonumber
\\
&= &\frac{r_1^{4-d}}{2(4-d)} \int_0^1 d\alpha \, \vert
\mathbf{y}_{\alpha} - \mathbf{y}_2 \vert^{2-d}\,.
\end{eqnarray}
Here, $\mathbf{y}_{\ell_1}$ is a point on the segment joining
$\mathbf{y}_1$ to $\mathbf{x}$, located a distance $\ell_1$ away from
$\mathbf{y}_1$. It is more convenient to replace the line-integration
over the dimensionful length $\ell_1$ ($0 \leq \ell_1 \leq r_1$) into
an integration over the dimensionless parameter $\alpha \equiv \ell_1 /
r_1$ ($0 \leq \alpha \leq 1$). This leads to the explicit expression
\begin{equation}
\label{eqD11}
\mathbf{y}_{\ell_1} \equiv \mathbf{y}_{\alpha} = \alpha \, \mathbf{x}
+ (1-\alpha) \, \mathbf{y}_1\,.
\end{equation}
The resummed line-integral expression (\ref{eqD10}) allows one to
define everywhere $\textsl{g}_{\text{loc} \, 1}$,
\textit{including in the domain} $r_1 > r_{12}$ where the original
series (\ref{eqD9}) was \textit{not} convergent. Having in hands such a
global definition of $\textsl{g}_{\text{loc} \, 1}$ then allows one to
compute its Laplacian, \textit{in the sense of distributions}, and to
see how it differs from $r_1^{2-d} \, r_2^{2-d}$. The calculation of
$\Delta \, \textsl{g}_{\text{loc} \, 1}$ is done by techniques similar
to the ones used in Ref.~\cite{Damour:1991ji}. One needs to rewrite
some terms in the form of $\alpha$-derivatives. For instance, several
of the terms appearing in $\Delta \, \textsl{g}_{\text{loc} \, 1}$ can
be rewritten as the line-integral
\begin{eqnarray}
\label{eqD12}
&&\int_0^1 d\alpha \, r_1^{2-d} \biggl[ \vert\mathbf{y}_{\alpha} -
\mathbf{y}_2 \vert^{2-d} + \alpha \, \frac{\partial}{\partial \alpha}
\, \vert\mathbf{y}_{\alpha} - \mathbf{y}_2 \vert^{2-d} \biggr]
\nonumber \\
&&\qquad =\int_0^1 d\alpha \, \frac{\partial}{\partial
\alpha} \Bigl[ \alpha \, r_1^{2-d} \, \vert\mathbf{y}_{\alpha} -
\mathbf{y}_2 \vert^{2-d} \Bigr] = r_1^{2-d} \, r_2^{2-d}\,,
\end{eqnarray}
where the last line-integral gave only the end contribution $\alpha =
1$ corresponding to $\mathbf{y}_{\alpha} = \mathbf{x}$. Besides the
terms yielding (\ref{eqD12}), \textit{i.e.}, the looked-for ``source''
of the complete $\textsl{g}$, the calculation of $\Delta \,
\textsl{g}_{\text{loc} \, 1}$ yields also the distributional source
(where $k
\equiv \tilde k / 4\pi$ entered through $\Delta \, r^{2-d} = -
\delta^{(d)} / k$)
\begin{equation}
\label{eqD13}
- \frac{r_1^{4-d}}{2(4-d) \, k} \int_0^1 d\alpha \, \alpha^2 \,
\delta^{(d)} \, (\mathbf{y}_{\alpha} - \mathbf{y}_2)\,.
\end{equation}
This is conveniently transformed by introducing $\beta \equiv 1 /
\alpha$ (with $1 \leq \beta \leq + \infty$) and
\begin{equation}
\label{eqD14}
\mathbf{y}_{\beta} \equiv (1-\beta) \, \mathbf{y}_1 + \beta \,
\mathbf{y}_2\,,
\end{equation}
which varies along a semi-infinite line going from $\mathbf{y}_2$ to
infinity along the direction $\mathbf{y}_2 - \mathbf{y}_1$,
\textit{i.e.}, \textit{away} from $\mathbf{y}_1$. This transformation
allows one to rewrite (\ref{eqD13}) in the more transparent form
\begin{equation}
\label{eqD15}
- \frac{r_{12}^{4-d}}{2(4-d) \, k} \int_1^{+\infty}d\beta \,
\delta^{(d)} \, (\mathbf{x} - \mathbf{y}_{\beta})\,.
\end{equation}

At this stage, we recognize in (\ref{eqD15}) a very simple source,
namely a \textit{uniform} distribution of ``mass'' along the half-line
along which $\beta$ runs. This allows one to easily compute the unique,
global (decaying at infinity) solution of the Poisson equation with
source (\ref{eqD15}) and to subtract it from $\textsl{g}_{\text{loc} \,
1}$ to get the unique, global $\textsl{g}$ in the form of two
line-integrals:
\begin{equation}
\label{eqD16}
\textsl{g} = \frac{r_1^{3-d}}{2(4-d)} \int_0^{r_1} d\ell_{\alpha} \,
\vert\mathbf{y}_{\alpha} - \mathbf{y}_2 \vert^{2-d} -
\frac{r_{12}^{3-d}}{2(4-d)} \int_{r_{12}}^{\infty} d\ell_{\beta} \,
\vert\mathbf{x} - \mathbf{y}_{\beta} \vert^{2-d}\,,
\end{equation}
where $d\ell_{\alpha} = \vert d \mathbf{y}_{\alpha} \vert = r_1 \,
d\alpha$ and $d\ell_{\beta} = \vert d \mathbf{y}_{\beta} \vert = r_{12}
\, d\beta$. In other words, (\ref{eqD16}) expresses $\textsl{g}$ as,
essentially, the difference between the Newtonian potentials generated
by two uniform line distributions: a segment joining $\mathbf{y}_1$ to
$\mathbf{x}$ and the half-line starting from $\mathbf{y}_2$ in the
direction away from $\mathbf{y}_1$. It is easily seen (modulo the
slight delicacy of the logarithmic divergence of the potential of a
semi-infinite line when $d \rightarrow 3$, \textit{i.e.}, the
occurrence of a $1/\varepsilon$ pole; see below) that the result
(\ref{eqD16}) yields, when $d \rightarrow 3$, the well-known result
(\ref{eqD2}). [Actually, this was the way one of us (T.D.) had derived
long ago for himself (\ref{eqD2}), unaware of its derivations in the
literature.]

The expression (\ref{eqD16}) has the advantage of being explicitly
regular (except at the point $\mathbf{x} = \mathbf{y}_1$) in the ball
$r_1 < r_{12}$. However, it has the default of treating
dissymmetrically the two points $\mathbf{y}_1$ and $\mathbf{y}_2$ (in
spite of the fact that the result (\ref{eqD16}) for $\textsl{g}$
\textit{is}, actually, symmetric under $1 \leftrightarrow 2$). One can
derive an exchange-symmetric expression for $\textsl{g}$ by modifying
the first step of our method. Instead of expanding the source
$r_1^{2-d} \, r_2^{2-d}$ in the neighborhood of $\mathbf{x} =
\mathbf{y}_1$, \textit{i.e.}, in a series of positive powers of $r_1$,
we can expand it in the \textit{neighborhood of infinity},
\textit{i.e.}, in a series of negative powers of $r_1$. Such an
expansion is directly related to the expansions used in
\cite{Damour:1991ji}, which led to the decomposition of
$\textsl{g}^{(d=3)}$ in two pieces denoted $k$ and $h$ there, where the
source of $h$ was a uniform mass distribution along the segment joining
$\mathbf{y}_1$ and $\mathbf{y}_2$. Let us briefly indicate the
successive steps of this new calculation of $\textsl{g}$. Instead of
the ``local'' expansion (\ref{eqD6}) (valid for $r_1 < r_{12}$), one
expands $r_2^{2-d}$ near infinity ($r_1 > r_{12}$) as
\begin{equation}
\label{eqD17}
r_2^{2-d} = r_1^{2-d} \sum_{\ell \geq 0} \left( \frac{r_{12}}{r_1}
\right)^{\ell} P_{\ell}^{(d)} (c_1)\,.
\end{equation}
Solving term by term $\Delta \, \textsl{g}_{\infty} = r_1^{2-d} \,
r_2^{2-d}$ ``near infinity'' by means of (\ref{eqD8}), and transforming
away the appearing $\ell$-dependent denominators by means of
$\int_{r_1}^{\infty} d\ell_1 \, \ell_1^{2-d-\ell} = r_1^{3-d-\ell} /
(\ell + d-3)$, leads to the following resummation of
$\textsl{g}_{\infty}$:
\begin{equation}
\label{eqD18}
\textsl{g}_{\infty} = - \frac{r_1^{4-d}}{2(4-d)} \int_1^{\infty}
d\alpha \,
\vert\mathbf{y}_{\alpha} - \mathbf{y}_2 \vert^{2-d}\,.
\end{equation}
Here, $\mathbf{y}_{\alpha}$ is still defined by (\ref{eqD11}), but the
parameter $\alpha$ now varies in $1 \leq \alpha \leq + \infty$ so that
(\ref{eqD18}) is the potential of a semi-infinite line. Computing the
distributional Laplacian of the particular solution
$\textsl{g}_{\infty}$, Eq.~(\ref{eqD18}), leads to the presence,
besides the looked-for source $r_1^{2-d} \, r_2^{2-d}$, of an
additional distributional source localized now along the segment
joining $\mathbf{y}_1$ to
$\mathbf{y}_2$, namely
\begin{equation}
\label{eqD19}
\frac{r_{12}^{4-d}}{2(4-d) \, k} \int_0^1 d\beta \, \delta^{(d)} \,
(\mathbf{x} - \mathbf{y}_{\beta})\,,
\end{equation}
where $\beta = 1/\alpha$ varies between 0 and 1 and where
$\mathbf{y}_{\beta}$ is again defined by (\ref{eqD14}). It is then
easy to subtract from $\textsl{g}_{\infty}$ (which tends, in $d=3$, to
the function $k (\mathbf{x} , \mathbf{y}_1 , \mathbf{y}_2)$ of
\cite{Damour:1991ji}) the Poisson integral of the source (\ref{eqD19})
(which is a uniform distribution along the segment
$\mathbf{y}_1$-$\mathbf{y}_2$ and which tends, in $d=3$, to the
function $\frac{1}{2} h (\mathbf{x} , \mathbf{y}_1 , \mathbf{y}_2)$ of
\cite{Damour:1991ji}) to get the following alternative expression for
$\textsl{g}$,
\begin{equation}
\label{eqD20}
\textsl{g} = - \frac{r_1^{3-d}}{2(4-d)} \int_{r_1}^{\infty}
d\ell_{\alpha} \,
\vert\mathbf{y}_2 - \mathbf{y}_{\alpha} \vert^{2-d} +
\frac{r_{12}^{3-d}}{2(4-d)} \int_0^{r_{12}} d\ell_{\beta} \,
\vert\mathbf{x} - \mathbf{y}_{\beta} \vert^{2-d}\,,
\end{equation}
or, equivalently,
\begin{equation}
\label{eqD20'}
\textsl{g} = - \frac{r_1^{4-d}}{2(4-d)} \int_1^{\infty} d \, \alpha \,
\vert\mathbf{y}_2 - \mathbf{y}_{\alpha} \vert^{2-d} +
\frac{r_{12}^{4-d}}{2(4-d)} \int_0^1 d \, \beta \, \vert \mathbf{x} -
\mathbf{y}_{\beta} \vert^{2-d}\,.
\end{equation}
The form (\ref{eqD20})-(\ref{eqD20'}) still does not look quite
symmetric between 1 and 2 but a moment of reflection will show that it
is.

The two methods above have expressed $\textsl{g}$ in terms of the
Newtonian potentials generated by half-lines or segments,
\textit{i.e.}, integrals of the type $\int d\ell \, \vert\mathbf{x}' -
\mathbf{y}_{\ell} \vert^{2-d}$ where $\mathbf{y}_{\ell}$ varies along a
straight line (but where $\mathbf{x}'$ might be $\mathbf{x}$ or
$\mathbf{y}_2$). Clearly, any such potential can be reduced (through
linear decompositions) to the Newtonian potential generated by a
\textit{half-line}. Let us then consider a generic half-line starting
at the point $\mathbf{r}_0$ and going to infinity in the direction
$\mathbf{n}$, and let us consider the Newtonian potential generated by
this half-line at the origin of the coordinate system (not located on
the half-line). Let us denote $\mathbf{r}_{\ell} = \mathbf{r}_0 + \ell
\, \mathbf{n}$, $r_\ell = \vert \mathbf{r}_{\ell} \vert$, $\mathbf{r}_0
= r_0 \, \mathbf{n}_0$, $c = \mathbf{n}_0 \cdot \mathbf{n}$ (cosine of
the angle $\theta$ between the radius vector from the origin,
\textit{i.e.}, the ``field point'', towards the beginning point of the
half-line and the direction of the half-line, away from its beginning).
Then it is easy to find that
\begin{equation}
\label{eqD21}
\int_0^{\infty} d\ell \, r_{\ell}^{2-d} = \varphi (c) \, r_0^{3-d}\,,
\end{equation}
where the function $\varphi (c)$ is given by the integral
\begin{equation}
\label{eqD22}
\varphi (c) \equiv \int_0^{\infty} \frac{d \, \lambda}{(1 + 2 \, c \,
\lambda + \lambda^2)^{\frac{d-2}{2}}}\,.
\end{equation}
The integral (\ref{eqD22}) converges for $d > 3$, has a pole $\propto
1/(d-3)$ as $d \rightarrow 3$, and can be expressed in terms of
hypergeometric functions, \textit{e.g.} $F_{2,1} \left[ \frac{1}{2} ,
\frac{d-2}{2} ; \frac{3}{2} ; z \right]$. It is, however, simpler to
keep the form (\ref{eqD22}).\footnote{The multipolar expansion of the
function $\varphi (c)$ reads
$$\varphi (c) = \sum_{\ell \geq
0}(-)^\ell\frac{2\ell+d-2}{(\ell+1)(\ell+d-3)}P^{(d)}_\ell (c)\,.
$$
On this expression one sees clearly the occurrence of the simple pole
of $\varphi (c)$ when $\varepsilon\equiv d-3\rightarrow 0$, which is
given by the ``monopolar'' term $\ell=0$ as
$$
\varphi (c) =
\frac{P^{(3)}_0(c)}{\varepsilon}+\mathcal{O}(\varepsilon^0) =
\frac{1}{\varepsilon}+\mathcal{O}(\varepsilon^0)\,.
$$}

Finally, using the half-line potentials (\ref{eqD21}) as building
blocks one can write our result (\ref{eqD20}) in the final,
$1\leftrightarrow 2$ symmetric, form
\begin{eqnarray}
\label{eqD23}
\textsl{g} (\mathbf{x} , \mathbf{y}_1 , \mathbf{y}_2) &=
&\frac{r_{12}^{3-d}}{2(4-d)} \, \Bigl[2 \, r_{\perp}^{3-d} \, \varphi
(0) - r_1^{3-d} \, \varphi (c_1) - r_2^{3-d} \, \varphi (c_2)\Bigr]
\nonumber \\ &-& \frac{r_1^{3-d} \, r_2^{3-d}}{2(4-d)} \, \varphi
(c_{12})\,.
\end{eqnarray}
The quantities entering (\ref{eqD23}) are those defined in
Fig.~\ref{fig8}, notably $c_1 \equiv \cos \theta_1$, $c_2 \equiv \cos
\theta_2$, $c_{12} \equiv \cos \theta_{12}$, with $r_{\perp}$ being the
orthogonal distance between the field point $\mathbf{x}$ and the
segment joining $\mathbf{y}_1$ to $\mathbf{y}_2$ [with associated
argument $c_{\perp} = \cos \frac{\pi}{2} = 0$ in $\varphi (c)$]. Note
the following properties of the function $\varphi (c)$,
\begin{subequations}\begin{eqnarray}
\label{eqD24}
&&\varphi (c) + \varphi (-c) = \frac{2 \, \varphi
(0)}{(1-c^2)^{\frac{d-3}{2}}}\,,\\
\label{eqD25}
&&\varphi (0) = \frac{1}{2} \, \frac{\Gamma \left( \frac{1}{2} \right)
\, \Gamma \left( \frac{d-3}{2} \right)}{\Gamma \left( \frac{d-2}{2}
\right)}\,, \quad \varphi (1) = \frac{1}{d-3}\,.
\end{eqnarray}\end{subequations}
The simplest way to prove (\ref{eqD24}) is to notice that the Newtonian
potential of an \textit{infinite} line can be written either as twice
that of two half-lines beginning at the orthogonal projection of the
field point on the original line, so that
\begin{equation}
\label{eqD26}
\int_{-\infty}^{+\infty} \frac{d\ell}{r_{\ell}^{d-2}} = 2 \,
\frac{\varphi (0)}{r_{\perp}^{d-3}}\,,
\end{equation}
or as that of two other half-lines obtained by a more arbitrary cut
(under an angle $\theta \ne \pi / 2$ and $c = \cos \theta$).

We can verify the $d \rightarrow 3$ limit of Eqs.~(\ref{eqD16}) and
(\ref{eqD23}) by using the following $\varepsilon\rightarrow 0$
expansion of the elementary function $\varphi (c)$, namely
\begin{equation}
\varphi (c) =
\frac{1}{\varepsilon} - \ln\left(\frac{1+c}{2}\right) +
\mathcal{O}\left(\varepsilon\right)\,.
\label{phic}
\end{equation}
To obtain (\ref{phic}) we notice that the finite part of $\varphi (c)$
when $\varepsilon\rightarrow 0$, which is
\begin{equation}
\varphi_0 (c) \equiv \lim_{\varepsilon\rightarrow 0} \left[\varphi
(c)-\frac{1}{\varepsilon}\right]\, ,
\label{phi0c}
\end{equation}
can be re-expressed in the form of the following sum of two convergent
integrals,
\begin{eqnarray}
\varphi_0 (c) &=& \int_0^1\frac{d\,\lambda}{\sqrt{1+2\, c
\,\lambda+\lambda^2}}+\int_0^1\frac{d\,\lambda}{
\lambda}\left(\frac{1}{\sqrt{1+2\, c\,
\lambda+\lambda^2}}-1\right)\nonumber\\ &=& -
\ln\left(\frac{1+c}{2}\right)\,.
\label{phi0c'}
\end{eqnarray}
Combining the expansion (\ref{phic}) with the basic
relations\,\footnote{Denoting $S\equiv r_1+r_2+r_{12}$ we have
$$\frac{1+c_1}{S}=\frac{r_1+r_{12}-r_2}{2r_1r_{12}}\,,
\qquad\frac{1+c_2}{S}=\frac{r_2+r_{12}-r_1}{2r_2r_{12}}\,,
\qquad\frac{1+c_{12}}{S}=\frac{r_1+r_2-r_{12}}{2r_1r_2}\,.$$
The perpendicular distance $r_\perp$ is given by
$$r_{\perp}=\frac{r_1r_2}{S}\sqrt{2(1+c_1)(1+c_2)(1+c_{12})}\ .$$}
associated with the triangle of Fig.~\ref{fig8}, our $d$-dimensional
expressions (\ref{eqD16}) or (\ref{eqD23}) are found to admit the
expansion
\begin{equation}
\textsl{g} = -\frac{1}{2\varepsilon} - \frac{1}{2} +
\ln\left(\frac{r_1+r_2+r_{12}}{2}\right) +
\mathcal{O}\left(\varepsilon\right),
\end{equation}
which indeed reduces to the three-dimensional result (\ref{eqD2})
modulo an additive constant linked to the $1/\varepsilon$ pole.

Nice as it is to have in hand an analytic expression for the
$d$-dimensional basic non linear potential $\textsl{g}$, its practical
utility in explicit computations of the $d$-dimensional equations of
motion is not evident because, contrary to the 3-dimensional expression
(\ref{eqD2}), the expression is not \textit{explicitly} regular along
the $\mathbf{y}_1-\mathbf{y}_2$ segment. [The regularity of
Eq.~(\ref{eqD23}) as $r_{\perp} \rightarrow 0$ comes by compensations
between the three terms in the bracket, using (\ref{eqD24}).] It would
need some transforming [using (\ref{eqD24}),
and/or using the other expressions derived from the previous form
(\ref{eqD16}), which are regular along the $\mathbf{y}_1-\mathbf{y}_2$
segment, but singular somewhere else] to write an explicit expression
which is regular everywhere, except at the two isolated points
$\mathbf{y}_1$ and $\mathbf{y}_2$.

Finally, let us just mention that the method explained above can, in
principle, be straightforwardly generalized to the computation of the
higher post-Newtonian potentials contained in the diagram of
Fig.~\ref{fig7}. For instance, in computing the $d$-dimensional analog
of (\ref{eqD3}), say $f = 2 \, \Delta^{-1} \, \textsl{g} = 2 \,
\Delta^{-2} (r_1^{2-d} \, r_2^{2-d})$, it is easy [by iterating
(\ref{eqD8})] to get the analog of (\ref{eqD9}). Then a more
complicated radial-integration operator (see, \textit{e.g.},
\cite{Damour:1993qi}) will allow one to resum the series to get a
line-integral expression for $f_{\text{loc} \, 1}$ or $f_{\infty}$. We
anticipate that a somewhat more delicate application of either $\Delta$
(to go back to $\textsl{g}$) or $\Delta^2$ (to go back to $2 \,
r_1^{2-d} \, r_2^{2-d}$) will yield additional line-distributed
sources. It should then be a simple matter to compute the Poisson, or
iterated Poisson, integral of these line-distributed sources. We leave
an explicit study of these details to future work.

\end{document}